\documentclass[journal,9pt]{IEEEtran}
\usepackage{amsmath,amsfonts}
\usepackage{algorithmic}
\usepackage{algorithm}
\usepackage{array}
\usepackage[caption=false,font=footnotesize,labelfont=rm,textfont=rm]{subfig}
\usepackage{textcomp}
\usepackage{stfloats}
\usepackage{url}
\usepackage{verbatim}
\usepackage{graphicx}
\usepackage{multirow}
\usepackage{multicol}
\usepackage{subfig}
\usepackage{cite}
\usepackage{color}
\usepackage{cases}
\usepackage{makecell}
\usepackage{enumitem}
\usepackage{threeparttable} 
\usepackage{booktabs}        
\begin{document}

\title{Empirical Study on Near-Field and Spatial Non-Stationarity Modeling for THz XL-MIMO Channel in Indoor Scenario}


\author{Huixin Xu, Jianhua Zhang,~\IEEEmembership{Senior Member,~IEEE}, Pan Tang, Hongbo Xing, \\ Chong Han,~\IEEEmembership{Senior Member,~IEEE}, Lei Tian, Qixing Wang, and Guangyi Liu
\thanks{Manuscript received April 19, 2021; revised August 16, 2021.}
\thanks{This work was supported in part by the National Key R\&D Program of China under Grant 2023YFB2904805, in part by the National Natural Science Foundation of China under Grant 62201086, in part by the Beijing Natural Science Foundation under Grant L243002, and in part by the BUPT-CMCC Joint Innovation Center.}
\thanks{Huixin Xu, Jianhua Zhang, Pan Tang, Hongbo Xing, and Lei Tian are with the State Key Lab of Networking and Switching Technology, Beijing University of Posts and Telecommunications, Beijing 100876, China (e-mail: xuhuixin@bupt.edu.cn; jhzhang@bupt.edu.cn; tangpan27@bupt.edu.cn; hbxing@bupt.edu.cn; tianlbupt@bupt.edu.cn).}
\thanks{Chong Han is with the Terahertz Wireless Communications (TWC) Laboratory, also with the Department of Electronic Engineering and the Cooperative Medianet Innovation Center (CMIC), Shanghai Jiao Tong University, Shanghai 200240, China (e-mail: chong.han@sjtu.edu.cn).}
\thanks{Qixing Wang and Guangyi Liu are with the China Mobile Research Institute, Beijing 100053, China (e-mail: wangqixing@chinamobile.com; liuguangyi@chinamobile.com).}
}

\markboth{Journal of \LaTeX\ Class Files,~Vol.~xx, No.~xx, August~2021}%
{Shell \MakeLowercase{\textit{et al.}}: A Sample Article Using IEEEtran.cls for IEEE Journals}

\maketitle

\begin{abstract}
Terahertz (THz) extremely large-scale MIMO (XL-MIMO) is considered a key enabling technology for 6G and beyond due to its advantages such as wide bandwidth and high beam gain. As the frequency and array size increase, users are more likely to fall within the near-field (NF) region, where the far-field plane-wave assumption no longer holds. This also introduces spatial non-stationarity (SnS), as different antenna elements observe distinct multipath characteristics. Conventional far-field stationary models with fixed path parameters fail to capture these variations. Therefore, this paper proposes a THz XL-MIMO channel model that accounts for both NF propagation and SnS, validated using channel measurement data. In this work, we first conduct THz XL-MIMO channel measurements at 100 GHz and 132 GHz using 301- and 531-element ULAs in indoor environments, revealing pronounced NF effects characterized by nonlinear inter-element phase variations, as well as element-dependent delay and angle shifts. Moreover, the SnS phenomenon is observed, arising not only from blockage but also from inconsistent reflection or scattering. Based on these observations, a hybrid NF channel modeling approach combining the scatterer-excited point-source model and the specular reflection model is proposed to capture nonlinear phase variation of different types of non-line-of-sight (NLoS) paths. For SnS modeling, amplitude attenuation factors (AAFs) are introduced to characterize the continuous variation of path power across the array. By analyzing the statistical distribution and spatial autocorrelation properties of AAFs, a statistical rank-matching-based method is proposed for their generation. Finally, the model is validated using measured data. Evaluation across metrics such as entropy capacity, condition number, spatial correlation, channel gain, Rician K-factor, and RMS delay spread confirms that the proposed model closely aligns with measurements and effectively characterizes the essential features of THz XL-MIMO channels.

\end{abstract}

\begin{IEEEkeywords}
Terahertz (THz), extremely large-scale massive MIMO (XL-MIMO), channel measurement, near-field (NF), spatial non-stationarity (SnS).
\end{IEEEkeywords}

\section{Introduction}
\label{Sec_1}
Terahertz (THz) communication is regarded as one of the promising technologies to satisfy the high data rate requirements of 6G and beyond wireless systems \cite{C1_6G_survey}. Compared to the sub-6 GHz and millimeter-wave (mmWave) bands, the THz bands offer continuous communication bandwidths of tens of gigahertz or more \cite{C1_THz_Rappaport,C1_THz_Survey}. This provides the necessary spectral resources to support peak data rates in future wireless communication systems, reaching up to terabits per second (Tbps)  \cite{C1_THz_ITU_RR}. However, THz signal propagation faces significant challenges, such as severe free-space path loss and molecular absorption loss, leading to strong signal attenuation, vulnerability to obstructions, and limited coverage \cite{C1_6G}. Fortunately, the shorter wavelength of THz signals enables the construction of extremely large-scale massive MIMO (XL-MIMO) in confined spaces. The implementation of XL-MIMO to form directionally controllable, high-gain narrow beams can overcome the strong attenuation of THz signals, enhancing their coverage range, and consequently increasing system transmission rates and capacity. Therefore, the integration of THz and XL-MIMO technologies is considered a promising solution to achieve the high-speed and wide-coverage requirements of 6G and beyond.

It is well known that channel properties determine the ultimate performance limit of wireless communications \cite{Molisch_book}. Accurate and realistic models of THz XL-MIMO are essential for system evaluation, network deployment, and performance testing \cite{C1_XL_MIMO}. Two main challenges arise in modeling THz XL-MIMO channels. First, according to the Rayleigh distance criterion \cite{C1_Rayleigh}, the shorter wavelength and larger array aperture of THz XL-MIMO increase the likelihood of user equipment (UE) being located within the near-field (NF) region, rendering the traditional far-field (FF) plane-wave assumption invalid \cite{C1_HC_WC}. Second, the substantial spatial span of XL-MIMO leads to spatially diverse channel characteristics observed across different antenna elements, resulting in spatial non-stationarity (SnS) \cite{C1_VR_3, C1_ICC_XHX}. Consequently, NF propagation and SnS are critical challenges that must be considered in practical THz XL-MIMO channel modeling.
\vspace{-1em}
\subsection{Related Work}
Extensive channel measurements have been conducted in recent years to investigate the propagation characteristics of THz XL-MIMO channels. Initially, THz MIMO channel measurements at 140/300/460 GHz were performed with a small number of antenna elements \cite{C1_THz_MIMO_Mea_1,C1_THz_MIMO_Mea_3,C1_THz_MIMO_Mea_4}, comparing power delay profiles (PDP) of MIMO sub-channels, and analyzing parameters such as channel capacity and path loss. Subsequently, the number of antenna elements used in measurements increased to hundreds and thousands. For instance, extensive channel measurements at 100 GHz with 360-element MIMO array were conducted in an indoor hall, analyzing path loss, delay spread, angular spread, and the Rician K-factor \cite{C1_THz_MIMO_Mea_8}. In our preliminary works \cite{C1_THz_MIMO_Mea_6}, measurements using a 2400-element uniform circular array at 100 GHz revealed NF propagation and SnS effects. Additionally, NF measurements at 300 GHz using a 4096-element array in an anechoic chamber observed variations in NF reception power and nonlinear phase evolution \cite{C1_THz_MIMO_Mea_5}. These studies reveal that most THz MIMO/XL-MIMO channel measurement campaigns focus primarily on large-scale or small-scale channel parameters. However, research on NF propagation and SnS for THz XL-MIMO remains largely observational, lacking measurement-based channel models.

\graphicspath{{picture/}}
\begin{figure*}[htbp]  
    \centering  
    \includegraphics[width=17cm]{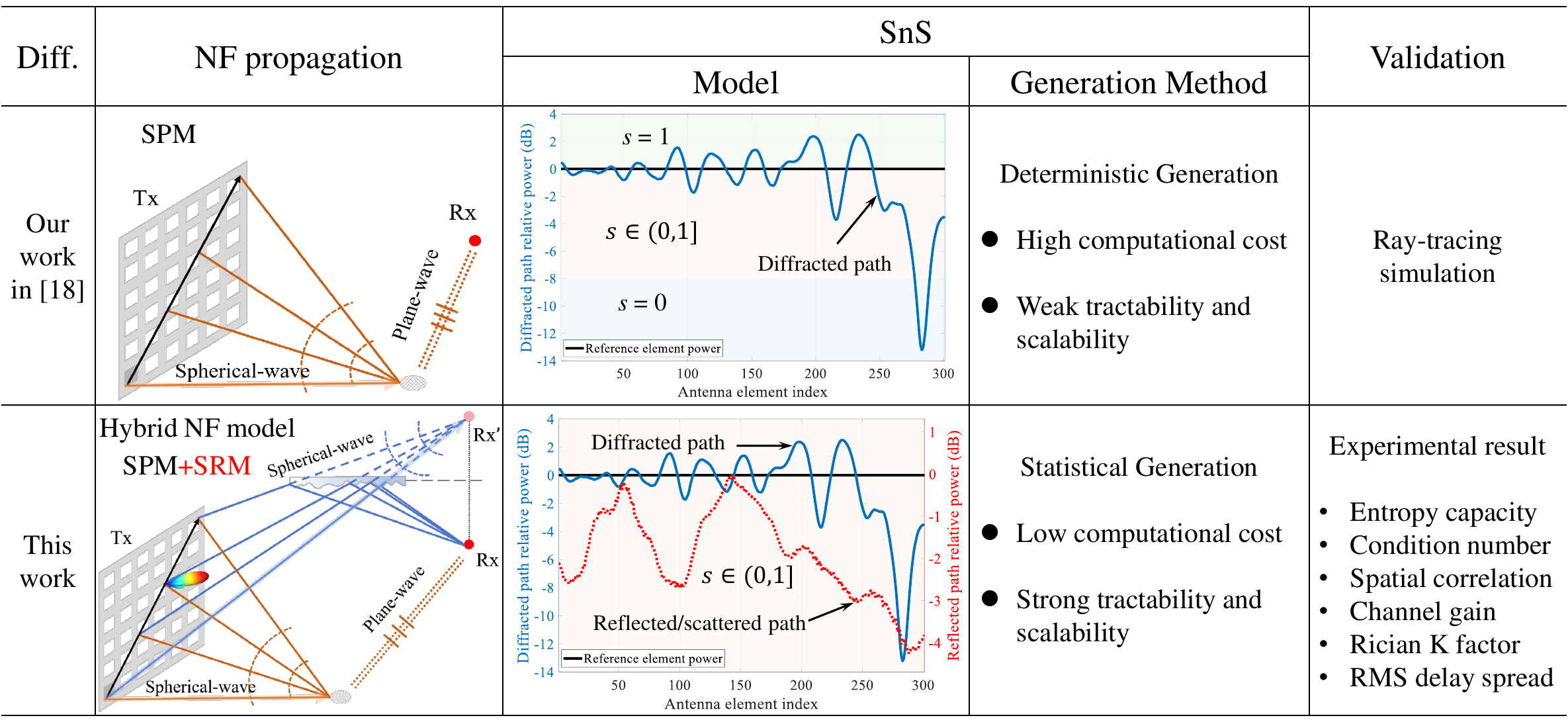}  
    \vspace{-0.6em}
    \caption{Comparison of XL-MIMO channel modeling approaches: our prior work \cite{C3_YZQ} vs. this work.} %
    \vspace{-1em}
    \label{Fig1}  
\end{figure*}

In NF propagation and SnS channel modeling, some existing research has proposed relevant models based on massive MIMO or XL-MIMO measurements from sub-6 GHz or mmWave bands, as well as THz-band simulations. For NF propagation channel modeling, a cluster-of-scatterers spatial channel model was proposed under the NF spherical-wave assumption and validated using 32$\times$32 uniform planar array measurements at 10 GHz \cite{C1_MIMO_Mea_1}. To unify the NF and FF channel models, a cross-field channel model was proposed for XL-MIMO, which refines the propagation path with its first and last bounces and differentiates the characterization of parameters in the NF and FF \cite{C1_HC_ARXIV}. For SnS modeling, the cluster visibility region (VR) model is widely adopted, characterizing SnS by defining regions where specific clusters remain visible to different parts of the array \cite{C1_VR_3,C1_VR_1, C1_cluster_LR,C1_GTY}. \textcolor{black}{This model has been extended to describe SnS by associating different scatterers or UEs with distinct subarrays, and has been applied to channel estimation and precoding design \cite{C1_VR_APP_1,C1_VR_APP_2,C1_VR_APP_3,C1_VR_APP_4}. Complementing these models, \cite{C1_SaLi} proposes a fading channel model for XL-MIMO systems that captures link-level SnS, focusing on line-of-sight (LoS)/non-line-of-sight (NLoS) conditions and shadowing effects across antenna elements. In contrast to these statistical approaches, \cite{C1_YZQ_RT} developed a fast deterministic ray-tracing channel modeling method for XL-MIMO that jointly models NF propagation and SnS, achieving reduced complexity through coarse-to-fine simulations across sparse antenna elements.}

\vspace{-1em}
\subsection{Discussion}

\textcolor{black}{Although NF propagation and SnS have been extensively studied in massive MIMO and XL-MIMO systems, they are typically addressed separately. In practical communication scenarios, NF and SnS effects often coexist and jointly influence the wireless channel. To address this, our earlier work proposed a unified NF and SnS channel model for mmWave massive MIMO systems \cite{C3_YZQ}, as illustrated in Fig. \ref{Fig1}. Specifically, NF propagation was modeled using a scatterer-excited point source model (SPM), where each scatterer acts as a spherical-wave emitter \cite{C3_YZQ}. For SnS modeling, we introduced amplitude attenuation factors (AAFs) to characterize the power variation of SnS paths across antenna elements—an effect that the VR model fails to capture, as they typically only indicate the presence or absence of paths without accounting for their varying strength.}

\textcolor{black}{Nevertheless, several limitations remain in existing XL-MIMO channel modeling approaches: 1) NF propagation modeling: Most current models describe NF wavefronts using the SPM model. While effective in capturing spherical wave behavior from scatterers, this approach becomes inadequate in environments where numerous specular reflections occur from physically large surfaces, such as walls and ground. Furthermore, many theoretical studies assume omnidirectional antenna elements, overlooking the significant influence of directional antennas commonly used in THz XL-MIMO systems \cite{C1_THz_Dir}. 2) SnS modeling: Although our previous work accounts for SnS power variation due to diffraction, it neglected other mechanisms such as reflection and diffuse scattering. In the THz band, the wavelength is often comparable to the size of environmental scatterers, which enhances diffuse scattering effects. The resulting inconsistent reflection and scattering can also contribute to SnS \cite{C1_THz_MIMO_Mea_6,C1_THz_SNS}. For instance, measurement results in Fig. \ref{Fig1} show that the power of an NLoS path reflected from concrete can fluctuate by several dB—behavior not captured by existing models. Furthermore, AAFs were deterministically generated using ray-tracing simulations \cite{C3_YZQ}, resulting in high complexity. No statistical method for generating AAFs has been established, limiting integration into statistical frameworks. 3) Model validation: To the best of our knowledge, no validation against THz XL-MIMO measurements has been performed, making it difficult to quantify the performance gap between NF SnS models and traditional FF, spatial stationarity (SS) models.}

\vspace{-0.8em}
\subsection{Contribution}
The contributions of this paper are summarized as follows.

\begin{itemize}
    \item \textcolor{black}{Extensive THz XL-MIMO channel measurements were conducted at 100 GHz and 132 GHz using 301-element and 531-element uniform linear arrays (ULAs) in indoor environments. The results reveal distinct NF propagation, characterized by nonlinear phase evolution and per-element delay and angle variations. Moreover, SnS behavior is clearly observed, caused not only by blockage but also by inconsistent reflection and scattering from finite-sized or rough surfaces.}
    
    \item \textcolor{black}{A joint NF and SnS channel modeling method is proposed for THz XL-MIMO systems. For NF propagation, a hybrid model combining the SPM and specular reflection model (SRM) is introduced to accurately capture the nonlinear phase evolution of NLoS paths. For SnS, AAFs are introduced to characterize the continuous variation of path power across the array. Based on the analysis of their statistical distribution and spatial autocorrelation, a rank-matching-based method is developed for generating spatially correlated AAFs in simulations.}
    
    \item \textcolor{black}{The proposed model is validated against measurement data by comparing simulation and measurement results across several key metrics, including entropy capacity, Demmel condition number, average spatial correlation of SnS channels, and statistical parameters such as channel gain, Rician K-factor, and RMS delay spread. The close agreement observed across all metrics confirms that the model accurately captures both NF propagation and SnS effects.}
    
\end{itemize}

The remainder of this paper is organized as follows. Section \ref{Sec_2} details the THz XL-MIMO channel measurements and initial experimental observations. Section \ref{Sec_3} presents a THz XL-MIMO channel model that integrates NF propagation and SnS. Section \ref{Sec_4} discusses model implementation and validation. Finally, conclusions are drawn in Section \ref{Sec_5}.

\section{Channel Measurements and Phenomenon Observations}
\label{Sec_2}
Channel measurements provide a direct and effective means of capturing propagation characteristics. \textcolor{black}{To comprehensively investigate NF propagation and SnS effects in THz XL-MIMO systems, four measurement cases were conducted at 100 GHz and 132 GHz in an indoor environment. These include: (1) reflection and scattering from different materials, (2) LoS blockage by a human body, (3) multiple receiver (Rx) positions without blockage, and (4) measurements with directional antenna elements. This section presents the THz channel sounder, measurement setup, and measurement scenario. The results clearly reveal variations in multipath power, phase, delay, and angle due to NF propagation. Additionally, two primary causes of SnS are observed: blockage and inconsistent reflection or scattering.}

\vspace{-0.6em}
\subsection{Channel Sounder and Measurement Setup}
\label{SubSec_2-1}

\textcolor{black}{Two distinct THz channel measurement systems were employed in this paper. The first system, based on a vector network analyzer (VNA) channel sounder, provides ultra-wideband measurement capability (90-110 GHz, 20 GHz bandwidth) achieving exceptional time-delay resolution (0.05 ns) with high dynamic range and phase stability. This configuration utilized 2001 frequency points enabling maximum detectable range of 30 m. Vertical-polarization omnidirectional antennas (5 dBi gain, 30$^\circ$ vertical half power beam width (HPBW)) were deployed at transmitter (Tx) and Rx sides. The Tx forms a 301-element XL-MIMO array via $\lambda$/2 horizontal displacement (1.364 mm step at 110 GHz). The Rayleigh distance of this array (111.6 m) exceeded all measured Tx-Rx distances. Detailed specifications are summarized in Table \ref{Tab_sounder}.}

The second system employed a correlation-based time-domain measurement platform operating at 132 GHz center frequency with 1.2 GHz bandwidth. Directional conical antennas (23 dBi, 14.6° HPBW) were used at the Tx, forming a 531-element virtual ULA through $\lambda$/2 vertical displacement (1.136 mm step). At the Rx, 10$^{\circ}$ rotation step directional scanning sounding was performed using pyramidal horn antennas (25.1 dBi, 9.9° HPBW). Furthermore, to eliminate system response, a back-to-back calibration was performed by directly connecting the Tx and Rx before measurements. Additional configuration details are available in \cite{C1_ICC_XHX}.

\begin{table}[]
\renewcommand\arraystretch{1.2}
\centering
\vspace{-0.5em}
\caption{Measurement Configuration}
\begin{tabular}{m{2cm}<{\centering}|m{0.8cm}<{\centering}|m{0.8cm}<{\centering}|m{0.8cm}<{\centering}|m{1.7cm}<{\centering}}
\hline
\multirow{2}{*}{Parameter} & \multicolumn{4}{c}{Values} \\  
\cline{2-5}
& Case 1 & Case 2 & Case 3 & Case 4 \\
\hline
Center frequency &  \multicolumn{3}{c|}{100 GHz} & 132 GHz\\
\hline
Bandwidth & \multicolumn{3}{c|}{20 GHz} & 1.2 GHz \\
\hline
Tx/Rx antenna type & \multicolumn{3}{c|}{omnidirectional antenna} & conical/pyramidal horn \\
\hline
Tx/Rx antenna gain & \multicolumn{3}{c|}{5 dBi} & 23.0 / 25.1 dBi  \\
\hline
Tx/Rx antenna H-HPBW & \multicolumn{3}{c|}{360$^{\circ}$} & 14.6$^{\circ}$ / 9.9$^{\circ}$ \\
\hline
Number of Tx elements & \multicolumn{3}{c|}{301} & 531  \\
\hline
Tx translation step & \multicolumn{3}{c|}{1.364 mm } & 1.13 mm  \\
\hline
Number of materials/Rx & 6 & 1 & 12 & 1  \\
\hline
Rayleigh distance & \multicolumn{3}{c|}{111.6 m } & 319.2 m \\
\hline
\end{tabular}
\label{Tab_sounder}
\end{table}

\vspace{-0.6em}
\subsection{Measurement Scenario}
\label{SubSec_2-2}

\begin{figure}[htbp]  
    \centering          
    \includegraphics[width=8.5cm]{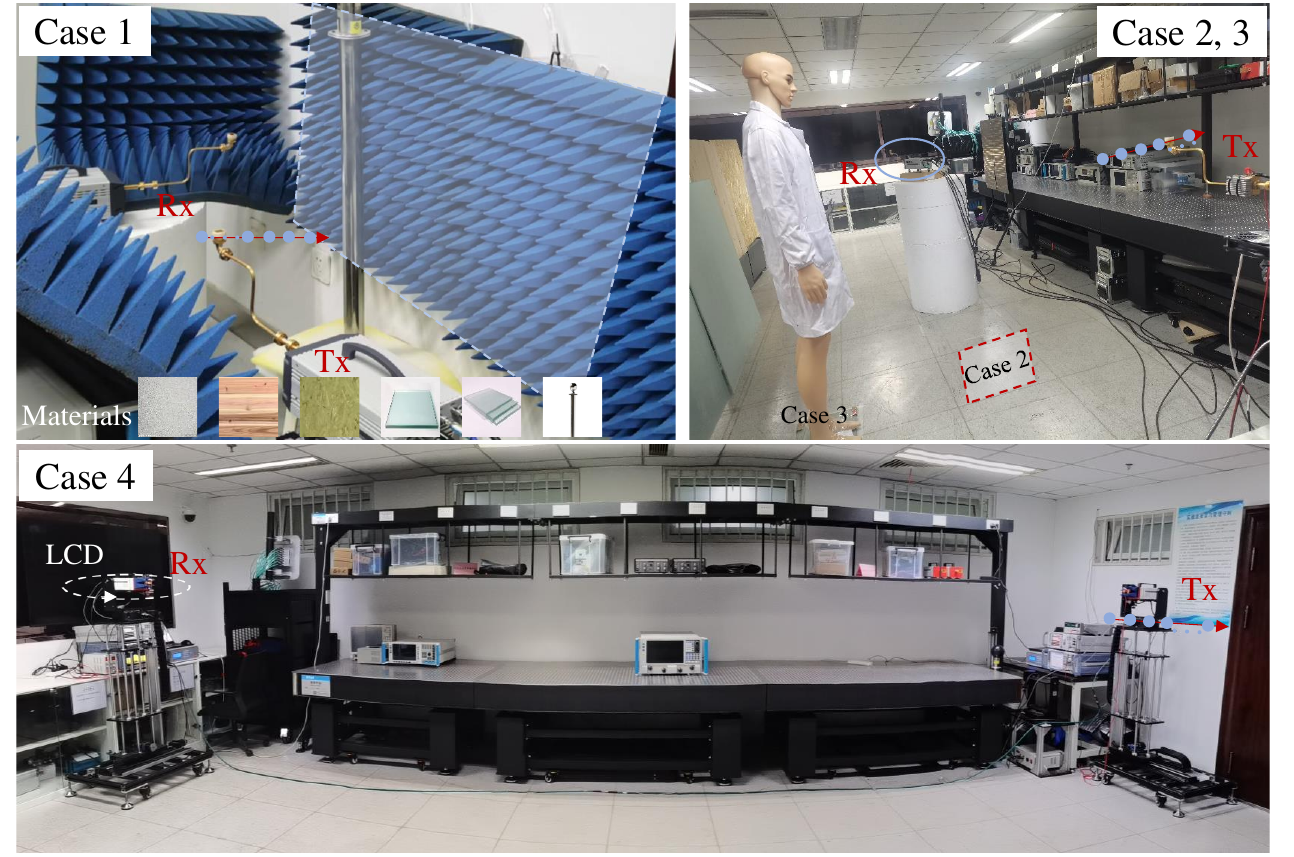} 
    \vspace{-0.5em}
    \caption{The photograph of the measurement scenario.} %
    \label{Fig_scenario}  
\end{figure}

\textcolor{black}{Case 1 was specifically designed to observe the phenomena of NF propagation and SnS in THz XL-MIMO systems. The environment was surrounded by microwave-absorbing materials to eliminate stray multipath effects, as shown in Fig. \ref{Fig_scenario}. Six types of scatterers were implemented: concrete (lime-painted), smooth wooden board, rough wooden board, smooth glass, frosted glass, and metal cylinder (3.5 cm diameter × 1.2 m height). The first five scatterers, constructed as 1.4 × 1.4 m\(^2\) square panels, were placed in shadowed areas marked in Fig. \ref{Fig_scenario} to observe LoS propagation and specular reflection-dominated NLoS propagation. The cylinders served as point-source scatterers. The minimum Tx element-Rx horizontal distance was set to 0.645 m, and the scatterer centers were aligned to the transceiver antenna elevation.}

\begin{figure}[htbp]  
    \centering          
    \includegraphics[width=8.5cm]{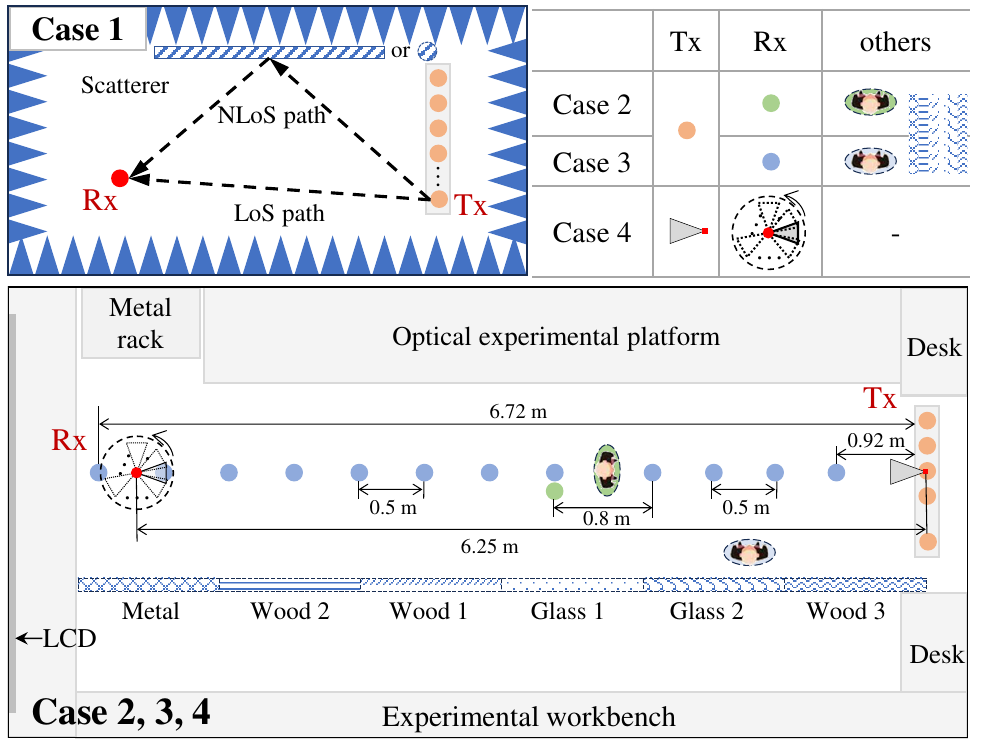}  
    \caption{Transmitters and receivers layout of the measurement scenario.} %
    \vspace{-1em}
    \label{Fig_layout}  
\end{figure}

\textcolor{black}{Cases 2-4 measurements were conducted in an 8.34 $\times$ 6.05 $\times$ 2.4 $\text{m}^3$ indoor laboratory with concrete walls (lime-painted), plasterboard ceiling, and tiled floor, as shown in Fig. \ref{Fig_scenario}. Notable reflectors included an optical platform, metal equipment rack, and a large LCD. The Tx was deployed near the wall to emulate a base station, while Rx positions varied across the cases. In Case 2, the Rx was placed 3.22 m from the Tx, with an anthropomorphic dummy blocking the LoS path to investigate SnS effects caused by blockage. In Case 3, twelve Rx positions were arranged along a straight line with 0.5 m spacing, except for a gap of 0.8 m between positions 4 and 5. No direct obstruction was present, and the dummy was placed near the wall’s edge to serve as a scatterer. In Case 4, a single Rx location was used with a Tx–Rx separation of 6.72 m to assess the impact of directional antenna elements. For Cases 2 and 3, a screen composed of various materials was installed to emulate a wall-like structure. Detailed measurement layouts for each case are illustrated in Fig. \ref{Fig_layout}.}

\graphicspath{{picture/}}

\graphicspath{{picture/}}

\vspace{-1em}
\subsection{Phenomenon Observations}
\label{Subsec_2-3}

\begin{figure}[htbp]  
    \centering
    \subfloat[]
    {\includegraphics[width=8cm]{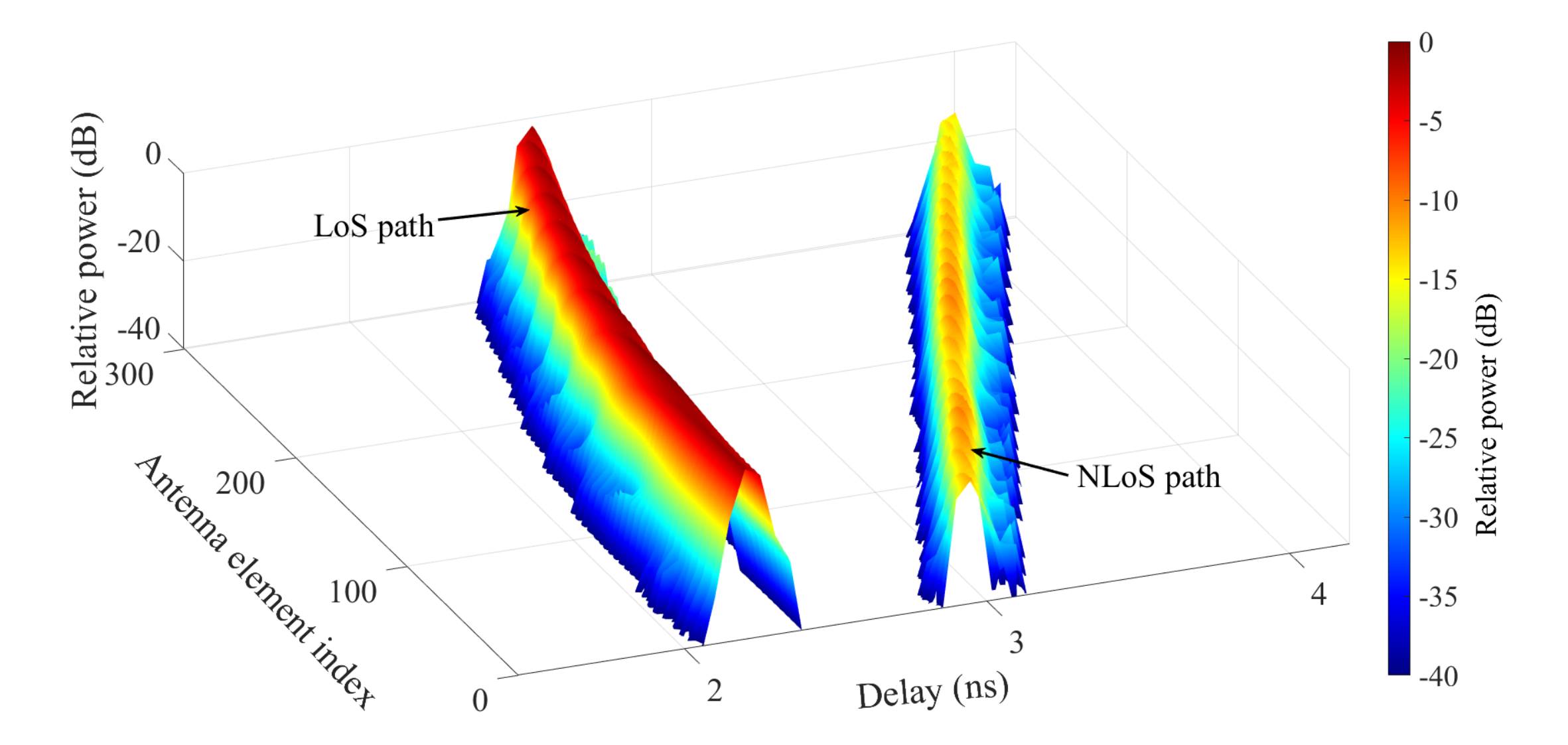 }  
    \label{Fig4_a}
    }
    \hfill
    \subfloat[]
    {\includegraphics[width=8cm]{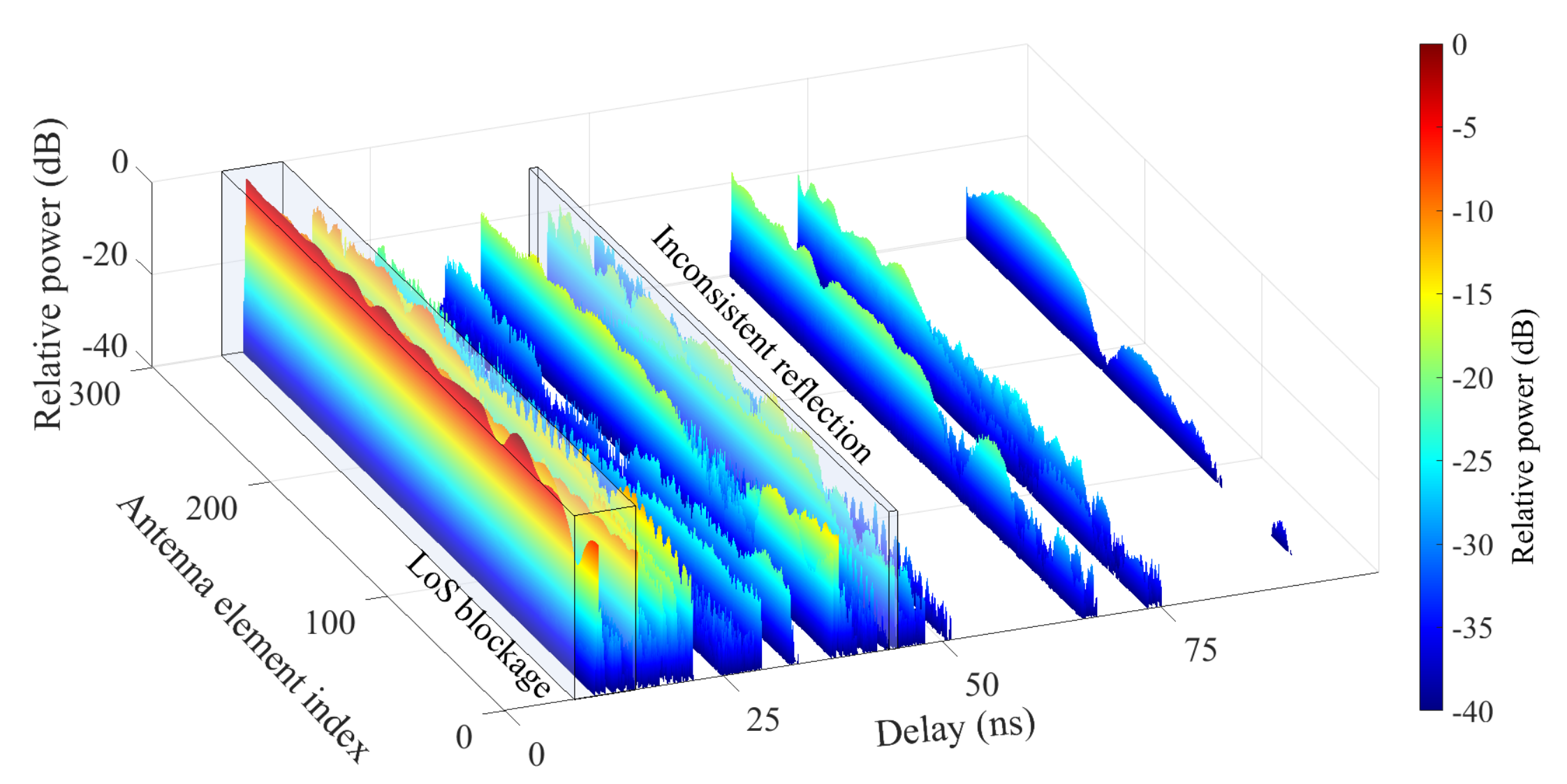 }  
    \label{Fig4_b}
    }
    \caption{Omnidirectional PDPs measured in XL-MIMO for (a) Case 1 (concrete) and (b) Case 2.} %
    \vspace{-1.5em}
    \label{Fig_XL-MIMO_PDP}  
\end{figure}

\begin{figure*}[htbp]  
    \centering  
    \subfloat[]{
    \includegraphics[width=7.5cm]{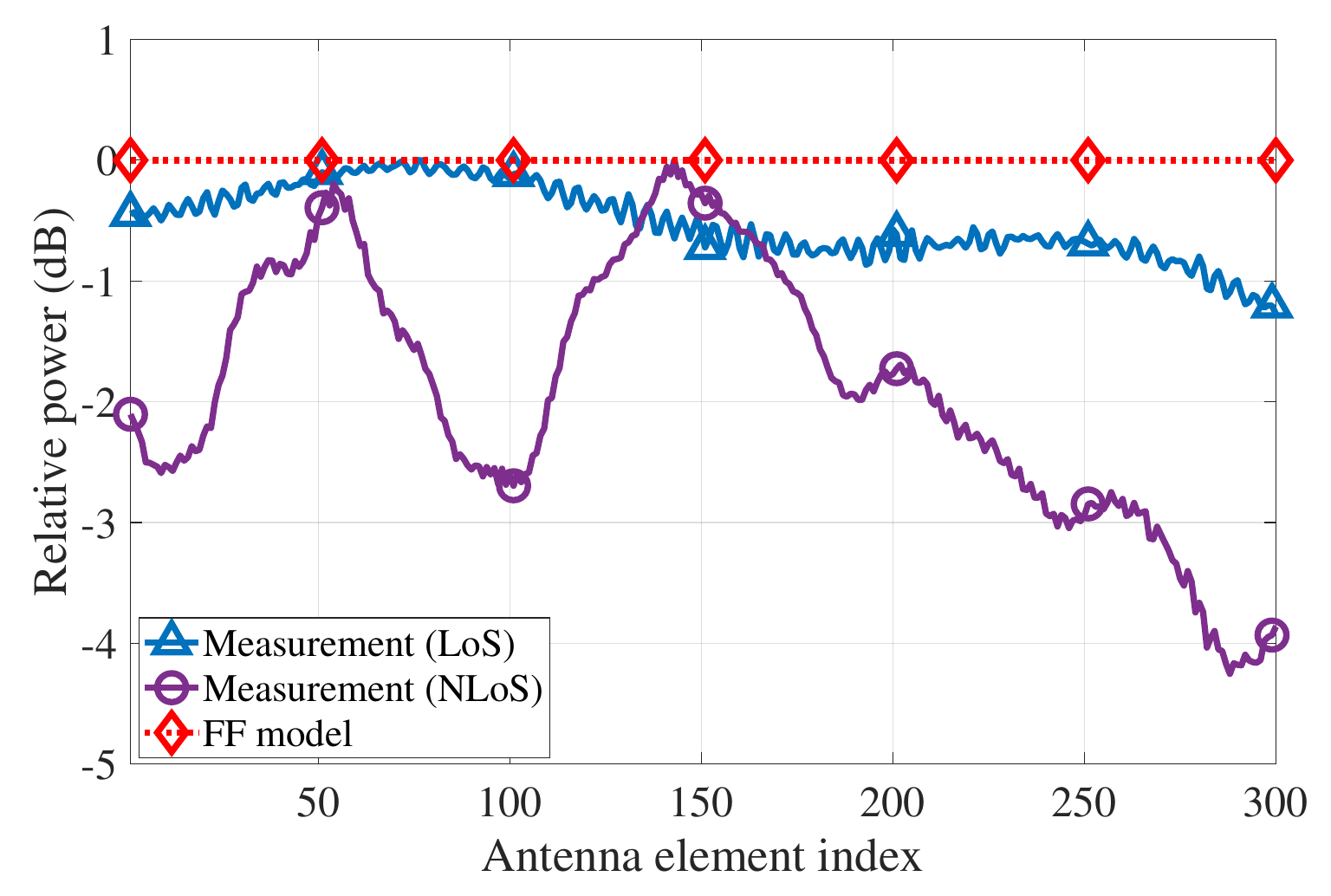}
    \label{Fig5_a}
    }
    \subfloat[]{
    \includegraphics[width=7.45cm]{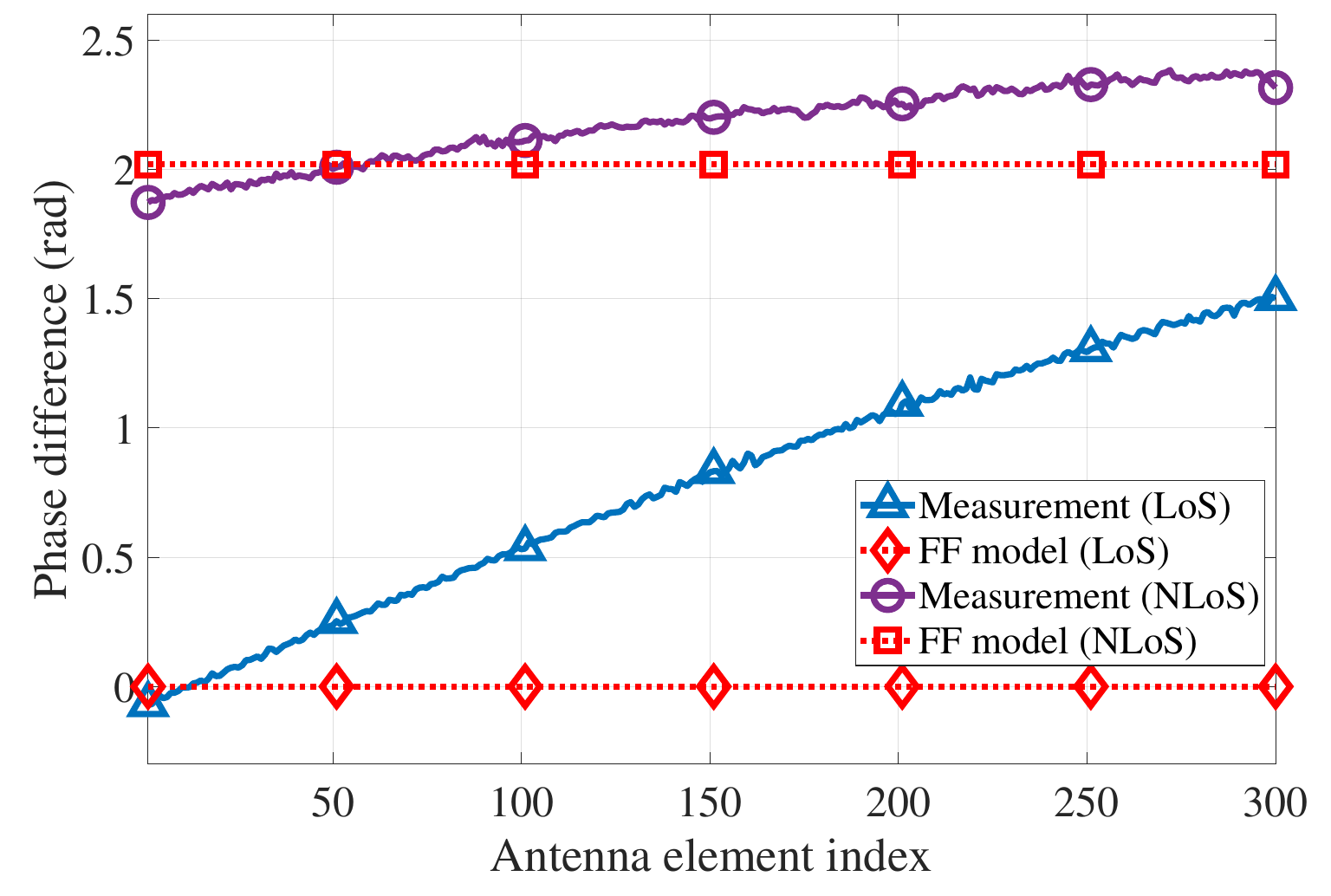}
    \label{Fig5_b}
    }
    \hfill
    \subfloat[]{
    \includegraphics[width=7.5cm]{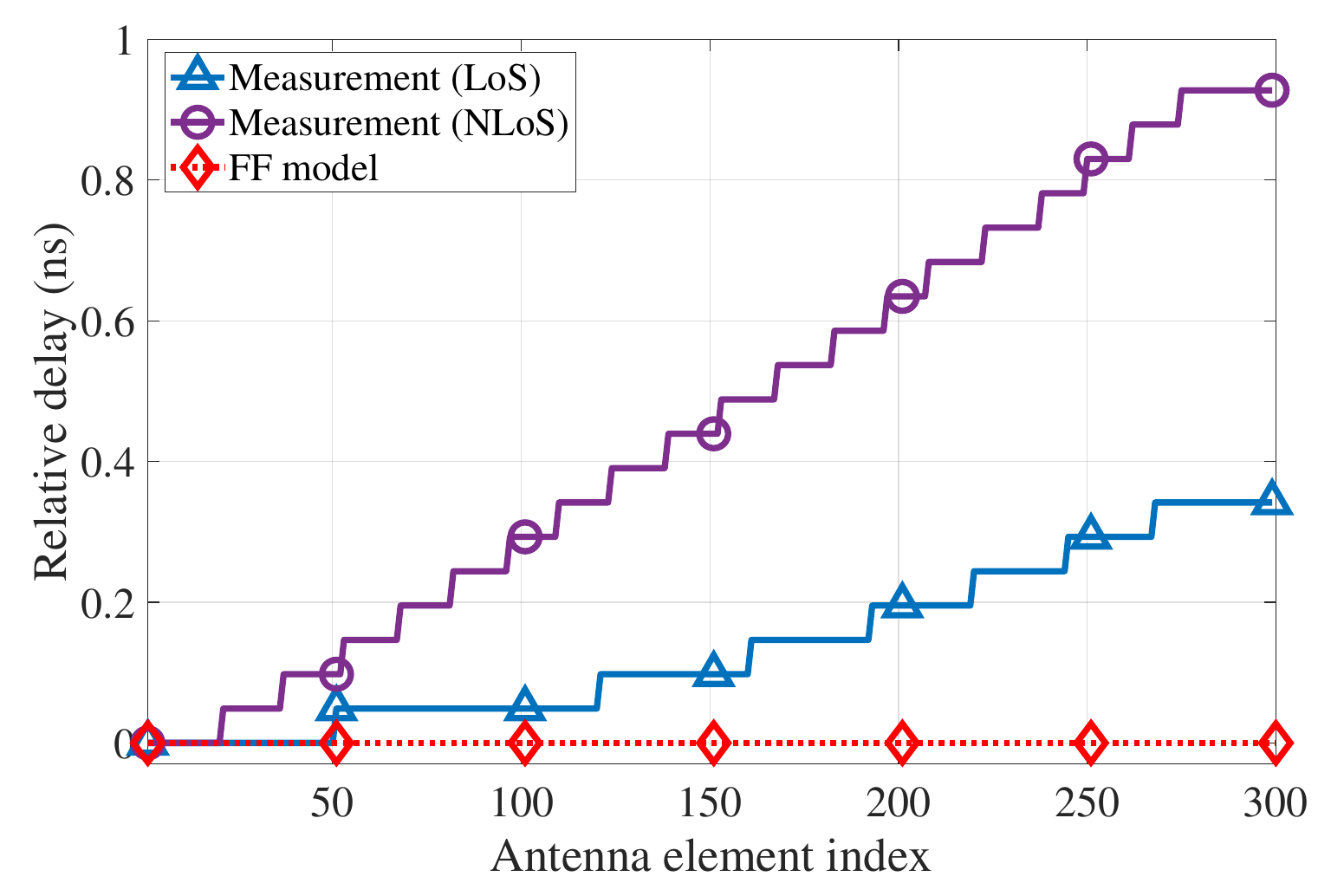}
    \label{Fig5_c}
    }
    \subfloat[]{
    \includegraphics[width=7.5cm]{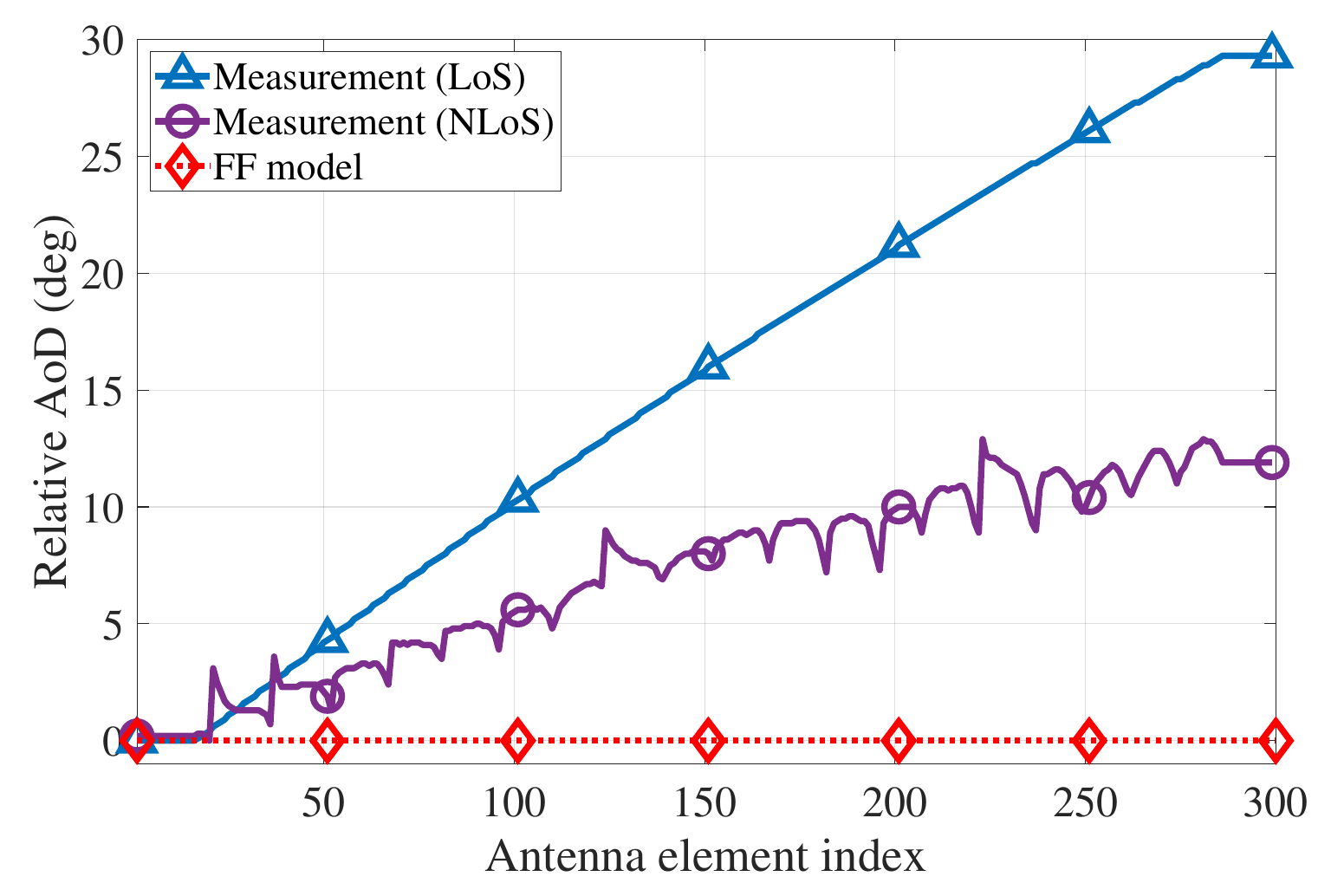}
    \label{Fig5_d}
    }
    \caption{Variation of NF multipath parameters across antenna elements in Case 1 (concrete): (a) Relative power, (b) Inter-element phase difference, (c) Relative delay, and (d) Relative AoD.}
    \vspace{-1.5em}
    \label{Fig_NF_observation}
\end{figure*}

\begin{figure}[htbp]  
    \centering  
    \subfloat[]{
    \includegraphics[width=4.3cm]{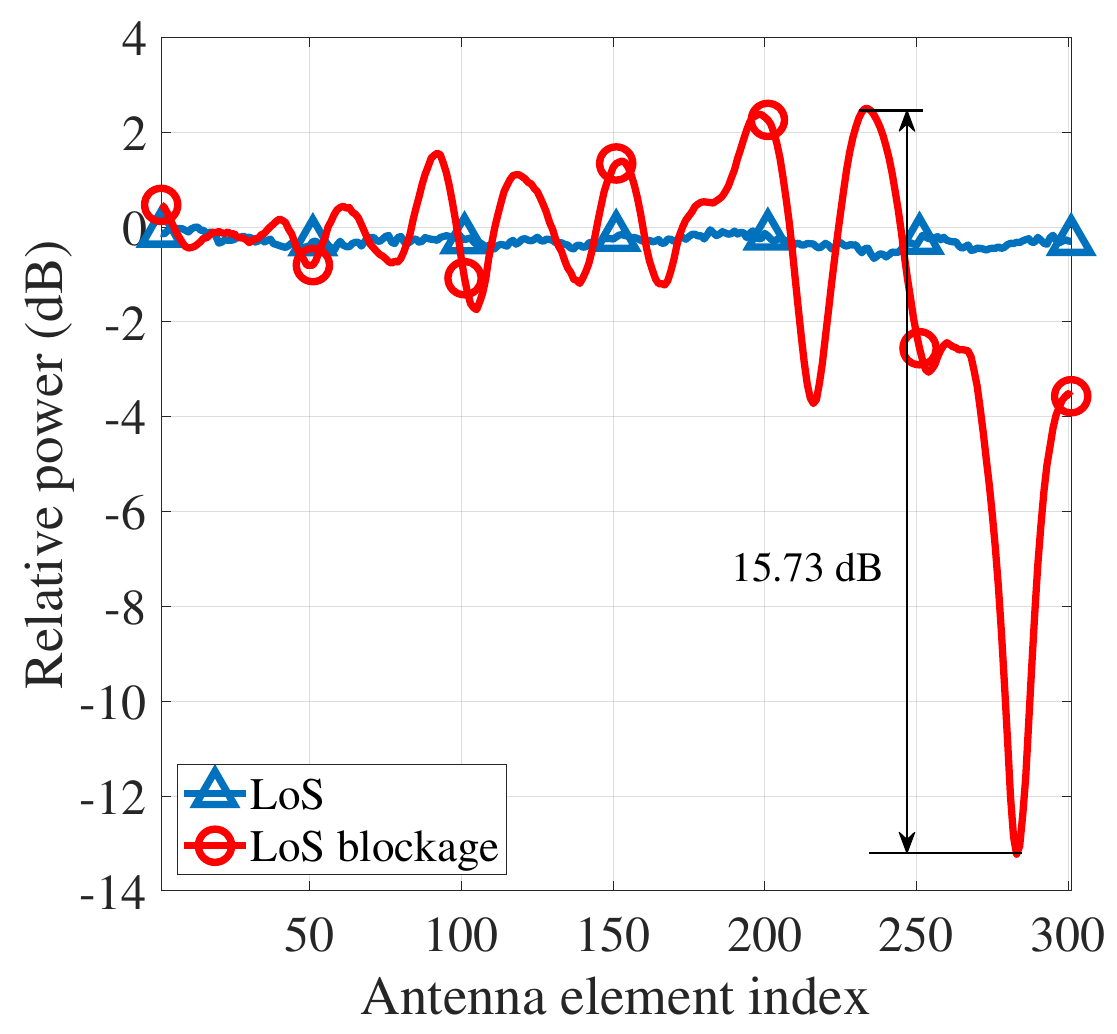}
    \label{Fig6_a}}
    \subfloat[]{
    \includegraphics[width=4.3cm]{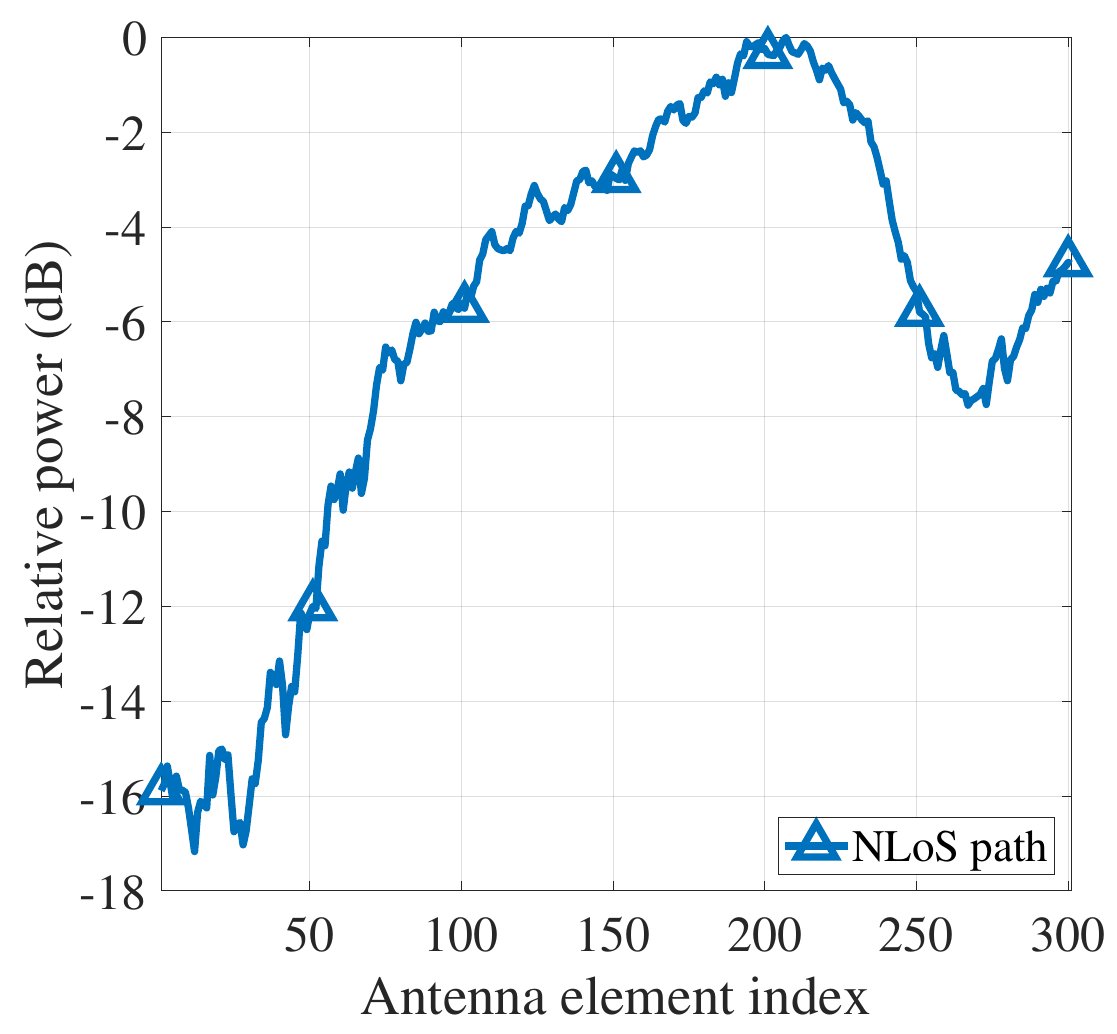}\label{Fig6_b}}
    \hfill
    \vspace{-1em}
    \subfloat[]{\includegraphics[width=4.4cm]{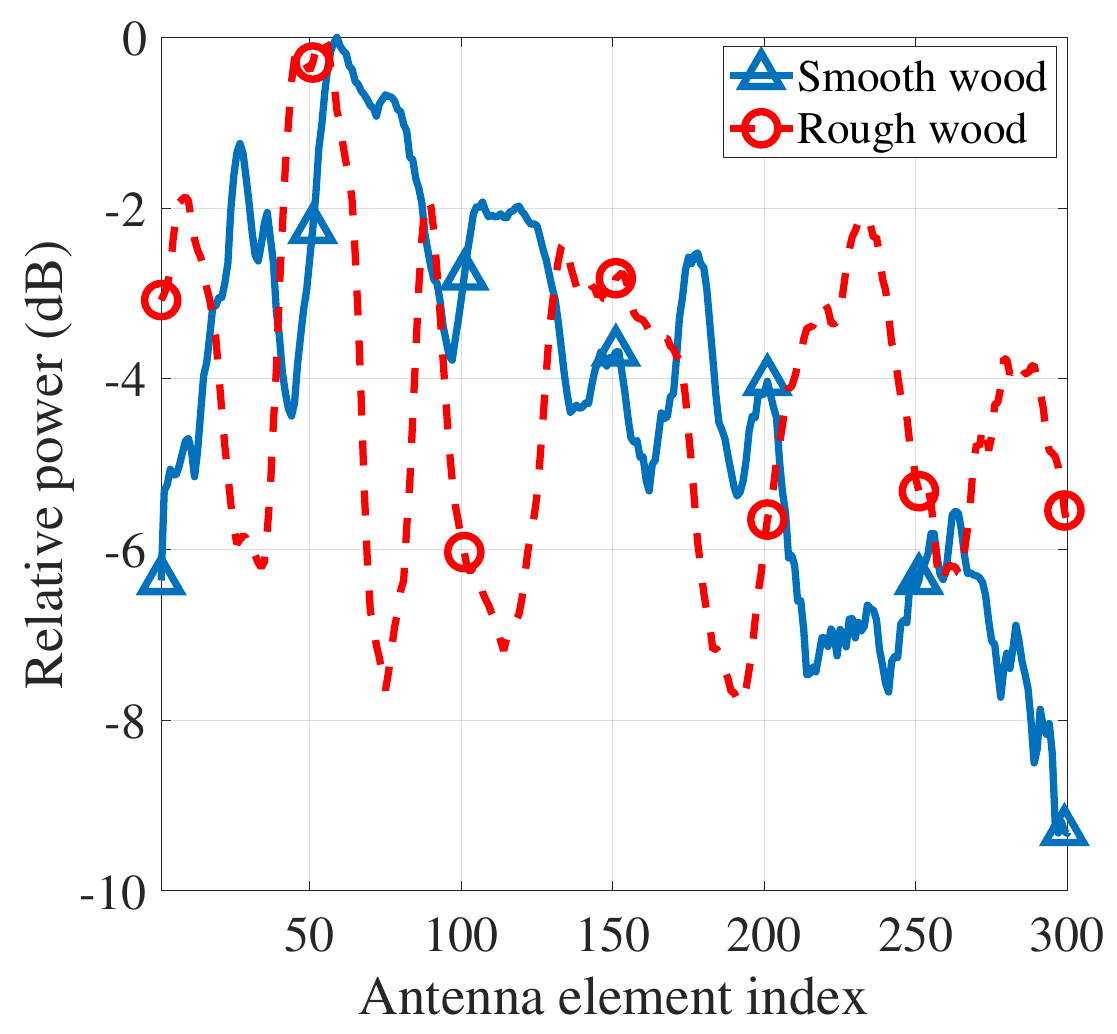} \label{Fig6_c}}
  \subfloat[]{\includegraphics[width=4.4cm]{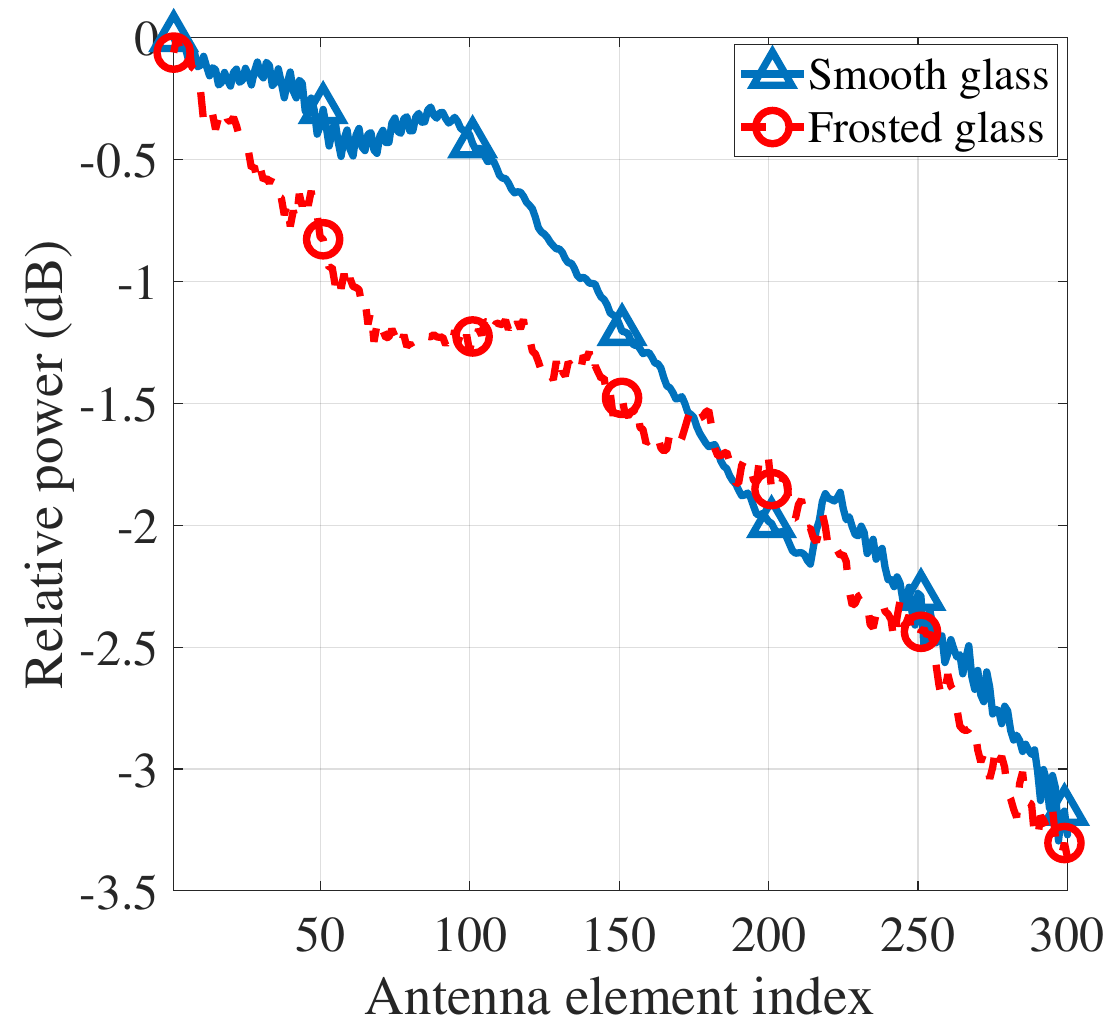} 
  \label{Fig6_d}}
  \hfill
  \vspace{-1em}
    \subfloat[]{
    \includegraphics[width=4.3cm]{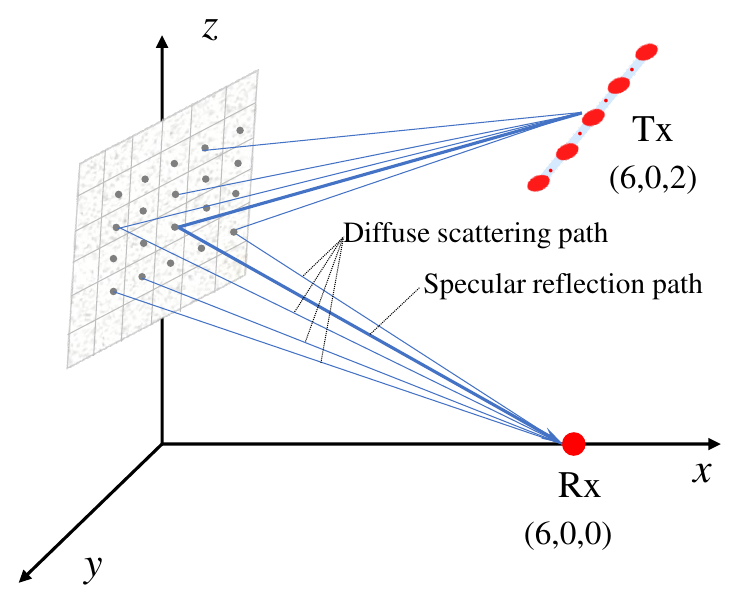}\label{Fig6_e}}
    \subfloat[]{
    \includegraphics[width=4.3cm]{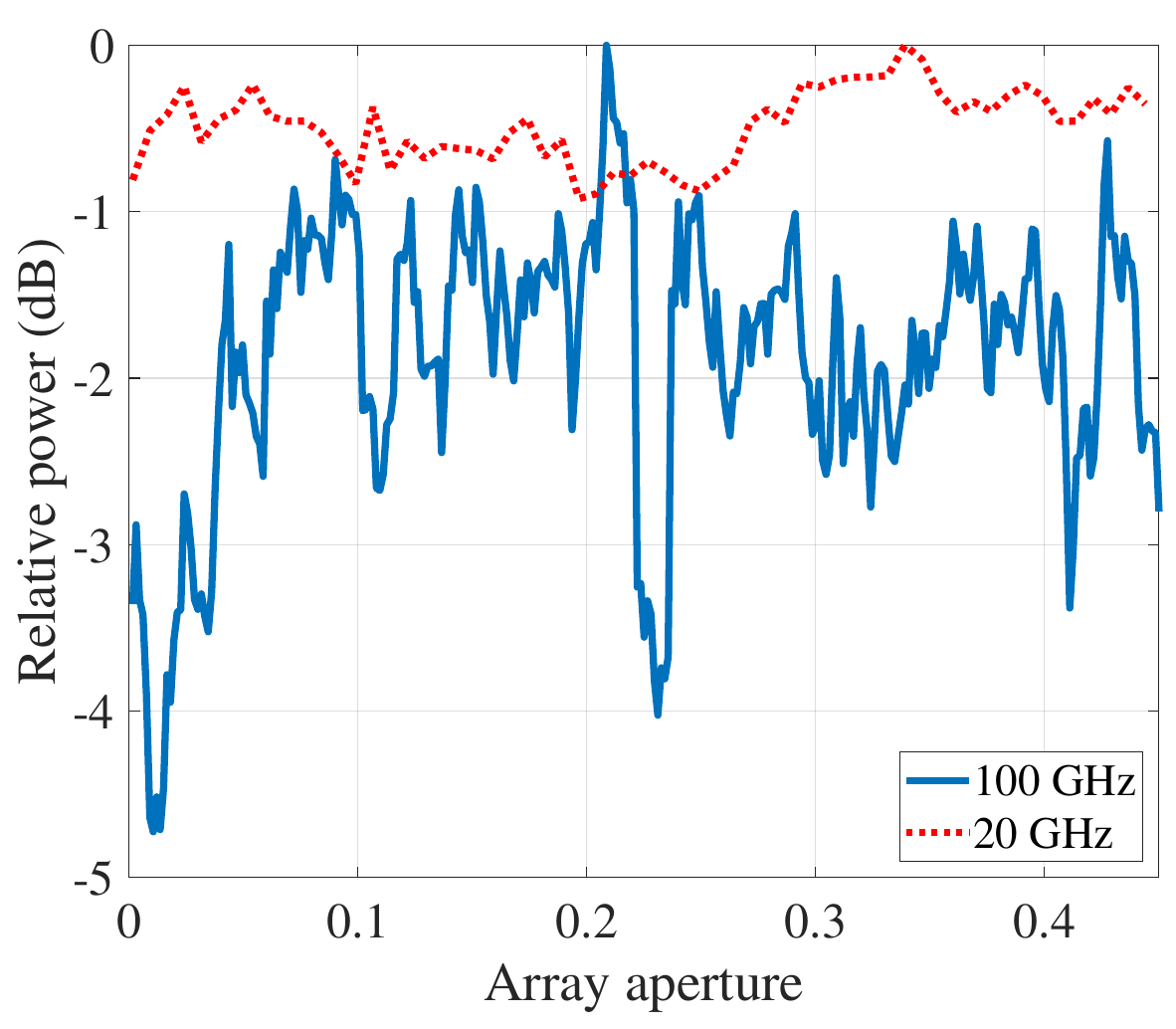}\label{Fig6_f}}
    \caption{Measured and simulated SnS phenomenon. Measured SnS caused by (a) human body blockage and (b) inconsistent reflection or scattering in Case 2.  Relative power variation of NLoS paths in Case 1: (c) wooden surfaces (smooth and rough) and (d) glass surfaces (smooth and frosted). Simulation based on the Beckmann–Kirchhoff model \cite{C3_Kurner}: (e) simulation configuration and (f) simulated power variation across antenna elements. The rough surface has a standard deviation height of 0.1 mm, correlation length of 1.5 mm, and complex refractive index of 1.92-j0.059.} 
    \vspace{-0.5cm}
    \label{Fig_SnS_observation}  
\end{figure}

From the THz XL-MIMO channel measurements, the omnidirectional channel impulse response (CIR) for each element was obtained. \textcolor{black}{Fig. \ref{Fig_XL-MIMO_PDP} shows the PDPs measured in Case 1 (concrete surface) and Case 2, where the x-, y-, and z-axes represent propagation delay, antenna element index, and relative power (dB), respectively. A 40 dB power dynamic range was applied,  consistent with later modeling and validation settings. In Fig. \ref{Fig4_a}, the LoS path and NLoS path reflected from concrete are clearly distinguishable in the delay domain. The propagation delay variation across antenna elements is visible due to the high delay resolution (0.05 ns) enabled by the 20 GHz bandwidth, capturing even 15 mm propagation distance differences. Fig. \ref{Fig4_b} illustrates the omnidirectional PDPs of XL-MIMO in Case 2. With a large-aperture array of 301 elements, significant variations and fluctuations in multipath power are observed across the array. The following discussion analyzes these observations in detail from the perspectives of NF propagation and SnS\footnote{\textcolor{black}{NF propagation and SnS are inherently interconnected. For clarity, this paper distinguishes between the two: NF refers to variations in multipath parameters across XL-MIMO elements due to spherical wave propagation, while SnS accounts for the multipath power variations caused by the distinct propagation environments experienced by different elements.}}, using Cases 1 and 2 as illustrative examples (Figs. \ref{Fig_NF_observation} and \ref{Fig_SnS_observation}).}

\graphicspath{{picture/}}

\graphicspath{{picture_v2/}}

1) \textit{Near-field Propagation}: When transitioning from FF plane-wave assumptions to NF spherical-wave modeling, multipath parameters such as power, phase, delay, and angle become dependent on the antenna element index. These variations are examined using the LoS and NLoS reflection paths observed in Case 1.
    \begin{itemize}
        \item \textit{\textbf{Power}}: Existing studies propose various criteria indicating that the differences in multipath power due to NF spherical wave diffusion may be negligible under certain conditions. Ref. \cite{C2_UPD} defines the Tx-Rx distance as the uniform power distance when the ratio of the minimum to maximum power among array elements reaches a specified threshold. When the threshold is set to 0.9, the uniform power distance is only 1.9 times the array aperture. This indicates that imply that noticeable amplitude variation across the array is typically observed only when the Tx–Rx distance is comparable to the array aperture. \textcolor{black}{For instance, in Case 1, where the array aperture is 0.409 m and the Tx–Rx distance is 0.645 m, the theoretically calculated maximum power difference across array elements is only 1.256 dB. This is consistent with the measurement results for the LoS path shown in Fig. \ref{Fig5_a}, which are not accurately captured by the FF model.} However, in practical scenarios, communication distances usually exceed this range. Therefore, amplitude differences between array elements under NF propagation can generally be considered negligible, similar to FF conditions. For the NLoS path, the combined effects of NF propagation and inconsistent reflections from scatterers lead to significant power variation across the array.
         
        \item \textit{\textbf{Phase}}: \textcolor{black}{Under the FF assumption, the relative phase of each antenna element with respect to a reference element (indexed by \(m = 0\)) is given by \(\varphi_m^{\rm{FF}} = \frac{2\pi}{\lambda} m\delta \sin \phi\), where \(m = 0, 1, \dots, M-1\) is the antenna element index, \(\delta\) is the antenna spacing, \(\lambda\) is the wavelength, and \(\phi \in (-\pi/2, \pi/2)\) is the angle of departure (AoD). This phase expression depends solely on the path angle and varies linearly with the element index. In contrast, under NF propagation, the relative phase is approximated as \(\varphi_m^{\rm{NF}} \approx \frac{2\pi}{\lambda} \left( -m\delta \sin \phi + {m^2 \delta^2 (1 - \sin^2 \phi)}/{2d} \right)\), where \(d\) is the distance from the reference antenna to the Rx. In this case, the phase depends on both the angle and the Tx–Rx distance, resulting in a nonlinear variation with respect to \(m\). To further capture this spatial variation, we define the inter-element phase difference as the difference between adjacent elements, \(\Delta \varphi_m = \varphi_m - \varphi_{m-1}\). Under the FF assumption, the inter-element phase difference remains constant and is given by \(\Delta \varphi_{\rm{FF}} = \pi \sin \phi\), independent of the element index. However, in the NF case, it becomes \(\Delta \varphi_{\rm{NF}} = -\pi \sin \phi + {\pi \lambda (1 - \sin^2 \phi)}/{4d}\), and clearly varies with the position along the array. Fig. \ref{Fig5_b} shows the measured inter-element phase differences for the LoS and NLoS paths in Case 1, which are obtained by unwrapping the measured path phases and differencing between adjacent elements. The measured results confirm that the inter-element phase differences vary significantly with the antenna index, indicating the breakdown of the FF assumption and the need to accurately model nonlinear phase variations in NF propagation.}
        
        \item \textit{\textbf{Delay}}:
        \textcolor{black}{Under FF conditions, all antenna elements are assumed to experience the same path delay. In contrast, under NF propagation, the path delay varies across antenna elements, as illustrated in Fig. \ref{Fig5_c}. This effect becomes more pronounced at THz bands, where the ultra-wide bandwidth significantly improves delay resolution, allowing even small variations in propagation distance to be captured. As a result, accurately modeling delay variation across the array is critical for the THz XL-MIMO channel.}
        \item \textit{\textbf{Angle}}: 
        In the FF region, all antenna elements experience nearly identical angles of arrival (AoA) and AoD. \textcolor{black}{In contrast, under NF conditions, the AoA/AoD becomes element-dependent and can vary significantly across the array\footnote{The variation of path angles with element index is obtained using a spatial-domain discrete Fourier transform method with a sliding window of 51 antenna elements.}, as shown in Fig. \ref{Fig5_d}, where the maximum angular variation reaches up to \(30^{\circ}\).} These angle variations lead to inconsistent directional gains across antenna elements, especially in THz XL-MIMO systems where directional antennas are used. Unlike ideal omnidirectional elements, directional elements exhibit gain differences due to angular misalignment between specific Tx elements and the Rx, resulting in noticeable power variations across array elements for the same path.
    \end{itemize}

\graphicspath{{picture/}}

2) \textit{Spatial Non-Stationarity}: The significant power variations caused by SnS may result from blockages and inconsistent reflections or scattering in the propagation environment.
    \begin{itemize}
        \item \textit{\textbf{Blockage}}: Objects with limited physical size may be unable to fully obstruct a large-aperture antenna array, meaning that not all antenna elements are simultaneously blocked from the Rx or scatterer. As shown in Fig. \ref{Fig6_a}, compared to the unblocked scenario, human body blockage causes power variations across antenna elements of up to 15.73 dB. In addition, diffraction effects near the edges of the blockage region produce power fluctuation fringes beyond the shadowed area.

        \item \textit{\textbf{Inconsistent Reflection or Scattering}}: In THz propagation environments, finite-sized reflectors/scatterers may no longer act as complete scatterers for the entire antenna array with such a large aperture, meaning that scatterers only affect certain array elements. As shown in Fig. \ref{Fig6_b}, an NLoS path in Case 2 exhibits power variations of up to 16 dB. Even when the same scatterer is encountered, variations in material properties and surface roughness can lead to differing reflection characteristics across elements. \textcolor{black}{For example, in Case 1, NLoS paths reflected by different materials exhibit distinct power variation patterns across the array, as illustrated in Figs. \ref{Fig6_c} and \ref{Fig6_d}. Variations in the power of glass-reflected NLoS paths are primarily governed by angle-dependent reflection coefficients, often resulting in a linear trend along the array. In contrast, reflections from wooden surfaces lead to stronger fluctuations with faster periodic variations\footnote{Power variations caused by NF propagation alone are relatively small-typically within 1.5 dB-indicating that SnS is the dominant contributor to the observed fluctuations in NLoS paths.}. Additionally, due to the short wavelength of THz signals, the size of environmental scatterers becomes comparable to the wavelength, and surfaces that appear smooth at lower frequencies can exhibit strong diffuse scattering at the THz band. This leads to increased randomness in multipath power across antenna elements, resulting in significantly enhanced SnS. As shown in Fig. \ref{Fig6_f}, our simulations based on the Beckmann–Kirchhoff model \cite{C3_Kurner} confirm that power variation across the array at 100 GHz is substantially greater than at 20 GHz under the same aperture size, indicating more pronounced SnS effects at higher frequencies.}
\end{itemize}

\section{THz XL-MIMO Channel Model}
\label{Sec_3}

Based on the observations from measurements, a THz XL-MIMO channel model that characterizes NF propagation and SnS is proposed in this section.

\vspace{-1em}
\subsection{General Model Framework}
\label{Subsec_3-2}

\begin{figure*}[htbp]  
    \centering  
    \includegraphics[width=16cm]{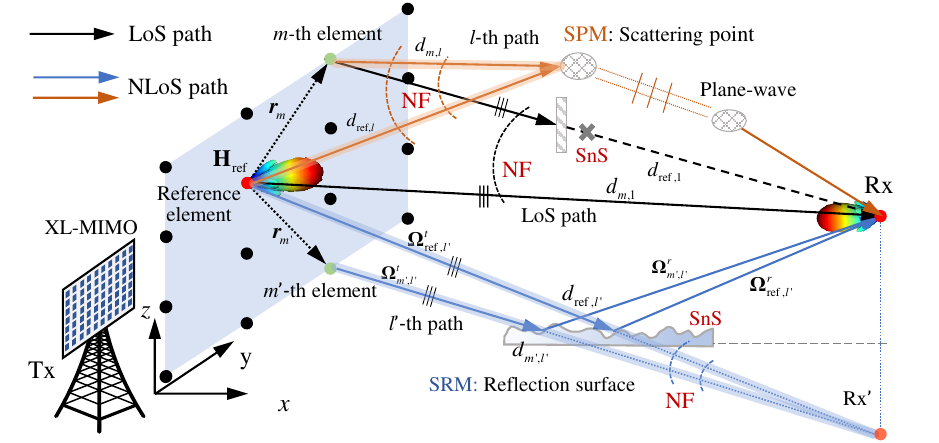}  
    \caption{A schematic diagram of the THz XL-MIMO NF propagation and SnS channel model, including a hybrid NF model (SPM+SRM) and two SnS causes (blockage, inconsistent reflection/scattering).} %
    \vspace{-1.2em}
    \label{Fig7_Framework}  
\end{figure*}

A THz XL-MIMO channel model can be constructed based on the channel response of a reference antenna element, while incorporating variations in channel parameters across the array. Consider the scenario illustrated in Fig.\ref{Fig7_Framework}, where the base station is equipped with $M$ antenna elements and the UE has a single antenna. The antenna radiation patterns at both the Tx and Rx are taken into account. In the propagation environment, multipath signals may encounter various interactions such as blockages, large reflective surfaces, and discrete scatterers. To incorporate both NF propagation and SnS characteristics, the THz XL-MIMO channel can be modeled as \cite{C3_YZQ}:
\begin{equation}
    \mathbf{H}_{\mathrm{NF\text{-}SnS}}(f) = \mathbf{A}_{\mathrm{NF}}(f) \odot \mathbf{S} \cdot \mathbf{H}_{\mathrm{ref}}(f),
    \label{equ_H_MIMO}
\end{equation}
\noindent
where \(\mathbf{H}_{\mathrm{NF\text{-}SnS}}(f) \in \mathbb{C}^{M \times 1}\) represents the XL-MIMO channel response vector at frequency \(f \in [f_L, f_U]\), and \(\odot\) denotes the Hadamard (element-wise) product. The matrices \(\mathbf{A}_{\mathrm{NF}}(f)\) and \(\mathbf{S} \in \mathbb{C}^{M \times L}\) represent the effects of NF propagation and SnS-induced AAFs, respectively. The term \(\mathbf{H}_{\mathrm{ref}}(f)\in \mathbb{C}^{L \times 1}\) denotes the channel frequency response of the reference antenna element, consisting of \(L\) resolvable multipath components, and is given by:
\begin{equation}
    \mathbf{H}_{\mathrm{ref}}(f) = \left[\alpha_1 e^{-j 2\pi f \tau_1}, \ldots, \alpha_l e^{-j 2\pi f \tau_l}, \ldots, \alpha_L e^{-j 2\pi f \tau_L} \right]^T,
    \label{equ_H}
\end{equation}
\noindent
where \(\alpha_l\) and \(\tau_l\) denote the complex amplitude and propagation delay of the \(l\)-th path, respectively.

\graphicspath{{picture/}}

\vspace{-1em}
\subsection{Near-Field Propagation}
\label{Subsec_3-2-1}

\begin{figure}[h]
  \centering
  \subfloat[]{\includegraphics[width=4.3cm]{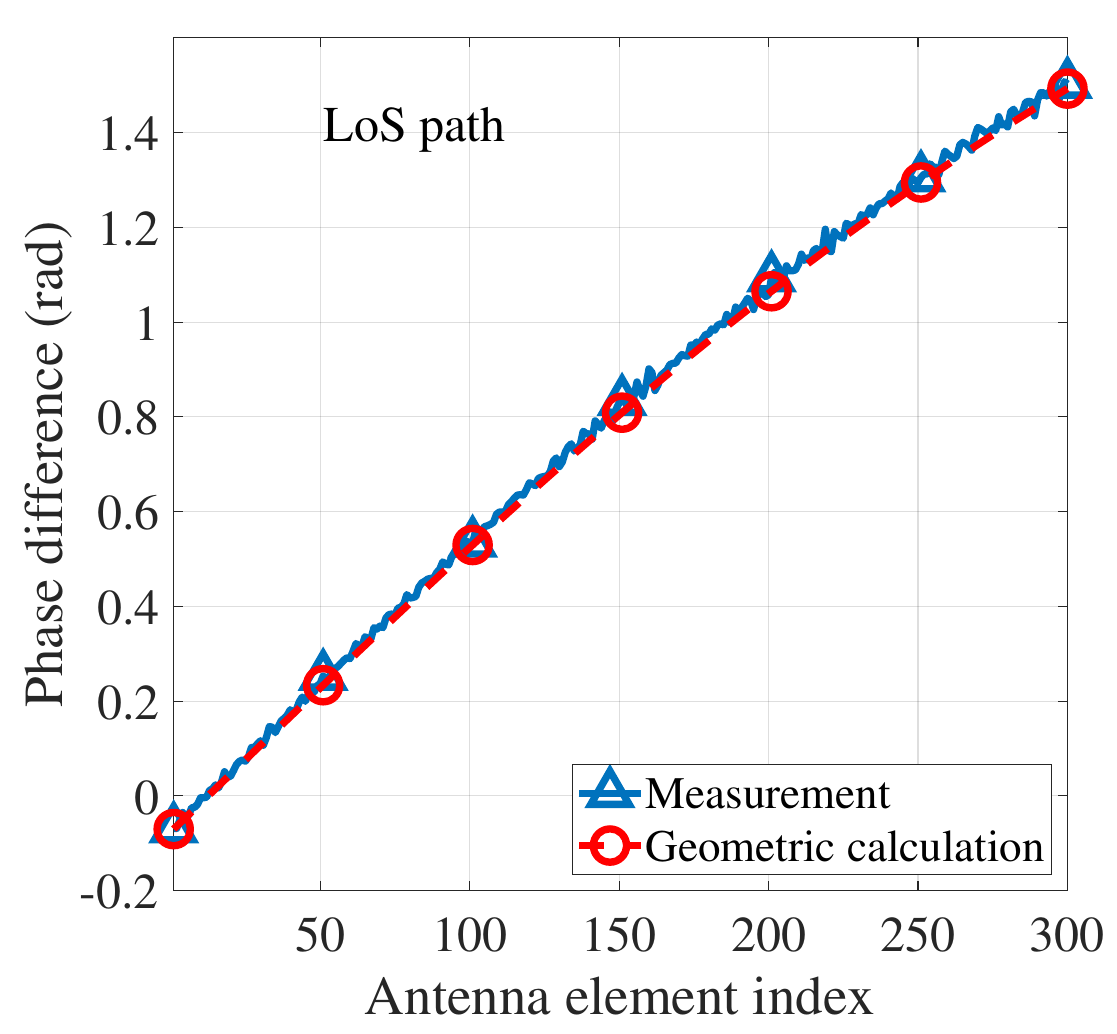} \label{Fig8_a}}
  \hfill
  \subfloat[]{\includegraphics[width=4.3cm]{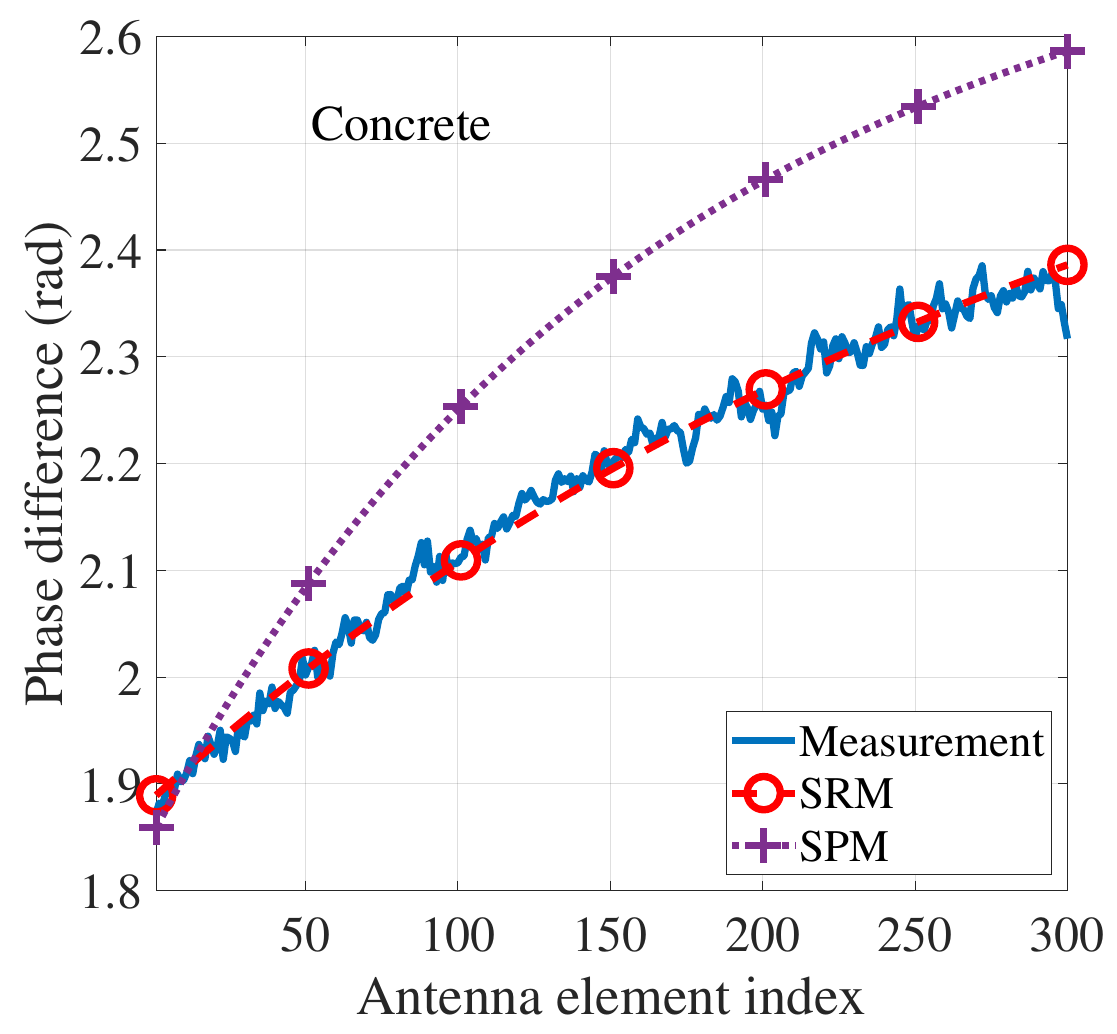} 
  \label{Fig8_b}}
  \\
  \vspace{-1em}
  \subfloat[]{\includegraphics[width=4.3cm]{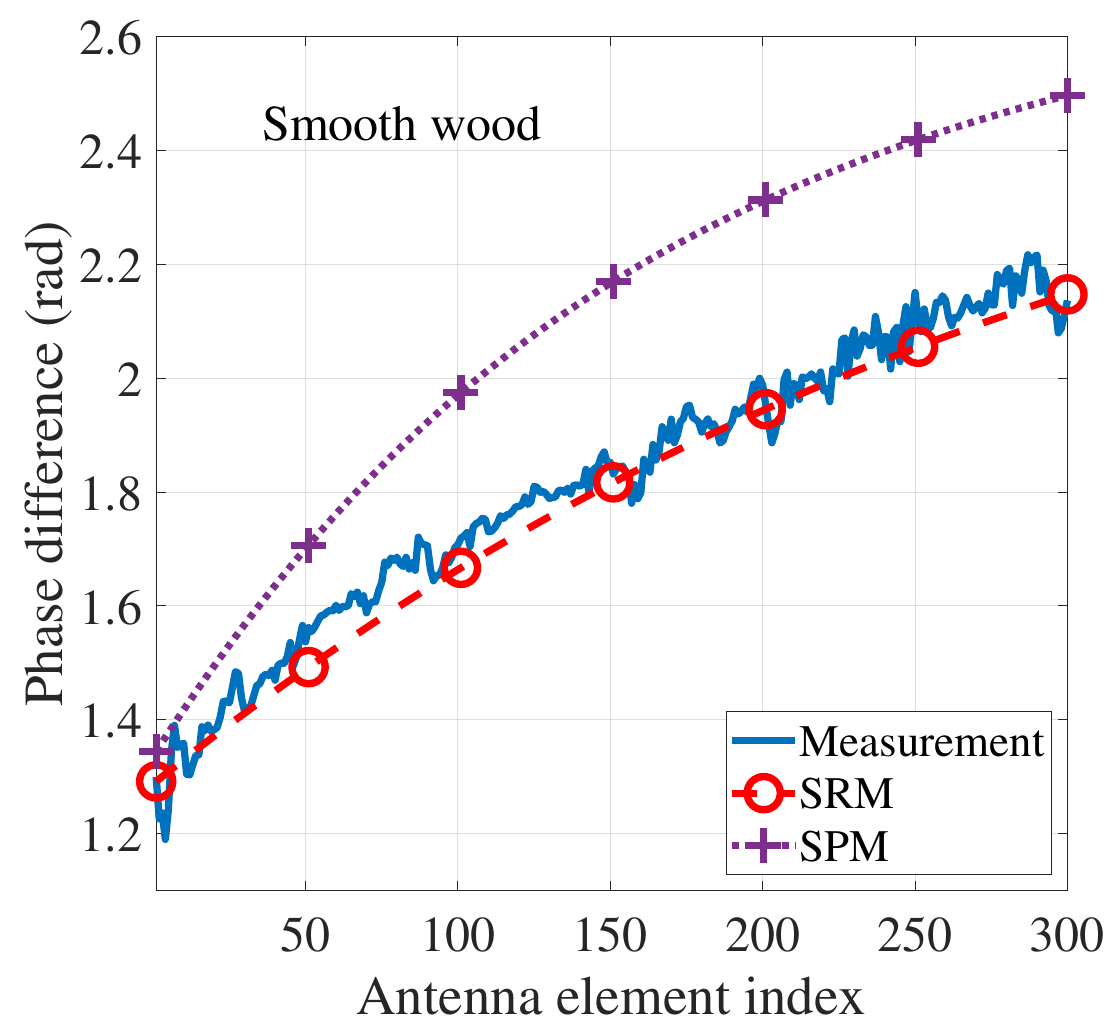} \label{Fig8_c}}
  \hfill
  \subfloat[]{\includegraphics[width=4.3cm]{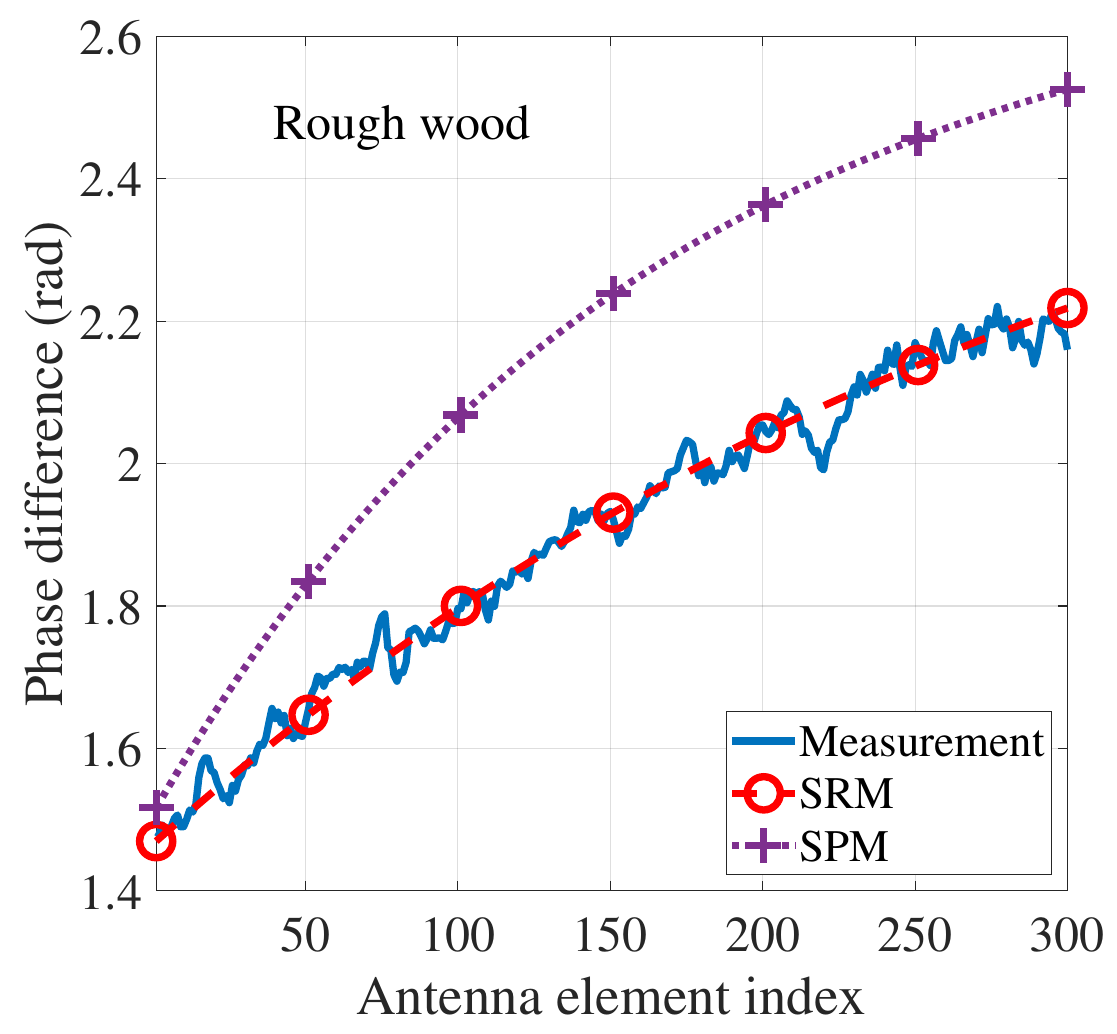} \label{Fig8_d}}
  \\
  \vspace{-1em}
  \subfloat[]{\includegraphics[width=4.3cm]{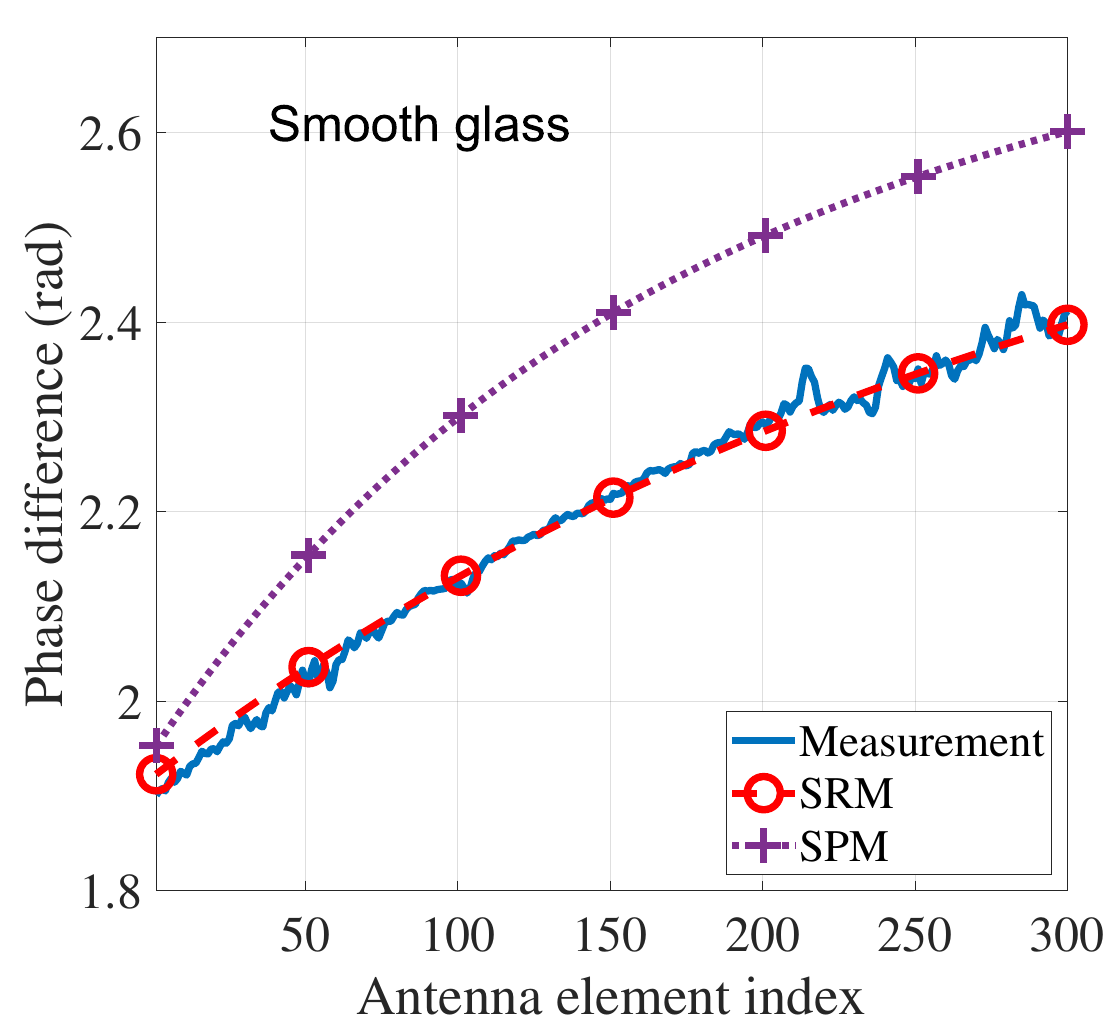} 
  \label{Fig8_e}}
  \subfloat[]{\includegraphics[width=4.3cm]{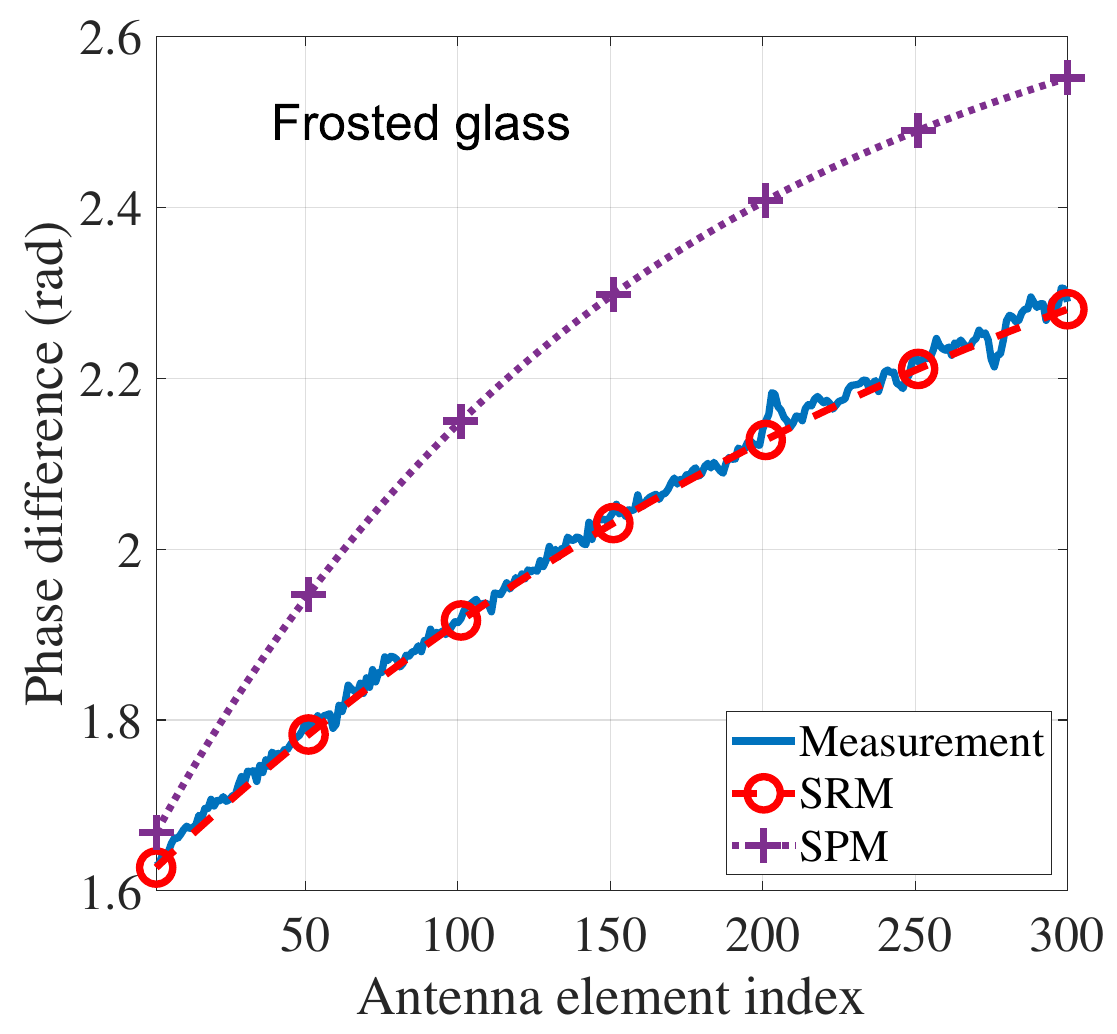} \label{Fig8_f}}
  \\
  \vspace{-1em}
  \subfloat[]{\includegraphics[width=7.4cm]{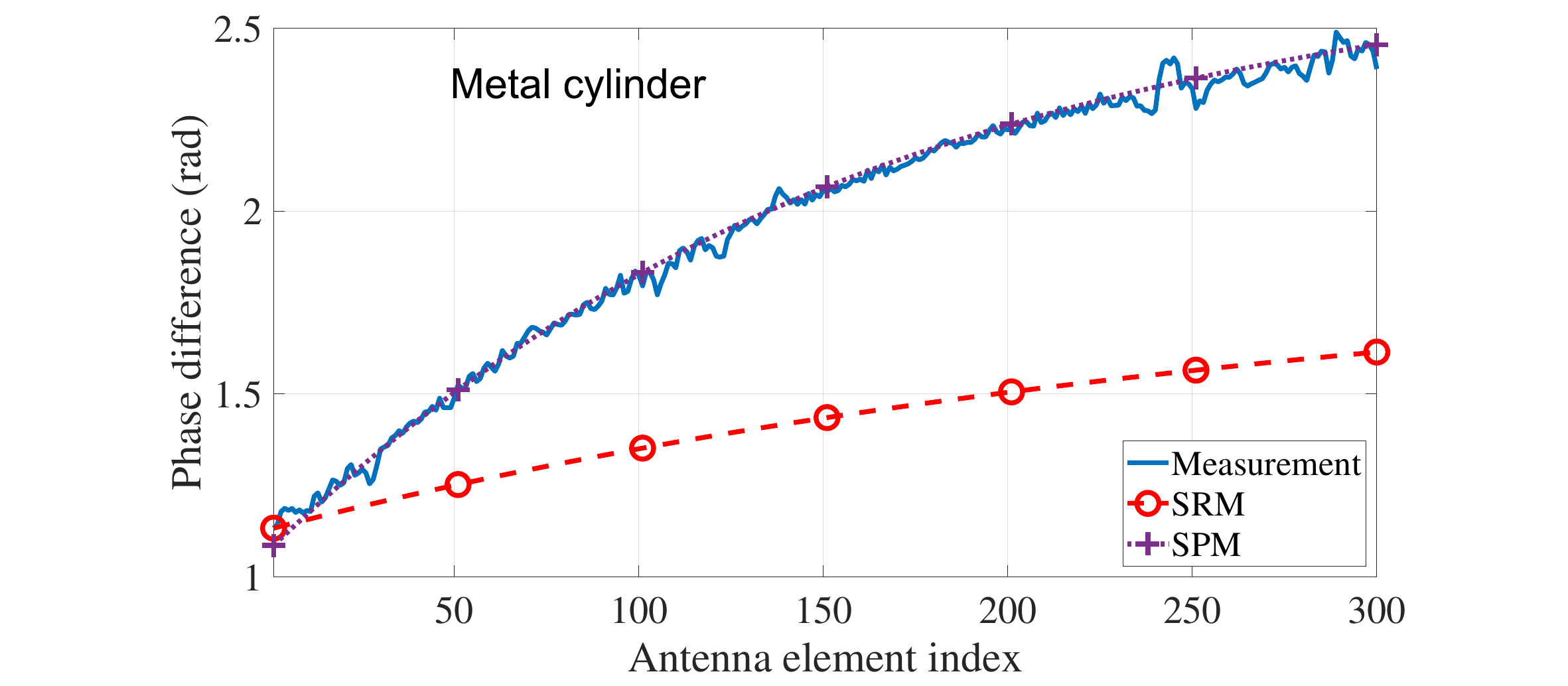} \label{Fig8_g}}
  \caption{Inter-element phase differences for different materials in Case 1. (a) LoS path; (b) NLoS path reflected by concrete; (c) NLoS path reflected by a smooth wooden surface; (d) NLoS path reflected by a rough wooden surface; (e) NLoS path reflected by a smooth glass; (f) NLoS path reflected by a frosted glass; (g) NLoS path scattered by a metallic cylinder.}
  \vspace{-0.5cm}
  \label{Fig_NF_Phase}
\end{figure}

NF propagation modeling can be categorized into two cases: LoS and NLoS paths. The modeling of the LoS path's spherical wavefront is relatively straightforward. The propagation parameters from each Tx antenna element to the Rx can be calculated individually based on their spatial positions. As shown in Fig.\ref{Fig8_a}, the measured inter-element phase differences for the LoS path closely match the theoretical values derived from corresponding propagation distances. Other element-wise path parameters can similarly be determined through geometric relationships and are omitted here for brevity. 

\textcolor{black}{In contrast, modeling NLoS paths is more challenging due to the complex wavefront behavior introduced by reflection and scattering. These characteristics depend not only on the scatterer's location but also on its impact on wavefront curvature. Existing mainstream approaches often model the scatterer as a point source, assuming that the signal originates from that point—a method referred to as scatterer-excited point-source model, as illustrated by the $l$-th path in Fig. \ref{Fig7_Framework}. However, since physical scatterers typically correspond to extended reflective surfaces, treating them as point sources may oversimplify the actual propagation process. To address this limitation, we propose a specular reflection model, in which certain NLoS paths are regarded as the result of specular reflections. In this model, the wavefront point source is assumed to emanate from the mirror image of the Rx with respect to the reflecting surface, rather than from the scatterer itself, as shown by the $l'$-th path in Fig. \ref{Fig7_Framework}.}

\textcolor{black}{To validate this model, experiments were conducted using five types of reflective materials with varying surface roughness and dielectric constants: concrete (lime-coated), smooth wooden board, rough wooden board, smooth glass, and frosted glass. As shown in Fig.\ref{Fig_NF_Phase}b–f, the measured inter-element phase differences for the NLoS paths were found to align well with the predictions of SRM. In contrast, SPM significantly overestimates the nonlinear phase variations across the array, resulting in inaccurate estimations of wavefront curvature. Moreover, the good agreement between SRM and measurements across surfaces with different roughness levels suggests that, despite the relatively short wavelength of THz signals (comparable to the dimensions of material surfaces), which enhances diffuse scattering and reduces specular energy, the specular component remains dominant. Therefore, NLoS paths arising from reflective surfaces can still be effectively modeled using SRM. However, for NLoS paths generated by small, irregularly shaped scatterers, SPM offers better accuracy. As shown in Fig. \ref{Fig8_g}, when a small metallic cylinder is used as the scatterer, the observed inter-element phase differences match the SPM prediction more closely, whereas the SRM underestimates the phase differences.}

\textcolor{black}{Consequently, a hybrid NF channel model is developed, in which different NLoS paths are modeled using either SRM or SPM, depending on their specific interaction with the environment. In SPM, the distance \( d \) is defined as the separation between the Tx element and the scatterer, whereas in SRM, \( d \) refers to the distance from the Tx element to the mirror image of the Rx point, corresponding to the total propagation distance of the path. In both cases, the \( l \)-th path propagation distances for the remaining array elements can be derived based on a reference element. Specifically,}
\begin{equation}
{d_{m,l}} = \left\| {d_{{\rm{ref}},l}}{{\bf{\Omega }}_{{{\rm{{ref}}},l}}^t - {{\textbf{\emph{r}}}_m}} \right\|,
\label{equ_d_ml}
\end{equation}
\noindent
where ${{\textbf{\emph{r}}}_m}$ represents the vector from the reference element to the $m$-th antenna element. \( d_{{\text{ref}},l} \) and \( \mathbf{\Omega}_{\text{ref},l}^t \) denote the propagation distance and the AoD direction vector of the \( l \)-th path at the reference Tx element, respectively. The direction vector ${\bf{\Omega }}$ is represented by a unit vector, uniquely determined by its azimuth angle $\phi$ and elevation angle $\theta$ as follows:
\begin{equation}
{\bf{\Omega }} = {\left[ {\sin\theta\cos\phi, \sin\theta\sin\phi, \cos\theta } \right]^T}.
\end{equation}

\textcolor{black}{Then, the amplitude, phase, delay, and angular parameters of the \( l \)-th NF path at the \( m \)-th element can be subsequently derived.
\begin{itemize}
    \item {Amplitude}: 
    \begin{equation}
        {\alpha_{m,l} = \alpha_{{\mathrm{ref}},l} \cdot \frac{d_{{\mathrm{ref}},l}}{d_{m,l}}}.
        \label{equ_NF_alpha}
    \end{equation}
    \item Phase:
    \begin{equation}
        {\varphi_{m,l} = \varphi_{\mathrm{ref},l} + \frac{2\pi f}{c} \left(d_{m,l} -d_{\mathrm{ref},l}\right)}.
    \end{equation}
    \item {Delay}:
    \begin{equation}
        \tau_{m,l} = \tau_{\mathrm{ref},l} + \frac{d_{m,l} - d_{\mathrm{ref},l}}{c}.
    \end{equation}
    \item {Angle}:
\begin{equation}
{{\bf{\Omega }}_{m,l}^t} = \frac{{{d_{{\rm{ref}},l}}{{\bf{\Omega }}_{{\rm{ref}},l}^t} - {{\textbf{\emph{r}}}_m}}}{{\left\| {{d_{{\rm{ref}},l}}{{\bf{\Omega }}_{{\rm{ref}},l}^t} - {{\textbf{\emph{r}}}_m}} \right\|}},
\label{Equ_omega_tx}
\end{equation}
\begin{equation}
{\bf{\Omega }}_{m,l}^r \approx \left\{ {\begin{array}{*{20}{c}}
{{\bf{\Omega }}_{m,l}^t - {\bf{\Omega}}_{{\rm{ref}},l}^t + {\bf{\Omega }}_{{\rm{ref}},l}^r,}&{{\rm{SRM}}},\\
{{\bf{\Omega }}_{{\rm{ref}},l}^r,}&{{\rm{SPM}}},
\end{array}} \right.
\label{Equ_omega_rx}
\end{equation}
\end{itemize}
\noindent
where \(\alpha_{\mathrm{ref},l}\), \(\varphi_{\mathrm{ref},l}\), \(\tau_{\mathrm{ref},l}\), \(\mathbf{\Omega}_{\mathrm{ref},l}^{t}\), and \(\mathbf{\Omega}_{\mathrm{ref},l}^{r}\) represent the amplitude, phase, delay, AoD vector, and AoA vector of the \(l\)-th path at the reference element, respectively.}

\textcolor{black}{
Accordingly, the $(m, l)$-th entry of ${\bf{A}}(f)$, denoted as ${a}_{m,l}(f)$, can be expressed as the transfer function difference between the $m$-th element and the reference element, i.e.,
\begin{equation}
\begin{aligned}
a_{m,l}^{{\rm{NF}}}(f) = & \dfrac{{d_{{\rm{ref}},l}}}  {{d_{m,l}}} \cdot \dfrac{F_t({\bf{\Omega }}_{m,l}^{t})}{F_t({\bf{\Omega }}_{{{\rm{ref}},l}}^{t})} \cdot \dfrac{F_r({\bf{\Omega }}_{m,l}^{r})}{F_r({\bf{\Omega }}_{{\rm{ref}},l}^{r})} \\
&\times \exp  \left(-  \dfrac{ j2\pi f} {c }  \left( {{d_{m,l}}} - {{d_{{\rm{ref}},l}}} \right) -j\varphi_{\mathrm{ref},l}\right),
\end{aligned}
\label{equ_NF}
\end{equation}
\noindent
where \(F_t(\cdot)\) and \(F_r(\cdot)\) denote the Tx and Rx antenna element radiation patterns, respectively.}

As discussed in Section \ref{Subsec_2-3}, the multipath power variations between XL-MIMO elements caused by spherical-wave propagation diffusion are minimal at practical Tx-Rx distances. Specifically, \({{d_{{\rm{ref}},l}}} / {{d_{m,l}}} \approx 1\). Under the FF assumption, \(\mathbf{\Omega}_{m,l}^{t} = \mathbf{\Omega}_{\mathrm{ref},l}^{t}\) and ${{{\bf{\Omega }}_{m,l}^r}} = {{{\bf{\Omega }}_{{\rm{ref}},l}^r}}$. The multipath phase is linearly related to the element index \( m \). Therefore, the \((m,l)\)-th entry of the FF matrix \(\mathbf{A}_{\mathrm{FF}}(f)\) can be expressed as
\begin{equation}
a_{m,l}^{\rm{FF}}(f) = \exp  \left( - \dfrac{j2\pi f \delta} { c} (m-1) \cos\phi \sin\theta -j\varphi_{\mathrm{ref},l}\right),
\label{equ_FF}
\end{equation}
where \(\delta\) denotes the antenna element spacing.

\subsection{Spatial Non-Stationarity}
\label{Subsec_3-2-2}

The SnS channel model primarily focuses on power variations across XL-MIMO antenna elements for each path. For the LoS path, power variations across elements are mainly attributed to partial blockage affecting specific antenna elements. In contrast, for NLoS paths, SnS may arise not only from blockage but also from inconsistent reflection or scattering, as illustrated in Fig. \ref{Fig_SnS_observation}. Given the wide variety of materials, geometries, and surface roughness of scatterers in realistic environments, accurately modeling SnS effects remains challenging. 

\textcolor{black}{To address this, a statistical modeling approach is adopted. Accordingly, all identified SnS paths across the 12 Rx positions in Case 3 were analyzed statistically. First, the path parameters \(\{\alpha, \tau\}\) for each antenna element were extracted from the CIRs. Then, a tracking algorithm was applied to monitor the variation of each path’s parameters across antenna elements. SnS paths were identified based on power variation thresholds. Specifically, to account for potential system jitter and NF-induced effects, a path was classified as SnS if its power variation across the array exceeded 3 dB. A total of 248 SnS paths were selected for analysis.} Due to significant differences in average power among different SnS paths, direct statistical analysis would be biased. Therefore, the amplitude of each SnS path was normalized, yielding AAF for each element along the path, defined as:
\begin{equation}
s_{m,l} = \frac{\alpha_{m,l} }{\max_{m'}(\alpha_{m',l}) }.
\label{equ_S}
\end{equation}
\noindent
where \(s_{m,l}\) denotes the normalized amplitude of the \(l\)-th path at the \(m\)-th antenna element, serving as its AAF. Conventional VR models characterize SnS using a binary indicator $s_{m,l}=0/1$, capturing only the presence or absence of a path at each element. However, such a binary approach fails to reflect the continuous nature of power variations across the array, which contradicts the physical principles of wireless propagation—multipath energy does not abruptly appear or vanish. To overcome this limitation, we adopt a statistical approach to model the continuous evolution of the AAF for each path. Rather than using discrete visibility states, the AAF is extracted from measurements and used to capture the smooth spatial transitions characteristic of SnS. In the following, we analyze the empirical characteristics of AAF and propose a statistical modeling framework based on its observed behavior.

\subsubsection{Statistical Characteristics of AAF}
\label{Subsec_3-2-2-1}

\textcolor{black}{
Based on the experimental observations, two key characteristics of SnS paths can be identified. First, different SnS paths exhibit varying power variation patterns and dynamic ranges across antenna elements, which leads to path-dependent AAFs. Second, power variation along a path is continuous, with no abrupt jumps or disappearances, implying that AAFs should also exhibit smooth transitions across the array. Accordingly, the statistical distribution and autocorrelation properties of the AAF are analyzed in the following.}

\begin{itemize}
    \item \textcolor{black}{\bf{Statistical distribution characteristic}}
\end{itemize}

\begin{figure}[htbp]  
    \centering  
    \vspace{-1em}
    \subfloat[]{\includegraphics[width=4.3cm]{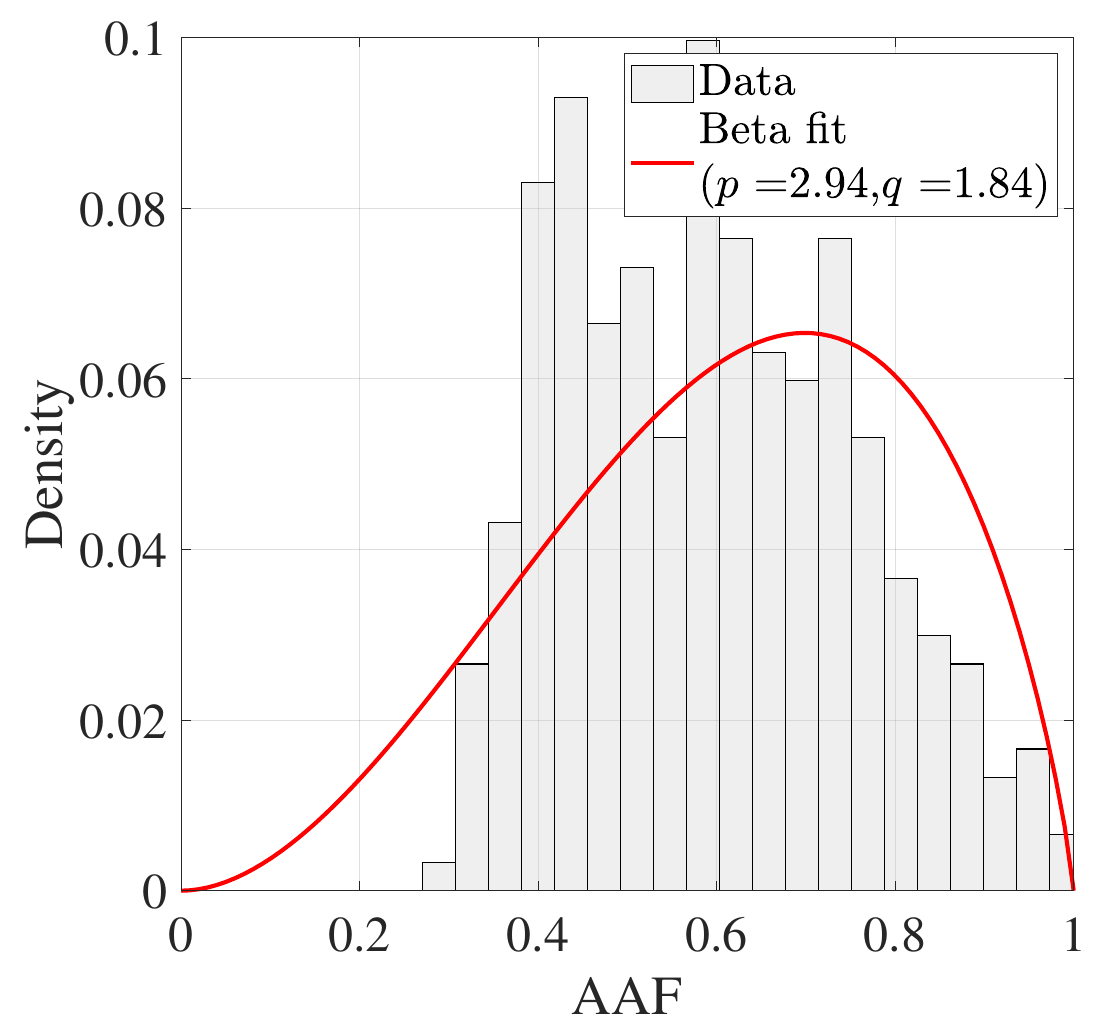} \label{Fig10_a}}
  \hfill
  \subfloat[]{\includegraphics[width=4.3cm]{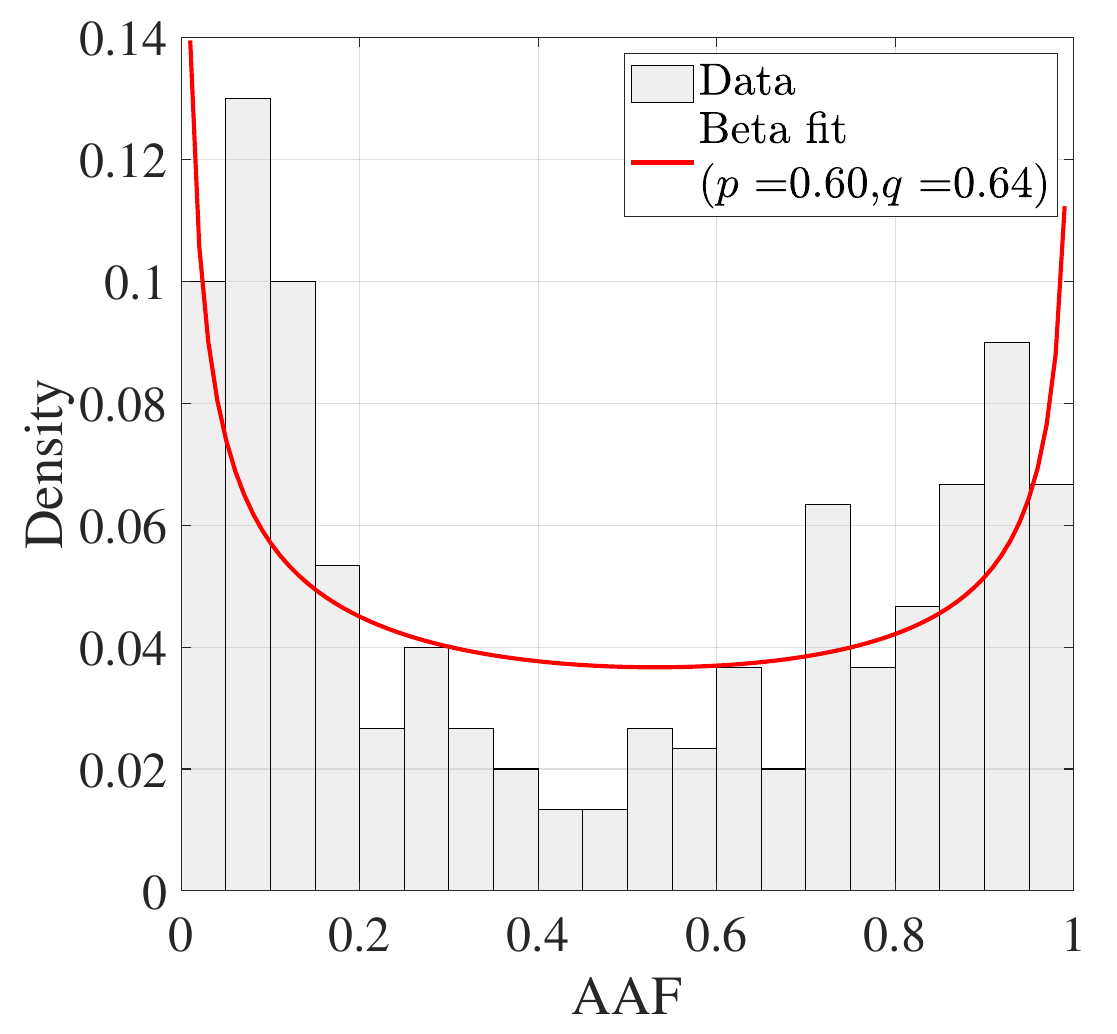} 
  \label{Fig10_b}}
    \caption{Probability density function of AAFs and corresponding Beta distribution fitting for (a) an NLoS path reflected by smooth wood in Case 1 and (b) an NLoS path in Case 3.} %
    \vspace{-0.6em}
    \label{Fig_AAF}  
\end{figure}

\textcolor{black}{The Beta distribution is employed to model the statistical behavior of each AAF sequence \( s_{:,l} \), as it is well-suitable for variables defined within the interval \([0,1]\). As illustrated in Fig. \ref{Fig_AAF}, the distribution of AAFs varies significantly across paths and conditions, often exhibiting strong skewness or even multimodal patterns. The flexibility of the Beta distribution stems from its two shape parameters, \( p \) and \( q \), which control the skewness and modality of the distribution. Specifically, the absolute values of \( p \) and \( q \) determine the sharpness of the distribution, while their relative ratio governs its asymmetry. This flexibility enables the Beta distribution to effectively capture the empirical variations observed in the AAFs.}
\graphicspath{{picture/}}

\begin{figure}[htbp]  
    \centering  
    \vspace{-1em}
    \subfloat[]{\includegraphics[width=4.3cm]{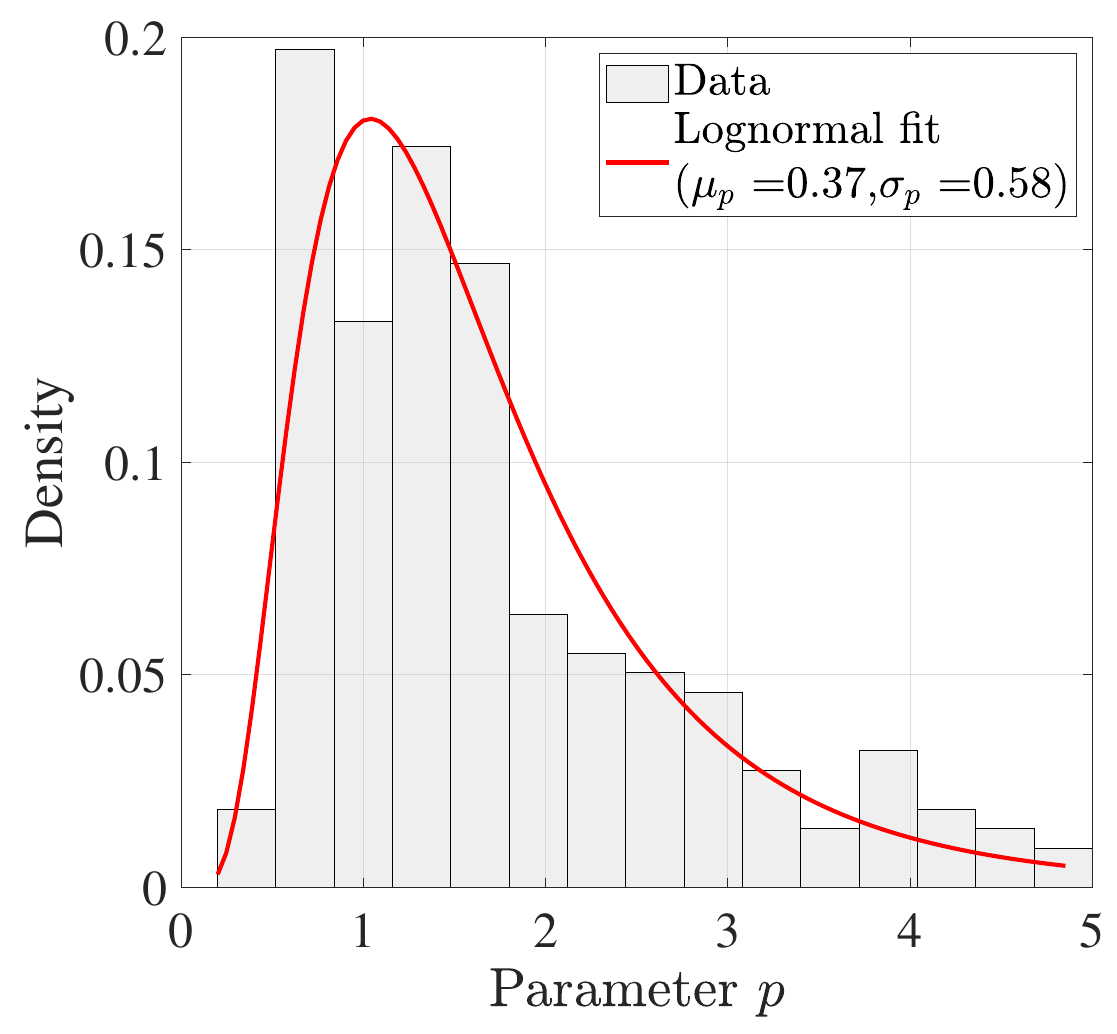} \label{Fig11_a}}
  \hfill
  \subfloat[]{\includegraphics[width=4.3cm]{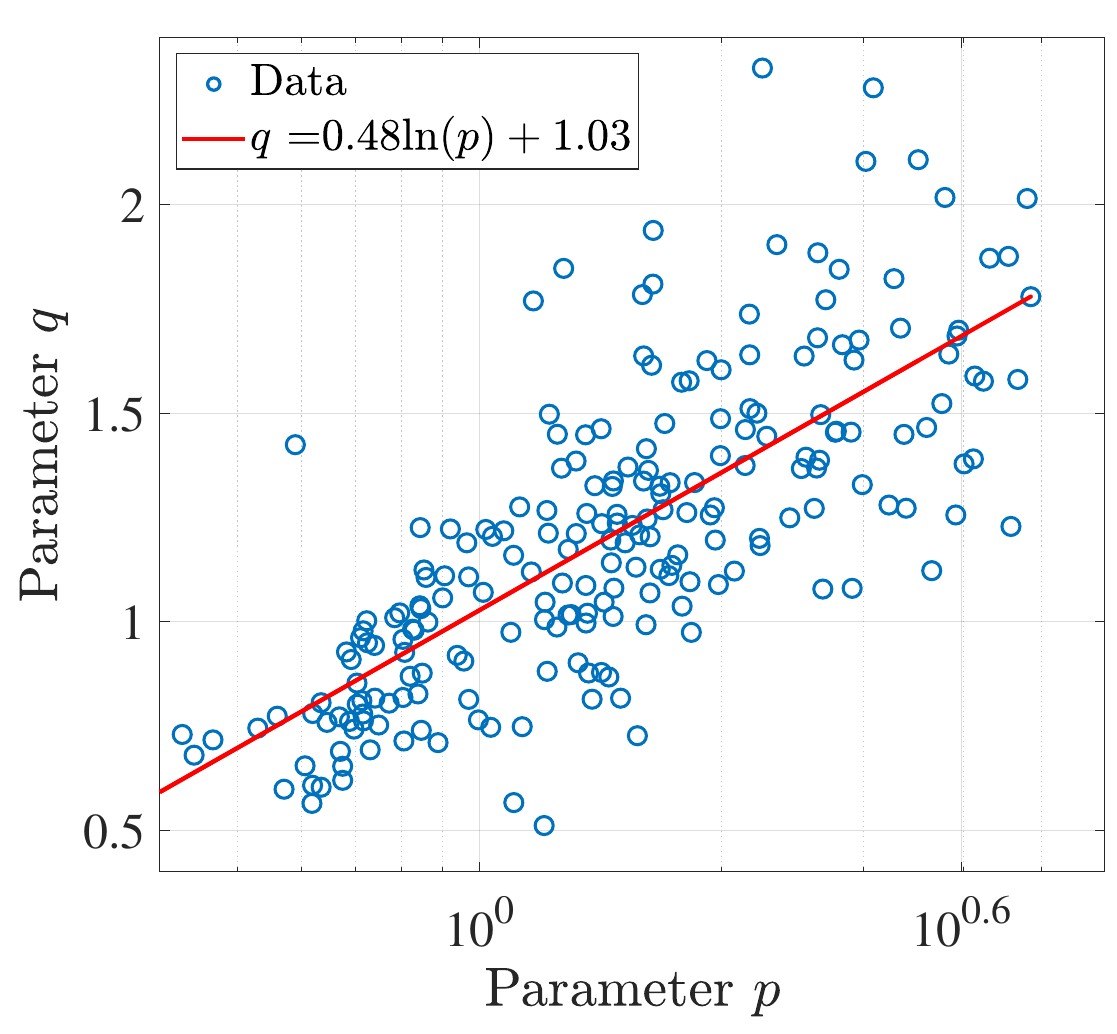} 
  \label{Fig11_b}}
    \caption{(a) Probability density function of the Beta distribution parameter \( p \) for the AAFs of SnS paths, along with its log-normal fit. (b) Relationship between the Beta distribution parameters \( p \) and \( q \).} %
    \vspace{-1em}
    \label{Fig11_S_CDF}  
\end{figure}

\textcolor{black}{Statistical analysis of all identified SnS paths reveals that the parameter \( p \) follows a log-normal distribution, with a mean and variance of 0.37 and 0.58, respectively, as shown in Fig. \ref{Fig11_a}. Furthermore, when \( p \) is represented on a logarithmic scale, a strong linear relationship with \(q\) emerges, as illustrated in Fig. \ref{Fig11_b}. This relationship can be approximated by the following model: \(q=0.48{{\ln}}(p)+{1.03}\).}

\graphicspath{{picture/}}

\begin{itemize}
    \item \bf{\textcolor{black}{Autocorrelation characteristic}}
\end{itemize}

The spatial autocorrelation of the AAF sequence characterizes the similarity of path amplitudes across antenna elements on a wavelength scale. It provides insight into how smoothly the path amplitude varies along the array. The spatial autocorrelation function (ACF) for the \(l\)-th path is defined as
\begin{equation}
{\mathrm{ACF}}_l \left( {\Delta x} \right) = \frac{\sum_{m=1}^{M-\Delta x} \left( s_{m,l} - \bar{s}_l \right) \left( s_{m+\Delta x,l} - \bar{s}_l \right)}{\sum_{m=1}^{M} \left( s_{m,l} - \bar{s}_l \right)^2},
\label{equ_rho}
\end{equation}
\noindent
where \( \bar{s}_l = \dfrac{1}{M} \sum_{m=1}^{M} s_{m,l} \) denotes the mean normalized AAF of path \(l\), and \(\Delta x = 0, \ldots, M-1\) is the antenna element index difference, corresponding to a physical spacing of \(\Delta x\delta\).

\begin{figure}[htbp]  
    \centering  
    \subfloat[]{\includegraphics[width=4.3cm]{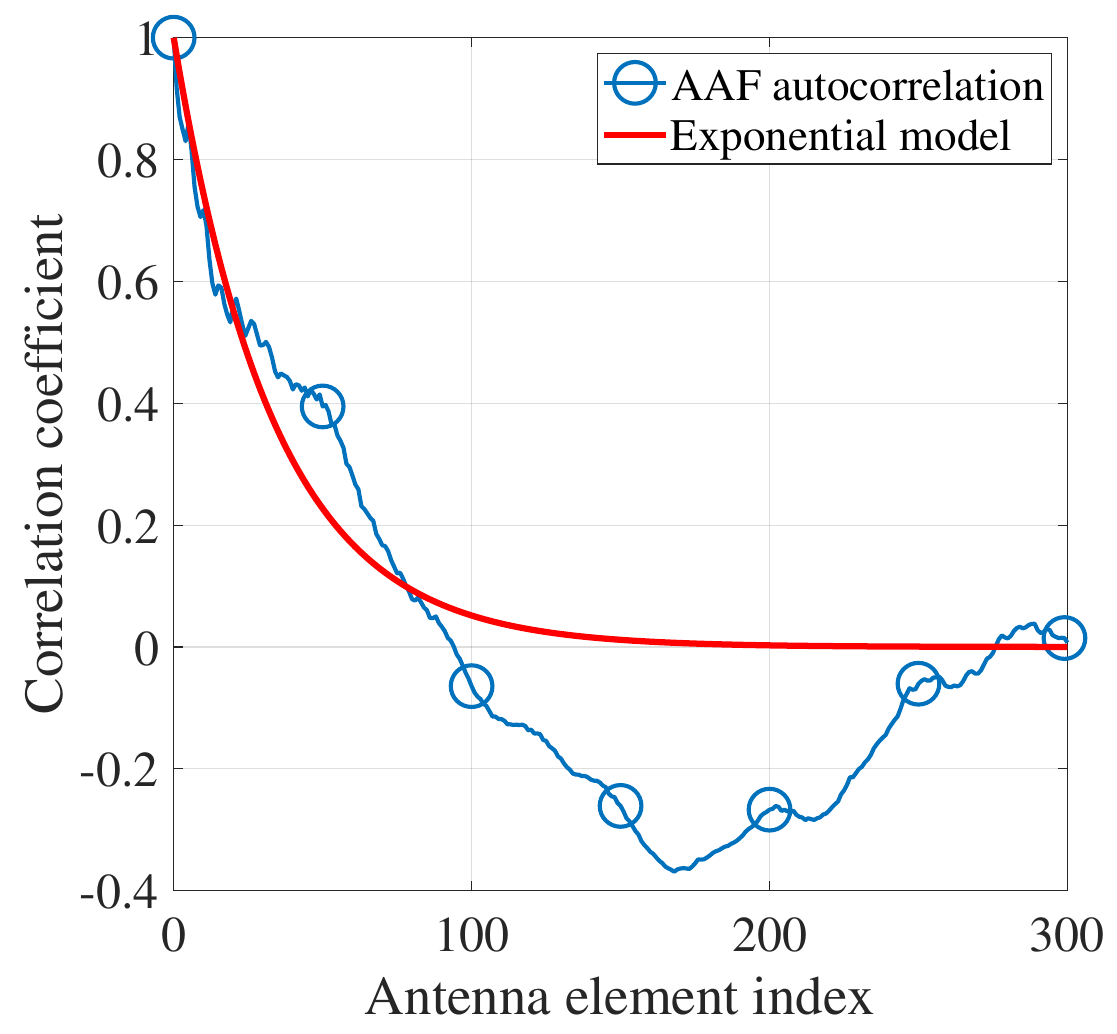} \label{Fig12_a}}
  \hfill
  \subfloat[]{\includegraphics[width=4.3cm]{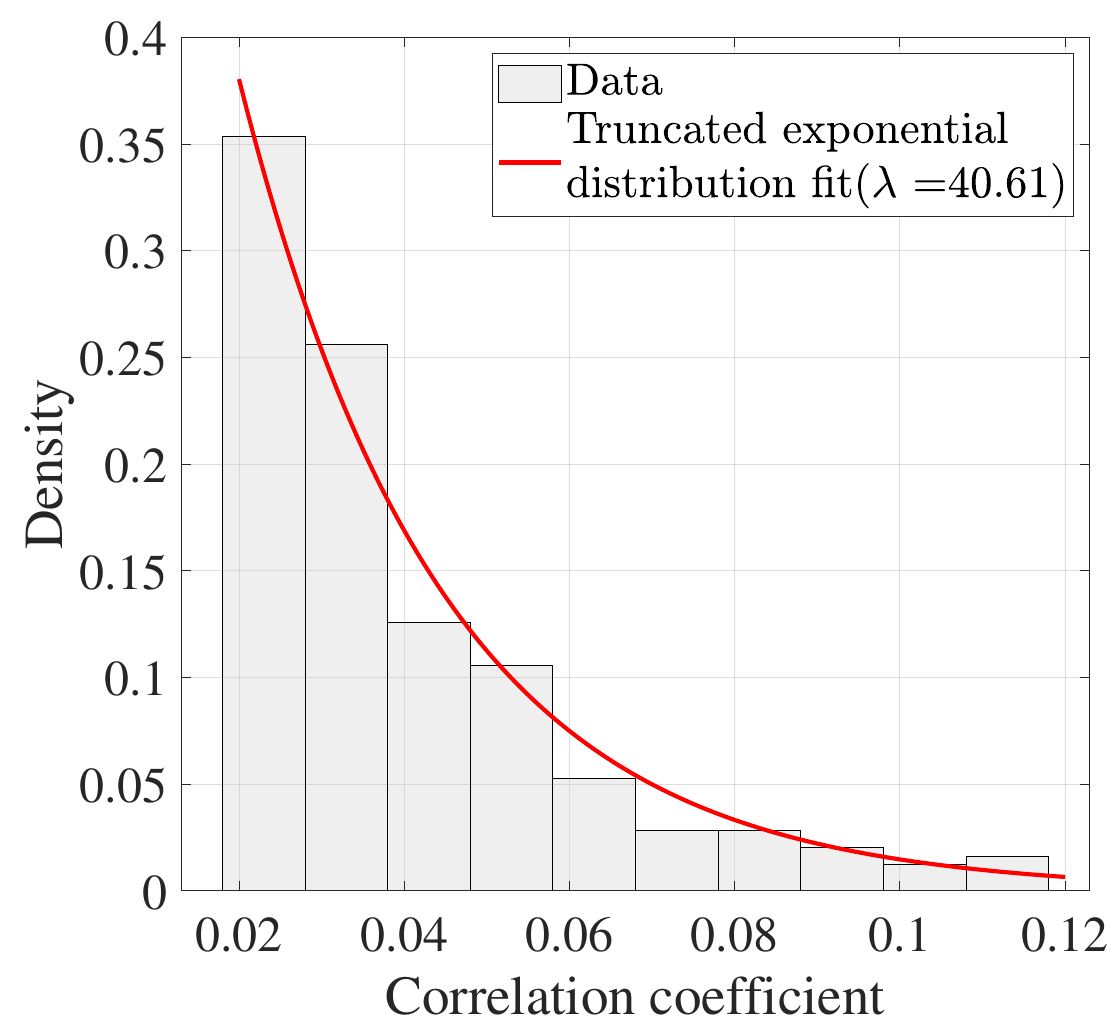} 
  \label{Fig12_b}}
    \caption{\textcolor{black}{(a) Spatial autocorrelation of AAF and corresponding exponential model fitting. (b) Statistical distribution of correlation coefficients and their truncated exponential fitting.}} %
    \label{Fig_Corr}
\end{figure}

\textcolor{black}{Empirical observations show that the spatial autocorrelation of AAFs for individual SnS paths decays exponentially with increasing element spacing, as shown in Fig. \ref{Fig12_a}. This behavior is accurately fitted by an exponential model \cite{C3_Corr_Sunshu}:
\begin{equation}
\rho(\Delta x) = \exp\left(-d_{\rm corr} \Delta x\right),
\label{equ_rho_fit}
\end{equation}
\noindent
where \( d_{\rm corr} \) represents the spatial autocorrelation decay coefficient, estimated by minimizing the mean square error between the measured ACF and the exponential model. Across all observed SnS paths, the fitted values of \( d_{\mathrm{corr} } \) follow a truncated exponential distribution, with a fitted parameter \( \lambda_{\mathrm{corr} } = 40.61 \), as illustrated in Fig. \ref{Fig12_b}. This right-skewed distribution indicates that most SnS paths exhibit high spatial correlation in their AAFs.} 

\graphicspath{{picture/}}

\subsubsection{Statistical Generation of AAF}
\label{Subsec_3-2-2-2}

For SS paths, the AAF is equal to 1 across all antenna elements. For SnS paths, the traditional VR model typically assumes that the path power remains constant within a certain region of the array and drops abruptly to zero outside of it \cite{C1_VR_3,C1_VR_1}. However, this binary approximation is insufficient to capture the gradual power variation observed across antenna elements when an SnS path is present. Instead of assigning binary values (0 or 1), the AAF should take continuous values between 0 and 1 to accurately reflect this behavior.

\begin{figure}[htbp]  
    \centering  
    \includegraphics[width=7.2cm]{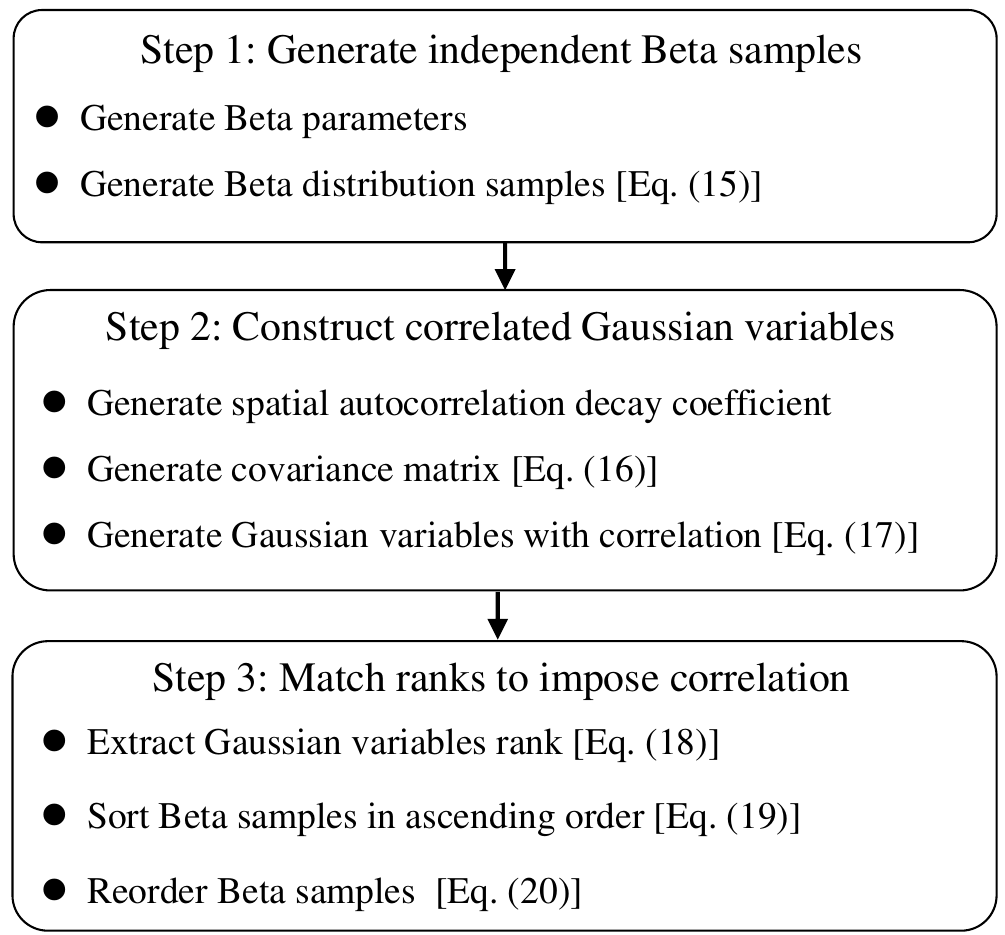} 
    \caption{Flowchart of the AAF generation procedure.} %
    \vspace{-1em}
    \label{Fig_AAF_Gen}  
\end{figure}

\textcolor{black}{To characterize SnS more accurately, we propose a statistical method for generating AAFs based on a rank-matching approach. The core idea is to first generate a sequence of independent random variables following the desired marginal distribution of the AAF. Then, spatial correlation is introduced via rank-based mapping, ensuring that the resulting sequence maintains both the target distribution and the spatial correlation structure across antenna elements. The overall procedure is illustrated in the flowchart shown in Fig. \ref{Fig_AAF_Gen}. The required model parameters are listed in Table \ref{Tab_AAF}\footnote{Distribution parameters for other deployment scenarios can be obtained through additional channel measurements or ray-tracing simulations tailored to specific environments.}. The detailed procedure for generating the AAF sequence of a single SnS path across the entire XL-MIMO array is described below.}

\graphicspath{{picture/}}

\begin{table}
\renewcommand\arraystretch{1.2}
    \centering
    \caption{Parameter Settings for the Proposed Statistical AAF Generation Method Based on Rank Matching}
    \centering
    \begin{tabular}{m{1.2cm}<{\centering}|m{2.4cm}<{\centering}|m{2cm}<{\centering}|m{1.5cm}<{\centering}}
    \hline
        Model Parameter & Distribution/Function & Parameter Values & Value Range\\ \hline
        \multirow{3}{*}[-2ex]{\thead{AAF  \\ distribution}} & \multicolumn{2}{c|}{ { \(s_{:,l} \sim {\rm{Beta}}(p,q)\)}}&  [0,1] \\ \cline{2-4}
        & \(p \sim {\rm{Logn}} (\mu_p,\sigma_p^2)\) & { \(\mu_p = 0.37\)  \(\sigma_p=0.58\)} & {[0.2,5]} \\ \cline{2-4}
        & \(q=\xi {\rm{ln}}(p)+{\gamma}\) & \(\xi\) = 0.48 \(\gamma = 1.03\) & - \\ \hline
        AAF Correlation & \({d_{\rm{corr}}} \sim {\rm{TruncExp}}(\lambda_{\rm{corr}})\) & \(\lambda_{\rm{corr}}=40.61\) &[0.018,0.12] \\ \hline         
    \end{tabular}
    \begin{tablenotes}
      \item[1] Notations: "Logn" and "TruncExp" represent lognormal and truncated exponential distributions, respectively.
    \end{tablenotes}
    \vspace{-2em}
    \label{Tab_AAF}
\end{table}

\textcolor{black}{
\textbf{Step 1:}  Generate independent Beta samples.
A set of \(M\) independent and identically distributed (i.i.d.) Beta random variables is generated.:
\begin{equation}
{{\mathbf{{x}}}} = {\left[ {{x_{1}},{x_{2}}, \cdot  \cdot  \cdot ,{x_{M}}} \right]^T} \sim {\rm{Beta}} \left( {{p},q} \right).
\label{equ_S_1}
\end{equation}
where the shape parameter \( p \)  is sampled from a log-normal distribution, \( p \sim \mathrm{Logn}(\mu_p, \sigma_p^2) \), while \( q \) follows a log-linear relationship with \(p\), given by \( q = \xi {\rm{ln}}(p)+\gamma \). This procedure ensures that the generated Beta samples match the observed marginal distribution of AAFs.}

\textcolor{black}{\textbf{Step 2:} Construct correlated Gaussian variables.}
\textcolor{black}{To embed spatial correlation into the AAF sequence, begin by constructing the covariance matrix \(\Sigma \in \mathbb{R}^{M \times M}\) with an exponential decay structure:}
\begin{equation}
    \Sigma_{i,j} = \rho(|i-j|) = \exp\left({-{d_{\rm{corr}}} |i - j|}\right), \quad 1 \leq i, j \leq M,
\end{equation}
where \(d_{\rm{corr}} \sim \mathrm{TruncExp}(\lambda_{\mathrm{corr}})\) controls the rate of spatial decorrelation. Apply Cholesky decomposition \(\boldsymbol{\Sigma} = LL^T\), and generate a standard Gaussian vector \(\mathbf{z} \sim \mathcal{N}(0, \mathbf{I})\). Transform \(\mathbf{z}\) to obtain a correlated Gaussian vector \(\mathbf{y}\) with zero mean and covariance \(\boldsymbol{\Sigma}\):
\vspace{-0.2em}
\begin{equation}
     \mathbf{{y}} = {L} \mathbf{{z}} \quad \Rightarrow \quad \mathbf{{y}} \sim \mathcal{N}(0, \boldsymbol{\Sigma}). 
\end{equation}
\vspace{-0.2em}
This transformation ensures that the elements of $\mathbf{{y}}$ exhibit the desired spatial correlation structure.

\textcolor{black}{\textbf{Step 3:} Match ranks to impose correlation.
Sort the correlated Gaussian vector \(\mathbf{y}\), and record the rank (i.e., the position in the sorted order) of each entry:
\begin{equation}
    R_{y_i} = \text{rank}(y_i) = \pi_y^{-1}(i), \quad i = 1, 2, \dots, M,
\end{equation}
where \( \pi_y^{-1} \) denotes the inverse permutation that maps the sorted indices back to their original positions in \(\mathbf{y}\). Next, sort the Beta samples in ascending order:
\begin{equation}
      {{x}}_{\text{sorted}} = \left\{ x_{(1)}, x_{(2)}, \dots, x_{(M)} \right\}, \\ x_{(1)} \leq x_{(2)} \leq \dots \leq x_{(M)}.
\end{equation}}

Finally, construct the correlated AAF sequence ${\textbf{\textit{s}}}_{:,l}$ by reordering the Beta samples according to the ranks of $\textbf{\textit{y}}$:
\begin{equation}
      {\textbf{\textit{s}}}_{:,l} = \left\{ x_{(R_{y_1})}, x_{(R_{y_2})}, \dots, x_{(R_{y_M})} \right\}.
\end{equation}

\textcolor{black}{This rank-based mapping approach, grounded in copula theory \cite{C3_Coupla}, preserves the marginal Beta distribution while accurately transferring the spatial dependence structure from the Gaussian vector to the AAF sequence. As Spearman’s rank correlation depends only on the order of values, this method ensures that the final AAFs maintain the target correlation without distorting their marginal distribution.}

It is worth noting that the AAF represents the normalized amplitude of the \( l \)-th path at each antenna element, relative to the maximum amplitude across the array. In contrast, Eq. (\ref{equ_H_MIMO}) models the amplitude variation with respect to a designated reference element. Accordingly, a conversion is required to align the normalization schemes. The AAF used in the channel model is expressed as:
\begin{equation}
    \hat{s}_{m,l} = {{{s}}}_{m,l} \cdot \dfrac{\max_m \alpha_{m,l}}{\alpha_{{\rm{ref}},l}}.
    \label{equ_SNS_end}
\end{equation}

For analytical convenience, the reference amplitude \( \alpha_{{\rm ref},l} \) is typically set equal to the maximum value \( \max_m \alpha_{m,l} \), in which case both normalization schemes become equivalent.

\section{Model Implementation and Validation}
\label{Sec_4}

\begin{figure}[htbp]  
    \centering  
    \includegraphics[width=7cm]{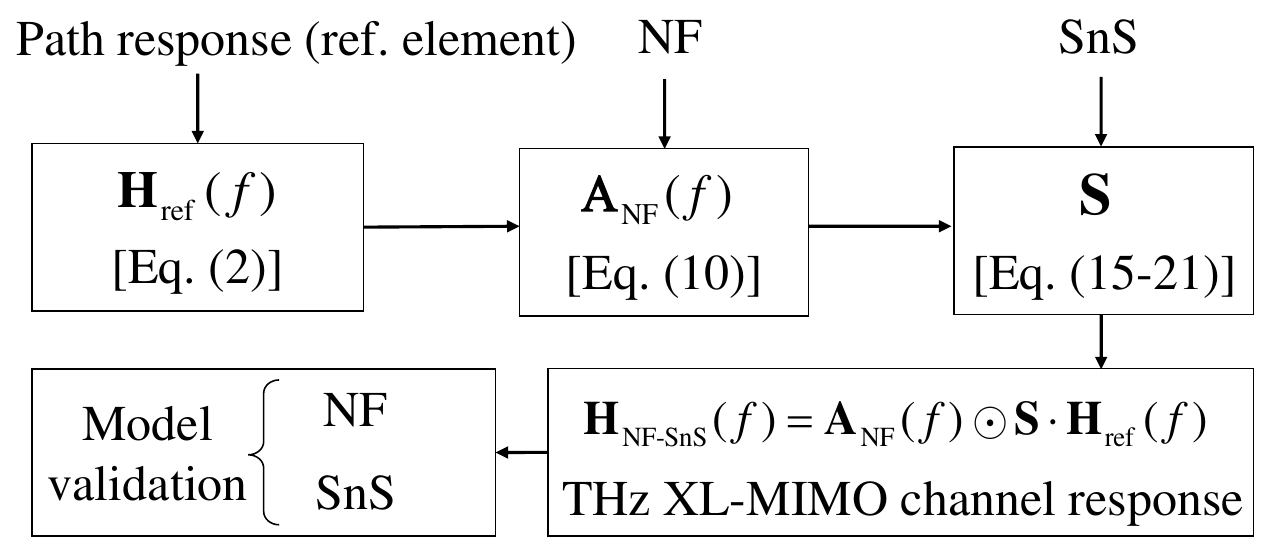} 
    \caption{The implementation and validation framework of the proposed model.} %
    \vspace{-0.4cm}
    \label{Fig_model}  
\end{figure}

In this section, we describe the procedure for generating THz XL-MIMO channels using the proposed model and validate its accuracy through measurement data, as illustrated in Fig. \ref{Fig_model}.

\graphicspath{{picture/}}

\vspace{-1em}
\subsection{Implementation}
The steps for implementing the proposed XL-MIMO channel model are summarized as follows
\begin{itemize}
\item [1)] 
Generate the reference element channel responses \({{{\bf{H}}_{\rm{ref}}}}(f)\): The frequency response of \(L\) paths is extracted from measurement data at a designated reference element. The extracted parameters include amplitude, phase, delay, propagation distance, and angles \(\left\{\alpha_{\text{ref},l}, \varphi_{\text{ref},l},\tau_{\text{ref},l},d_{\text{ref},l}, \Omega_{t,\text{ref},l}, \Omega_{r,\text{ref},l}\right\}\).
\item [2)] 
Generate the NF spherical wave matrix \({\bf{A}}(f)\):
Based on the XL-MIMO array geometry, the variations in amplitude, phase, delay, and angle caused by NF spherical wavefronts are computed according to (\ref{equ_NF_alpha})–(\ref{Equ_omega_rx}), using the reference element as a baseline. The corresponding NF response matrix \( \mathbf{A}(f) \) is then derived from Eq. (\ref{equ_NF}).

\item [3)] 
Generate the AAF matrix \({\bf{S}}\): For SS paths, the AAF is set to 1 across all antenna elements. For SnS paths, the AAF is statistically generated according to the Beta distribution and spatial correlation model in (\ref{equ_S_1})–(\ref{equ_SNS_end}).
\item [4)] 
Construct the THz XL-MIMO channel response: Substitute the reference response \( \mathbf{H}_{\mathrm{ref}}(f) \), the NF matrix \( \mathbf{A}(f) \), and the AAF matrix \( \mathbf{S} \) into (\ref{equ_H_MIMO}) to obtain the final THz XL-MIMO channel response.
\end{itemize}

\vspace{-1em}
\subsection{Validation}

The proposed model will be validated from both NF propagation and SnS aspects. A practical and intuitive validation approach is to compare the statistical parameters calculated from the simulated channel with those obtained from measurements. This validation approach is widely adopted in existing channel modeling literature \cite{C1_VR_1,C4_Veri_Rapp}, with typical metrics including entropy capacity, condition number, channel spatial correlation, channel
gain, Rician K-factor, and root mean square (RMS) delay spread. The discrepancy between the cumulative distribution functions (CDFs) of the simulated and measured results is quantified using the Cramér–von Mises (CvM) statistic, with the results summarized in Tables \ref{Tab_Veri_NF} and \ref{Tab_Veri_SNS}. Smaller CvM values indicate higher distribution similarity. The results show that the proposed NF–SnS model closely matches the measurement data across all evaluated metrics, thereby validating its accuracy and reliability. 
\subsubsection{Near-field propagation validation}

\textcolor{black}{We employ the channel entropy capacity, Demmel condition number and path gain as validation metrics. We first define the multi-user channel response matrix as:
\begin{equation}
    \mathbf{H}(f) = \left[\mathbf{H}_{\mathrm{NF\text{-}SnS}}^1(f), \dots, \mathbf{H}_{\mathrm{NF\text{-}SnS}}^n(f) ,\dots,\mathbf{H}_{\mathrm{NF\text{-}SnS}}^N(f) \right]^T,
    \label{equ_MIMO}
\end{equation}
where \(N\) denotes the number of single-antenna UEs. \( \mathbf{H}(f) \in \mathbb{C}^{N \times M} \) represents the frequency-domain channel responses from the array to $N$ UEs. Specifically, \( \mathbf{H}_{\mathrm{NF\text{-}SnS}}^n(f) \in \mathbb{C}^{M \times 1} \) denotes the channel response vector of the \(n\)-th UE under the proposed NF–SnS channel model. The entry \( \mathbf{H}_{n,m}(f_k) \) denotes the \((n,m)\)-th element of the channel matrix at frequency point \( f_k \), corresponding to the response between the \(m\)-th antenna element at the base station and the \(n\)-th UE.
}
\begin{itemize}
    \item \textcolor{black}{Entropy capacity}
\end{itemize}

\textcolor{black}{Assuming that the channel state information at the Tx is unavailable and water-filling power allocation is not applied, the entropy capacity is calculated as:
\begin{equation}
C = \frac{1}{{{K}}}\sum\limits_{k = 1}^{{K}} {\sum\limits_{i = 1}^I {{{\log }_2}\left( {1 + \frac{\Gamma }{{M\eta }}\sigma _{i,k}^2} \right)} } ,
\end{equation}
where \( K = 2001 \) is the number of frequency points, \( \Gamma = 15 \) dB denotes the signal-to-noise ratio (SNR). Singular value decomposition is performed on \( \mathbf{H}(f_k) \), and \( \sigma_{i,k} \) denotes the \(i\)-th singular value \cite{C3_PSS}. The number of nonzero singular values is given by \( I = \min(M,N) \). The average channel gain \( \eta \) is defined as:}

\textcolor{black}{
\begin{equation}\eta  = \frac{1}{{M{N}{K}}}\sum\limits_{k = 1}^{{K}} {\sum\limits_{n = 1}^{{N}} {\sum\limits_{m = 1}^M {{{\left| {{{\bf{H}}_{n,m}}\left( {{f_k}} \right)} \right|}^2}} } }. \end{equation}
}

\begin{figure}[htbp]  
    \centering  
     \subfloat[]
    {\includegraphics[width=8cm]{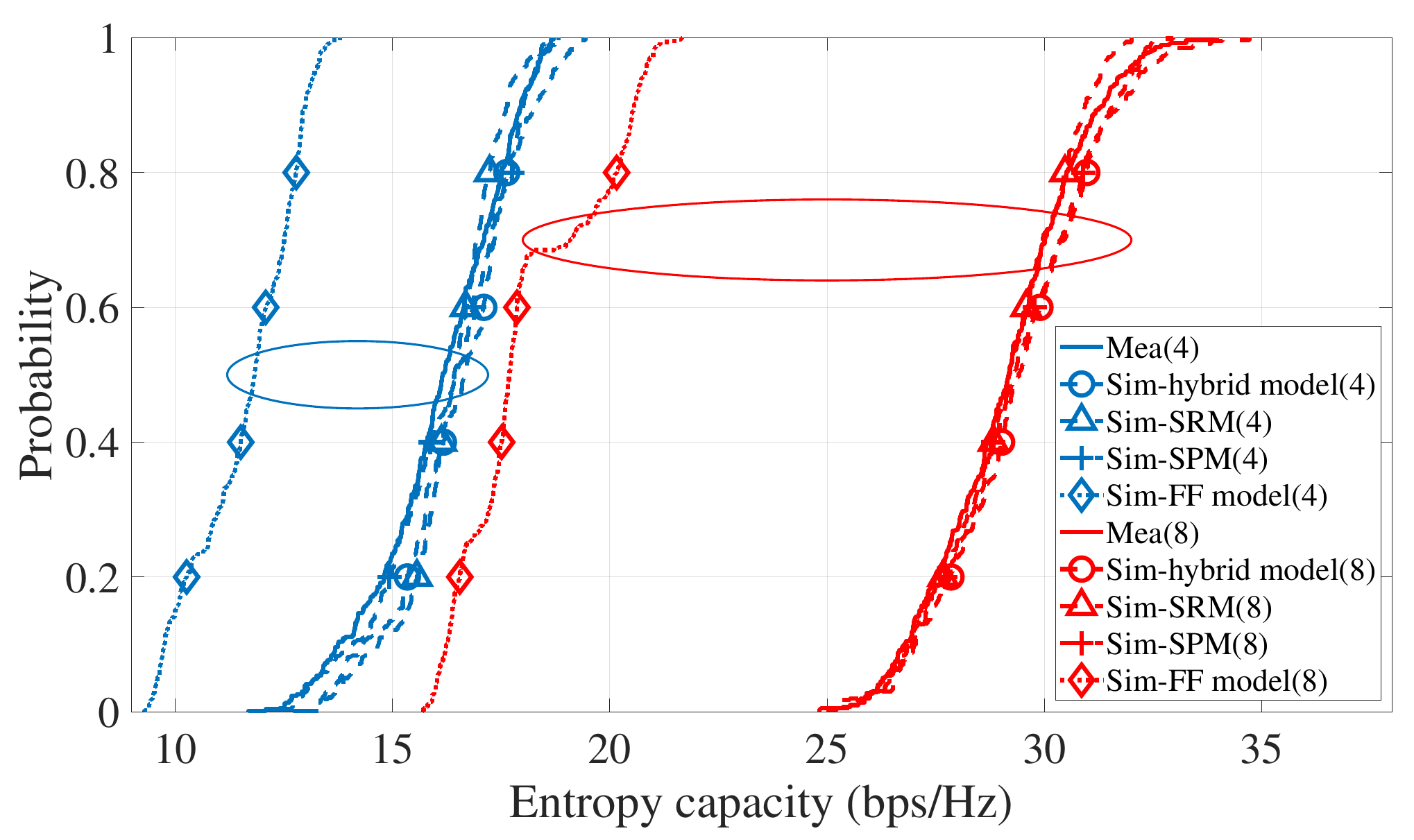 }  
    \label{Fig17_a}
    } 
    \\
    \vspace{-1em}
    \subfloat[]
    {\includegraphics[width=8cm]{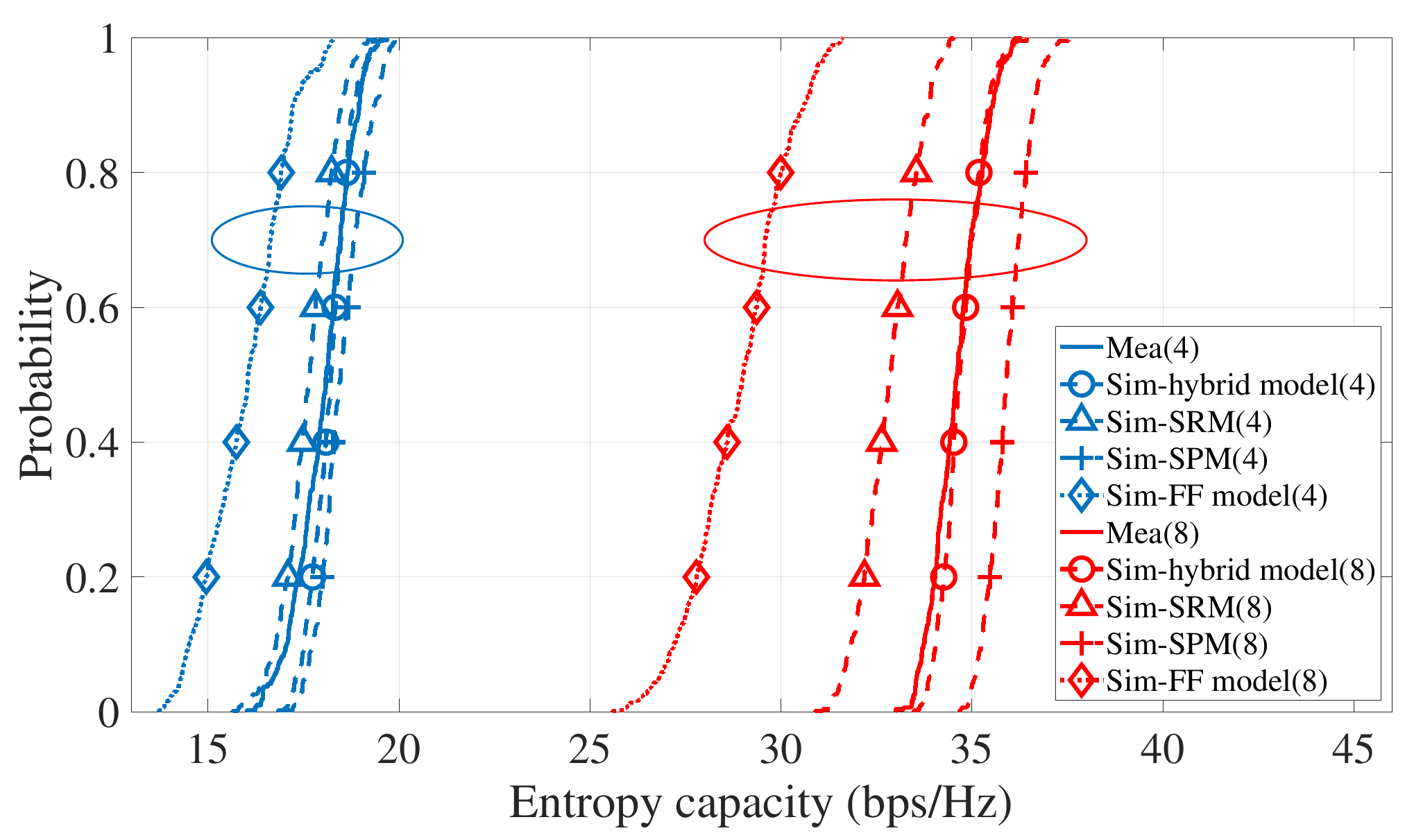 }  
    \label{Fig17_b}
    } 
    \caption{Comparison of entropy capacity for \(N = 4, 8\): Measured, hybrid NF, SRM, SPM, and FF models in (a) LoS and (b) NLoS conditions.} %
    \vspace{-0.3em}
    \label{Fig_Veri_capacity}  
\end{figure}

\textcolor{black}{In the simulation, \( N = 4, 8 \) UEs are randomly selected from measured or simulated datasets, and the entropy capacity is computed accordingly. This selection and computation process is repeated 800 times to obtain CDF. Fig. \ref{Fig17_a} compares the entropy capacity between simulated and measured channels under different UE numbers. It is observed that the FF-based simulation significantly underestimates the entropy capacity. This highlights the limitation of FF models in capturing the true spatial capacity in indoor NF scenarios. In contrast, the hybrid NF model, along with the SRM- and SPM-based NF simulations, yields results that closely match the measurements, confirming the effectiveness of the NF modeling approach. However, due to the dominant power contribution of the LoS path in THz channels, the differences in NLoS path modeling among the three NF models are less readily distinguishable. To further investigate this, we remove the LoS path components from the measured responses and recompute the entropy capacities for each model, as shown in Fig. \ref{Fig17_b}.
After removing the LoS contributions, it can be observed that the traditional SPM model overestimates the entropy capacity due to exaggerated nonlinear phase variations in certain NLoS paths, while the SRM model underestimates it by underrepresenting such variations. In contrast, the proposed hybrid NF channel model achieves a much closer match with the measurements.}

\graphicspath{{picture/}}

\begin{itemize}
    \item \textcolor{black}{Condition number}
\end{itemize}

\textcolor{black}{The Demmel condition number is a widely used metric to assess the structure of multi-antenna channel matrices. It quantifies the energy distribution across spatial dimensions and indicates whether spatial multiplexing or diversity is more favorable. A low value suggests well-balanced, orthogonal channels suitable for multiplexing, while a high value implies spatial correlation or imbalance, favoring diversity. It is defined as:}

\begin{figure}[htbp]  
    \centering  
     \subfloat[]
    {\includegraphics[width=8cm]{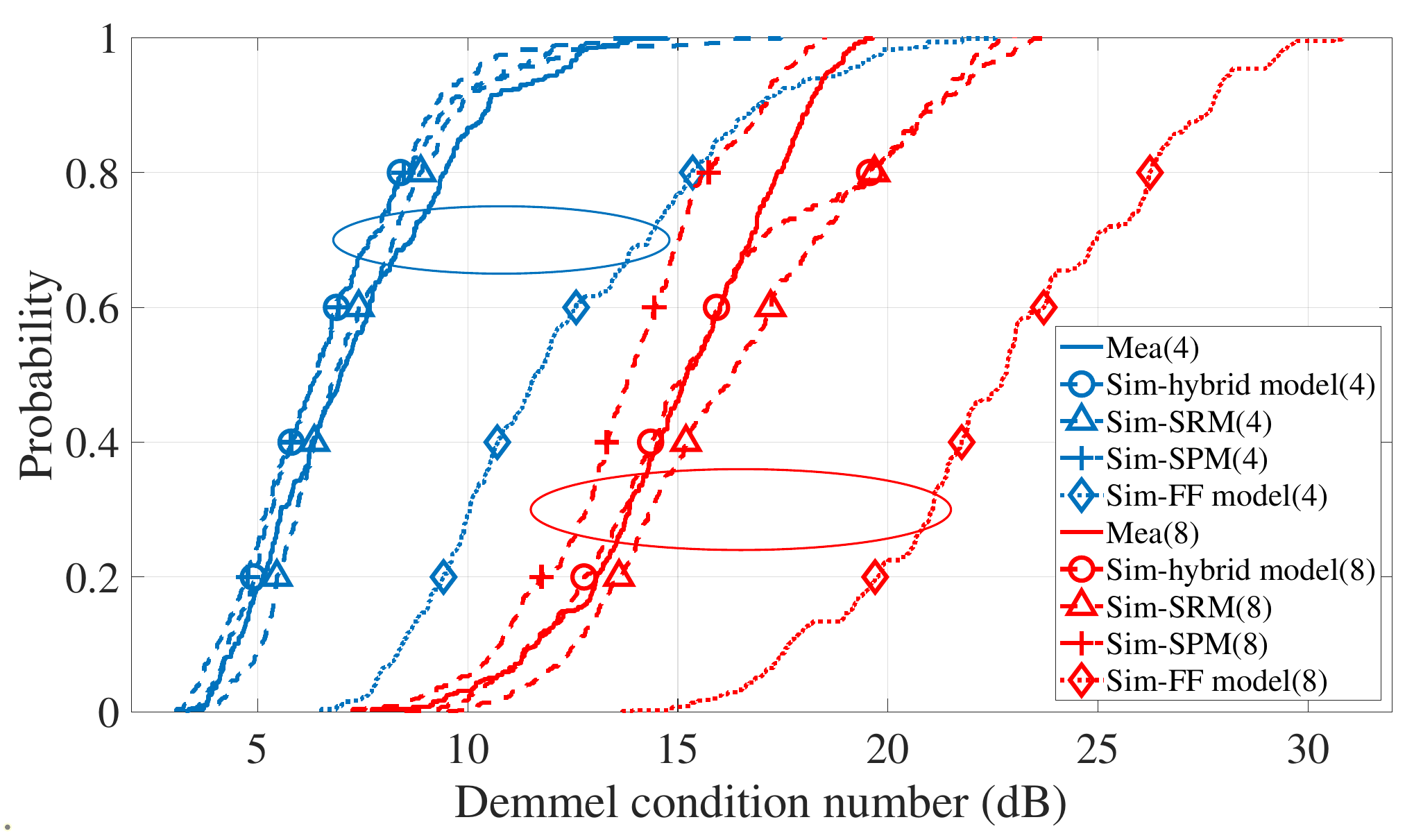 }  
    \label{Fig18_a}
    } 
    \\
    \vspace{-1em}
    \subfloat[]
    {\includegraphics[width=8cm]{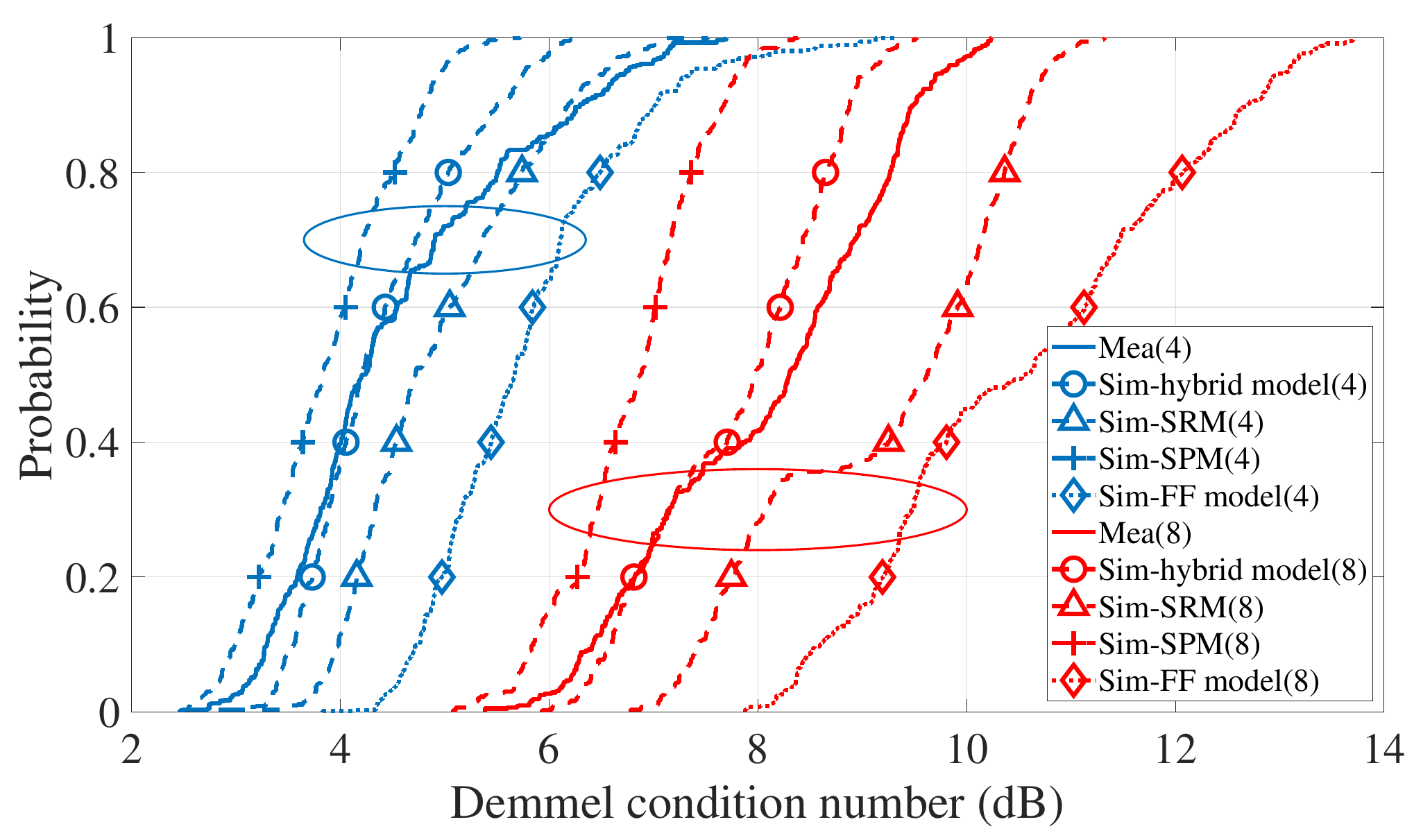 }  
    \label{Fig18_b}
    } 
    \caption{Demmel condition number comparison for \(N = 4, 8\): measured, hybrid NF, SRM, SPM, and FF models in (a) LoS and (b) NLoS conditions.} %
    \vspace{-1em}
    \label{Fig_Condition_num}  
\end{figure}

\textcolor{black}{
\begin{equation}
    \kappa  = \frac{1}{{{K}}}\sum\limits_{k = 1}^{{K}} {\frac{{{{\left\| {{\bf{H}}\left( {{f_k}} \right)} \right\|}_F}}}{{{\sigma _{\min }}\left( {{\bf{H}}\left( {{f_k}} \right)} \right)}}},
\end{equation}
where \( \sigma_{\min}(\cdot) \) is the smallest singular value, and \(\left\| \cdot \right\|_F\) is the Frobenius norm. Fig. \ref{Fig18_a} shows a comparison of condition numbers between the simulated and measured channels under varying numbers of UEs.}

\textcolor{black}{Among the models, the FF model consistently and significantly overestimates the condition number, reflecting an overly pessimistic assessment of spatial multiplexing capability. In comparison, the condition numbers obtained from the hybrid NF model show a closer alignment with the measurements, whereas the SPM model tends to underestimate them, and the SRM model tends to overestimate them.} \textcolor{black}{Similarly, a comparison of the Demmel condition number across different models after removing the LoS path is presented in Fig. \ref{Fig18_b}. It can be observed that the proposed hybrid NF model exhibits good agreement with the measurement results.}

\graphicspath{{picture/}}

\begin{table}[]
\renewcommand\arraystretch{1.2}
\centering
\caption{Cramér–von Mises distance between measured and simulated distributions of entropy capacity and Demmel condition number, using hybrid NF, SRM, SPM, and FF channel models.}

\begin{tabular}{m{0.5cm}<{\centering}|m{1.6cm}<{\centering}|m{1cm}<{\centering}|m{1cm}<{\centering}|m{1.1cm}<{\centering}|m{1.1cm}<{\centering}}
\hline
\multirow{2}{*}{Con.}  &{Parameter} & \multicolumn{2}{c|}{Capacity (bps/Hz)}  & \multicolumn{2}{c}{Condition number (dB)}\\
\cline{2-6}
 & UE number &4 & 8  & 4 & 8 \\ 
     \hline
    \multirow{4}{*}{LoS} &   Hybrid model & 0.02 & 0.01 & 0.03& 0.12 \\ \cline{2-6}
     \multirow{4}{*}{} &  SRM &  0.04 & 0.01 & 0.03& 0.23 \\ \cline{2-6}
    \multirow{4}{*}{}   &  SPM & 0.02 & 0.01 & 0.06& 0.18  \\ \cline{2-6}
     \multirow{4}{*}{}  &  FF model &  2.92 & 9.17 & 1.99 & 4.25  \\ \hline

         \multirow{4}{*}{NLoS} &   Hybrid model & 0.02 & 0.01 & 0.02& 0.04 \\ \cline{2-6}
     \multirow{4}{*}{} &  SRM &  0.06 & 1.01 & 0.08& 0.33 \\ \cline{2-6}
    \multirow{4}{*}{}   &  SPM & 0.08 & 0.67 & 0.09& 0.48  \\ \cline{2-6}
     \multirow{4}{*}{}  &  FF model &  1.06 & 4.72 & 0.49 & 1.09  \\ \hline
\end{tabular}
\label{Tab_Veri_NF}
\end{table}

\begin{itemize}
    \item Path gain 
\end{itemize}

\begin{figure}[htbp]  
    \centering  
    \includegraphics[width=8.2cm]{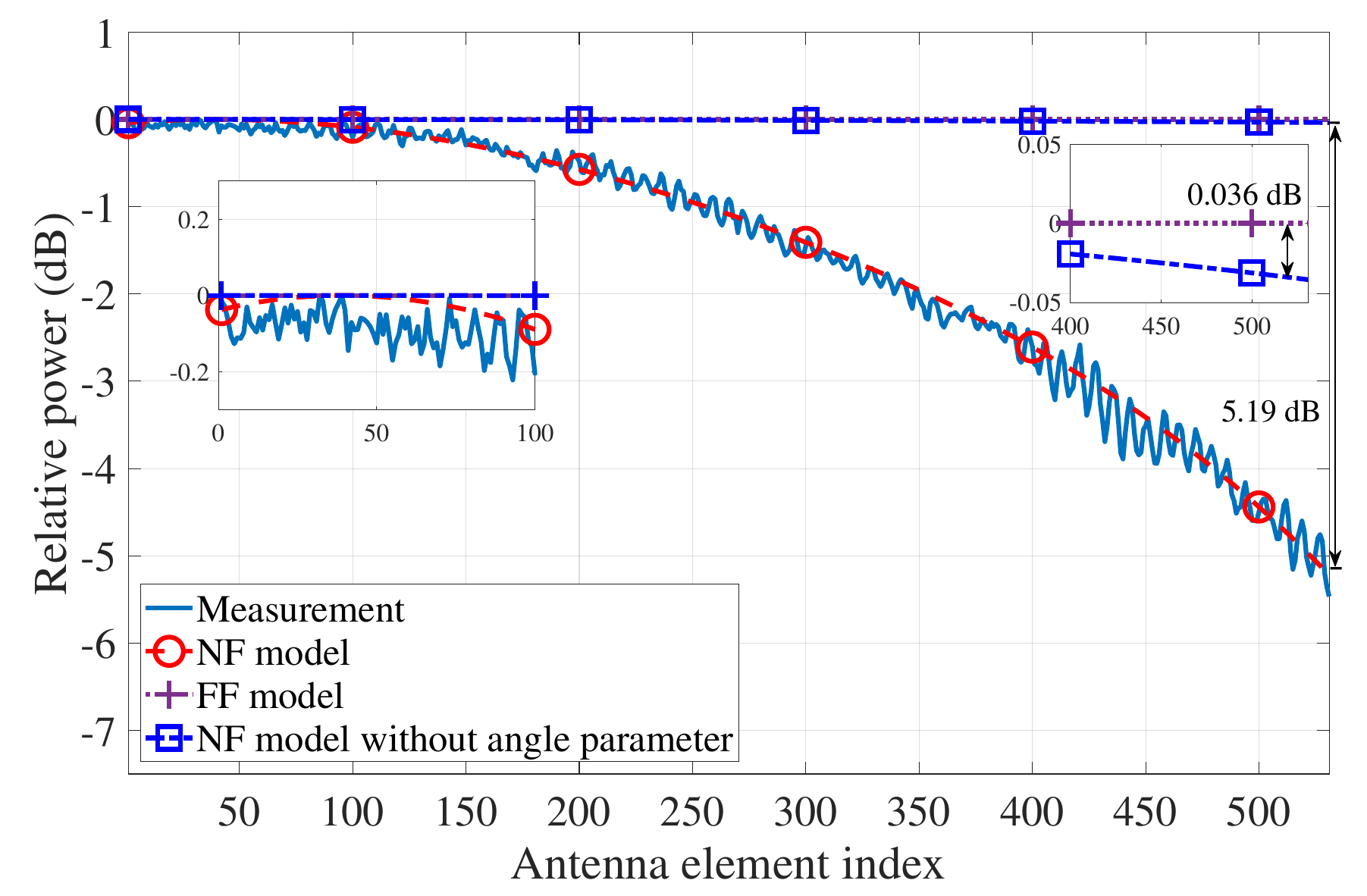}  
    \caption{Validate the proposed model from the NF propagation perspective by comparing the LoS path measurement results with the NF model, FF model, and NF model without angle parameter.} %
    \vspace{-1em}
    \label{Fig15_LoS}  
\end{figure}

The necessity of accounting for element radiation patterns and the spatial variation of multipath angles across the array in the proposed NF model is demonstrated using the LoS path in Case 4. Under the FF assumption, power differences between antenna elements are typically negligible,. However, in the NF region with spherical wavefronts, the power of each path must be calculated individually for each antenna element. Fig. \ref{Fig15_LoS} compares the measured LoS path power with three models: the proposed NF model, the FF model, and the NF model that omits angular parameters. The results show that the proposed NF model closely matches the measurements, while the FF model fails to capture the observed power variations. Omitting angular parameters or antenna patterns in the NF model reduces its accuracy to that of the FF model, leading to significant deviations. This discrepancy arises from substantial AoA and AoD variations across the array, which cause differing directional responses among antenna elements. These observations underscore the necessity of accurately modeling the combined effects of directional antenna responses and angle variations across the array in THz XL-MIMO systems under NF propagation.

\graphicspath{{picture/}}

\graphicspath{{picture/}}
\begin{figure}[htbp]  
    \centering  
    \includegraphics[width=8.2cm]{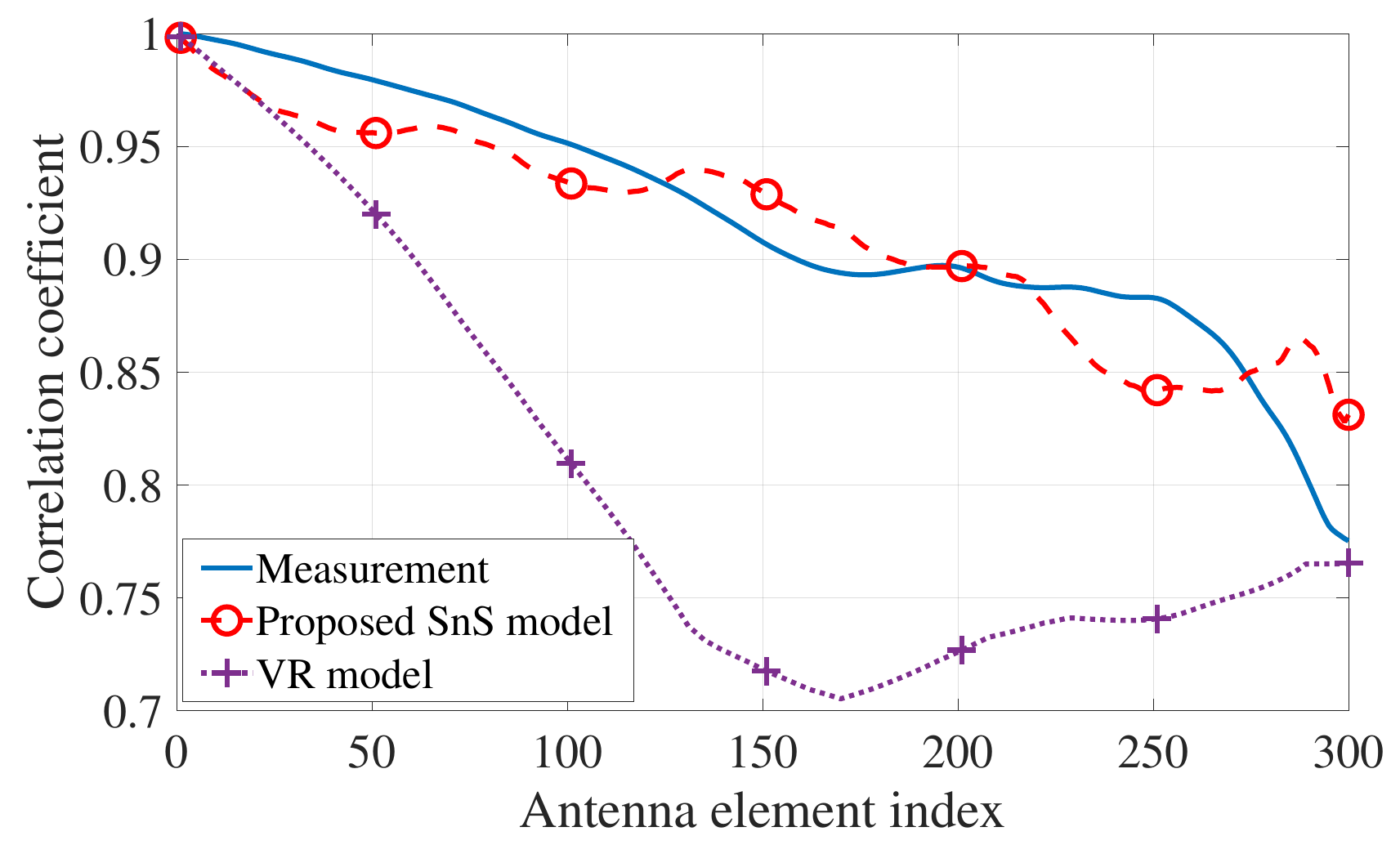}
    \caption{Comparison of average spatial correlation between the simulated and measured SnS channels.} %
    \label{Fig_Veri_SnS_Corr}
    \vspace{-0.5em}
\end{figure}

\subsubsection{Spatial non-stationary validation}

The SnS model can be validated by evaluating the average channel correlation, channel gain, Rician K-factor, and RMS delay spread metrics.

\begin{itemize}
    \item Average spatial correlation
\end{itemize}

\textcolor{black}{
The accuracy of the generated AAF is evaluated by analyzing the average spatial correlation of the SnS channel, which is defined as:
\begin{equation}
    \bar\rho(\Delta x) = \frac{1}{M - \Delta x}\;\sum_{i=1}^{M-\Delta x}
\mathrm{corr}\bigl(\mathbf{H}_{i,:}^{\text{SnS}}(f),\;\mathbf{H}_{i+\Delta x,:}^{\text{SnS}}(f)\bigr),
\end{equation}
where \( \bar\rho(\Delta x) \) denotes the average correlation between antenna elements separated by \(\Delta x\) positions. The matrix \( \mathbf{H}^{\text{SnS}}(f) \in \mathbb{C}^{M \times L} \) represents the path-wise amplitude response of all \(L\) paths across the \(M\) antenna elements, and is computed as:
\begin{equation}
 \mathbf{H}^{\text{SnS}}(f) = \left| \mathbf{S} \cdot \mathrm{diag} \left(\mathbf{H}_{\text{ref}}(f) \right) \right|.
\end{equation}
where \(\mathrm{diag}(\cdot)\) transforms the \(L \times 1\) vector \(\mathbf{H}_{\text{ref}}(f)\) into an \(L \times L\) diagonal matrix. The operator \(\mathrm{corr}(\cdot)\) computes the Pearson correlation coefficient between the amplitude vectors of two antenna elements across all paths. As shown in Fig. \ref{Fig_Veri_SnS_Corr}, the simulated SnS channel exhibits spatial correlation trends that closely match the measured results. In contrast, the conventional VR model assumes binary visibility and fails to account for continuous variations in path amplitude. Consequently, it tends to underestimate the overall spatial correlation, leading to reduced statistical consistency with measurement data.}

\begin{itemize}
    \item Channel gain, Rician K-factor, and RMS delay spread.
\end{itemize}

\graphicspath{{picture/}}
\begin{figure*}[htbp]  
    \centering  
    \subfloat[]{
    \includegraphics[width=6.1cm]{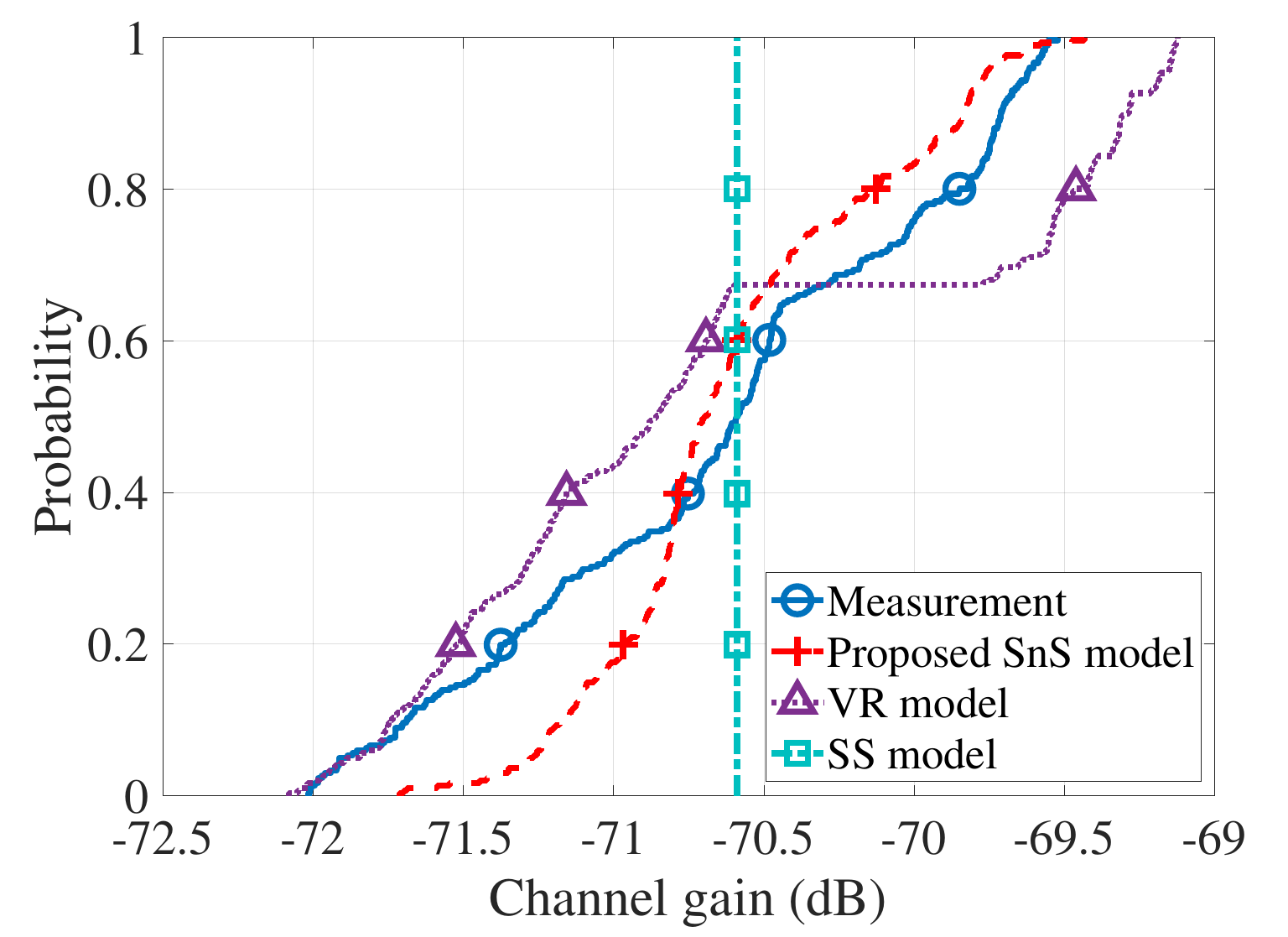}
    \label{Fig_sub_Veri_a}
    }
    \subfloat[]{
    \includegraphics[width=6.1cm]{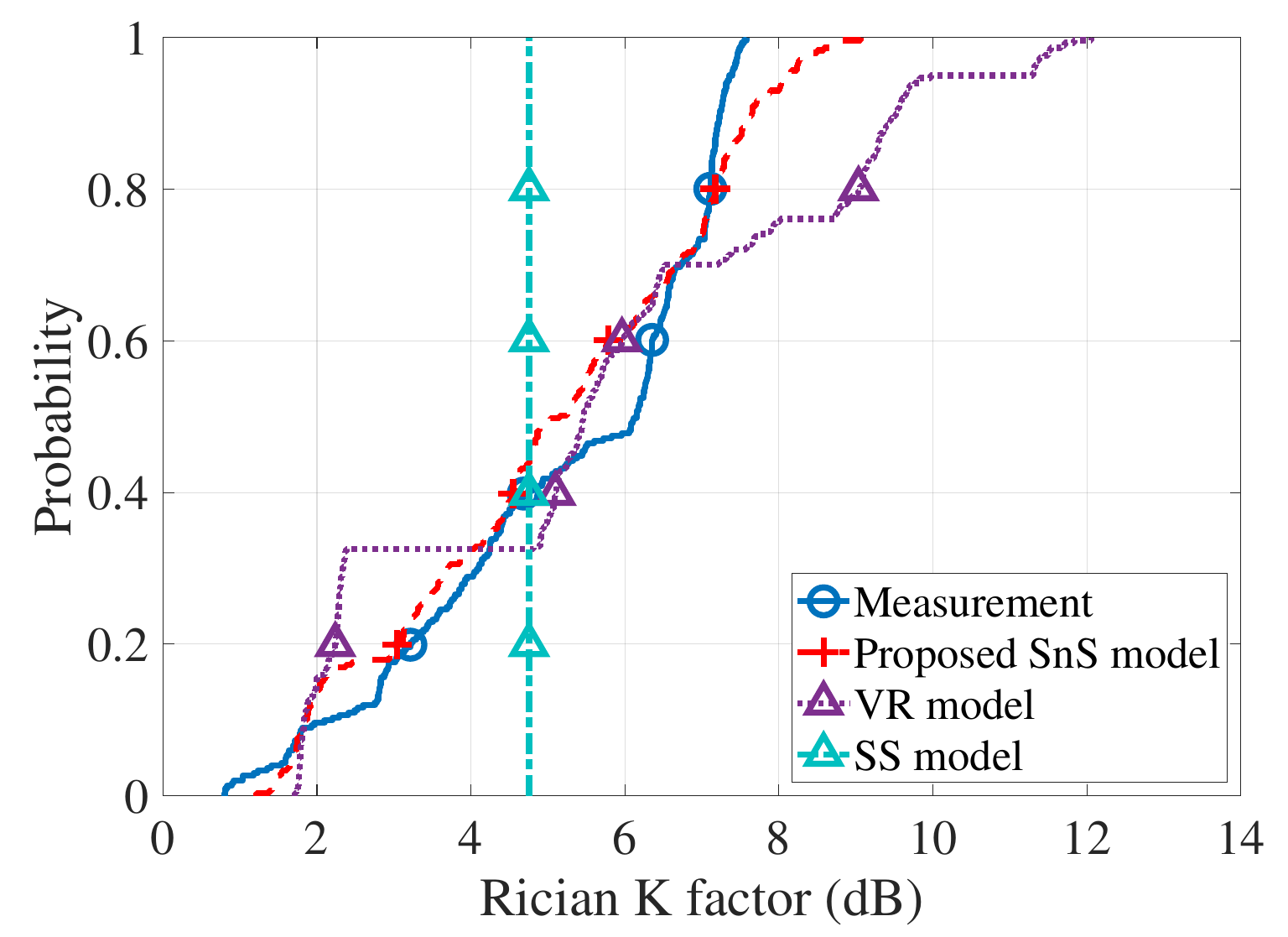}
    }
    \subfloat[]{
        \includegraphics[width=6.1cm]{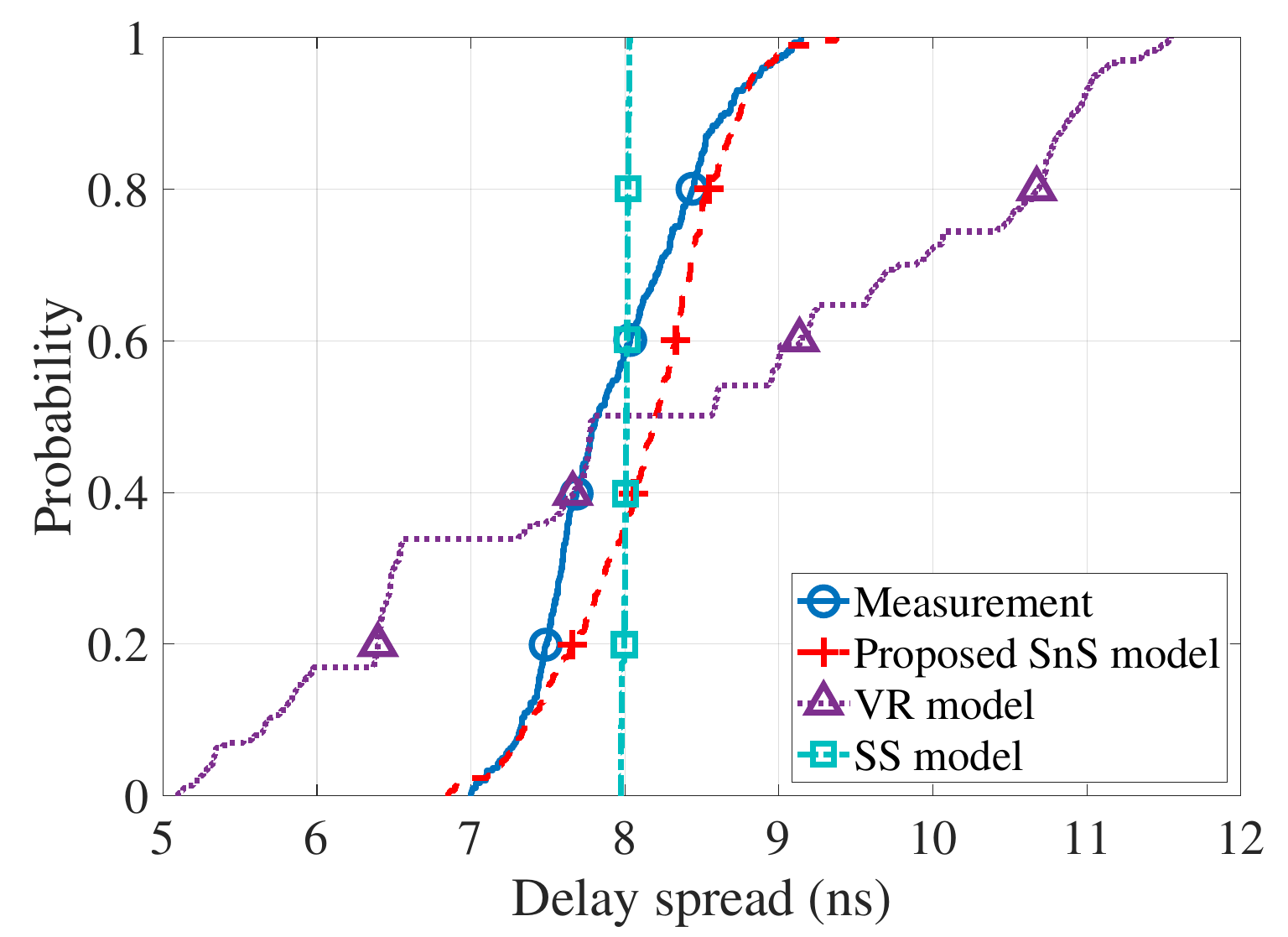}}
    \caption{CDFs of channel statistical parameters across all XL-MIMO elements in Case 3, comparing measurement results with model simulations: (a) Channel gain, (b) Rician K-factor, (c) RMS delay spread.}
    \vspace{-1em}
    \label{Fig_Veri_LSP}  
\end{figure*}

These parameters respectively characterize the average channel power attenuation, the power ratio between the LoS path and the remaining NLoS components, and the delay dispersion of multipath signals. As shown in Fig. \ref{Fig_Veri_LSP}, the proposed SnS model achieves close agreement with the measured results across all three metrics, whereas both the VR and SS models exhibit noticeable discrepancies. The corresponding quantitative results, summarized in Table \ref{Tab_Veri_SNS}, further confirm the improved accuracy of the proposed model. A closer examination reveals that the VR model not only shows larger deviations from the empirical distributions but also produces abrupt statistical discontinuities—such as sudden jumps in channel gain near -70.5 dB or in the K-factor around 3 dB—which are rarely observed in real-world measurements. This observation suggests that the VR model may be unsuitable for accurately modeling practical channel behavior. As for the SS model, it assumes equal path power across the XL-MIMO array, leading to fixed metric values that do not vary with antenna position. This assumption fails to capture the dispersion observed in the measurement data.

\begin{table}[]
\renewcommand\arraystretch{1.2}
\centering
\caption{Cramér–von Mises distance between measured and simulated distributions of channel gain, K-factor, and delay spread.}
\begin{tabular}{m{1.6cm}<{\centering}|m{1.8cm}<{\centering}|m{1.8cm}<{\centering}|m{1.8cm}<{\centering}}
\hline
Parameter & Channel gain (dB) &Rician K-factor (dB) & Delay spread (ns) \\
     \hline
        Proposed SnS & 0.027 & 0.030 & 0.035  \\ \hline
        VR model &  0.042 & 0.177 & 0.385  \\ \hline
        SS model& 0.178 & 0.710 & 0.143  \\ \hline
\end{tabular}
\label{Tab_Veri_SNS}
\end{table}

\section{Conclusion}
\label{Sec_5}

This paper presents an empirical and modeling study of THz XL-MIMO channels with a focus on NF propagation and SnS characteristics. First, channel measurements were conducted \textcolor{black}{at 100 GHz with 301-element} and at 132 GHz with 531-element linear arrays. \textcolor{black}{The results reveal that, due to NF propagation, MPC parameters—including power, delay, and angle—vary across antenna elements, and the phase exhibits a nonlinear trend along the array. The measurements also highlight that, apart from blockage, SnS can arise from inconsistent reflection and scattering effects.} Then, a THz XL-MIMO channel modeling framework is established. Within this framework, \textcolor{black}{a hybrid NF model is developed by combining the SPM and SRM, enabling accurate characterization of the nonlinear phase evolution of NLoS paths under different propagation mechanisms. For SnS modeling, an AAF defined over the range [0,1] is introduced to represent the continuous power variation of SnS paths across the array. The statistical distribution and spatial autocorrelation properties of the AAF are thoroughly analyzed, and a rank-matching-based statistical AAF generation method is proposed.} Finally, the implementation and validation of the proposed model are detailed. Results show strong agreement between simulations and measurements in terms of \textcolor{black}{entropy capacity, Demmel condition number, path gain, average spatial correlation of SnS channels, and various statistical metrics, including channel gain, Rician K-factor, and RMS delay spread.} Both measurement and simulation results confirm that accurately modeling NF spherical-wave propagation and SnS characteristics is essential for achieving high-fidelity THz XL-MIMO channel modeling.

\bibliographystyle{IEEEtran}
\bibliography{paper}

\begin{thebibliography}{10}
\providecommand{\url}[1]{#1}
\csname url@samestyle\endcsname
\providecommand{\newblock}{\relax}
\providecommand{\bibinfo}[2]{#2}
\providecommand{\BIBentrySTDinterwordspacing}{\spaceskip=0pt\relax}
\providecommand{\BIBentryALTinterwordstretchfactor}{4}
\providecommand{\BIBentryALTinterwordspacing}{\spaceskip=\fontdimen2\font plus
\BIBentryALTinterwordstretchfactor\fontdimen3\font minus \fontdimen4\font\relax}
\providecommand{\BIBforeignlanguage}[2]{{%
\expandafter\ifx\csname l@#1\endcsname\relax
\typeout{** WARNING: IEEEtran.bst: No hyphenation pattern has been}%
\typeout{** loaded for the language `#1'. Using the pattern for}%
\typeout{** the default language instead.}%
\else
\language=\csname l@#1\endcsname
\fi
#2}}
\providecommand{\BIBdecl}{\relax}
\BIBdecl

\bibitem{C1_6G_survey}
J.~Zhang, J.~Lin, P.~Tang \emph{et~al.}, ``{Channel measurement, modeling, and simulation for 6G: A survey and tutorial},'' \emph{arXiv preprint arXiv:2305.16616}, 2023.

\bibitem{C1_THz_Rappaport}
T.~S. Rappaport, Y.~Xing, O.~Kanhere \emph{et~al.}, ``{Wireless communications and applications above 100 GHz: opportunities and challenges for 6G and beyond},'' \emph{{IEEE Access}}, vol.~7, pp. 78\,729--78\,757, Jun. 2019.

\bibitem{C1_THz_Survey}
C.~Han, Y.~Wang, Y.~Li \emph{et~al.}, ``{Terahertz wireless channels: a holistic survey on measurement, modeling, and analysis},'' \emph{IEEE Commun. Surv. Tutor.}, vol.~24, no.~3, pp. 1670--1707, third quarter 2022.

\bibitem{C1_THz_ITU_RR}
\BIBentryALTinterwordspacing
``{Radio regulations},'' World Radiocommunication Conference, 2020. [Online]. Available: \url{https://www.itu.int/pub/R-REG-RR}
\BIBentrySTDinterwordspacing

\bibitem{C1_6G}
J.~Zhang, P.~Tang, L.~Yu \emph{et~al.}, ``{Channel measurements and models for 6G: current status and future outlook},'' \emph{Front. Inf. Technol. Electron. Eng.}, vol.~21, no.~1, pp. 39--61, Mar. 2020.

\bibitem{Molisch_book}
A.~F. Molisch, \emph{Wireless communications}, 2nd~ed.\hskip 1em plus 0.5em minus 0.4em\relax John Wiley \& Sons, 2012.

\bibitem{C1_XL_MIMO}
P.~Tang, J.~Zhang, H.~Miao \emph{et~al.}, ``{XL-MIMO channel measurement, characterization, and modeling for 6G: a survey},'' \emph{Front. Inf. Technol. Electron. Eng.}, 2024, accepted.

\bibitem{C1_Rayleigh}
A.~Yaghjian, ``{An overview of near-field antenna measurements},'' \emph{IEEE Trans. Antennas and Propag.}, vol.~34, no.~1, pp. 30--45, Jan. 1986.

\bibitem{C1_HC_WC}
C.~Han, Y.~Chen, L.~Yan \emph{et~al.}, ``{Cross far- and near-field wireless communications in terahertz ultra-large antenna array systems},'' \emph{IEEE Wirel. Commun.}, vol.~31, no.~3, pp. 148--154, Jun. 2024.

\bibitem{C1_VR_3}
J.~Flordelis, X.~Li, O.~Edfors \emph{et~al.}, ``{Massive MIMO extensions to the COST 2100 channel model: modeling and validation},'' \emph{IEEE Trans. Wireless Commun.}, vol.~19, no.~1, pp. 380--394, Jan. 2020.

\bibitem{C1_ICC_XHX}
H.~Xu, P.~Tang, J.~Zhang \emph{et~al.}, ``{An empirical study on near-field, spatial non-stationarity, and beam misalignment characteristics of THz XL-MIMO channels at 132 GHz},'' in \emph{Proc. IEEE Int. Conf. Commun. Workshops (ICC Workshops)}, Jun. 2024, pp. 744--749.

\bibitem{C1_THz_MIMO_Mea_1}
S.~Singh, T.~Le, and H.~Tran, ``{Measurement of 2x2 LoS terahertz MIMO channel},'' in \emph{in Proc. IEEE Wireless Commun. Netw. Conf. (WCNC)}, May 2020, pp. 1--5.

\bibitem{C1_THz_MIMO_Mea_3}
N.~Khalid and O.~B. Akan, ``{Wideband THz communication channel measurements for 5G indoor wireless networks},'' in \emph{Proc. IEEE Int. Conf. Commun. (ICC)}, May 2016, pp. 1--6.

\bibitem{C1_THz_MIMO_Mea_4}
S.~Priebe, M.~Kannicht, M.~Jacob \emph{et~al.}, ``{Ultra broadband indoor channel measurements and calibrated ray tracing propagation modeling at THz frequencies},'' \emph{J. Commun. Netw.}, vol.~15, no.~6, pp. 547--558, Dec. 2013.

\bibitem{C1_THz_MIMO_Mea_8}
Y.~Lyu, Z.~Yuan, P.~Zhang \emph{et~al.}, ``{Large virtual antenna array-based empirical channel characterization for sub-THz indoor hall scenarios},'' \emph{{IEEE Trans. Antennas Propag.}}, pp. 1--1, 2024, early access.

\bibitem{C1_THz_MIMO_Mea_6}
J.~Zhang, J.~Lin, P.~Tang \emph{et~al.}, ``{Deterministic ray tracing: a promising approach to THz channel modeling in 6G deployment scenarios},'' \emph{IEEE Commun. Mag.}, vol.~62, no.~2, pp. 48--54, Feb. 2024.

\bibitem{C1_THz_MIMO_Mea_5}
Y.~Wang, S.~Sun, and C.~Han, ``{Far- and near-field channel measurements and characterization in the terahertz band using a virtual antenna array},'' \emph{IEEE Commun. Lett.}, vol.~28, no.~5, pp. 1186--1190, Feb. 2024.

\bibitem{C3_YZQ}
Z.~Yuan, J.~Zhang, Y.~Ji \emph{et~al.}, ``{Spatial non-stationary near-field channel modeling and validation for massive MIMO systems},'' \emph{IEEE Trans. Antennas Propag.}, vol.~71, no.~1, pp. 921--933, Jan. 2023.

\bibitem{C1_MIMO_Mea_1}
G.~Jing, J.~Hong, X.~Yin \emph{et~al.}, ``{Measurement-based 3-D channel modeling with cluster-of-scatterers estimated under spherical-wave assumption},'' \emph{{IEEE Trans. Wirel. Commun.}}, vol.~22, no.~9, pp. 5828--5843, Sep. 2023.

\bibitem{C1_HC_ARXIV}
Y.~Wang, C.~Han, S.~Sun \emph{et~al.}, ``{Cross far-and near-field channel measurement and modeling in extremely large-scale antenna array (ELAA) systems},'' \emph{arXiv preprint arXiv:2405.16893}, 2024.

\bibitem{C1_VR_1}
J.~Li, B.~Ai, R.~He \emph{et~al.}, ``{On 3D cluster-based channel modeling for large-scale array communications},'' \emph{IEEE Trans. Wireless Commun.}, vol.~18, no.~10, pp. 4902--4914, Oct. 2019.

\bibitem{C1_cluster_LR}
J.~Chen, X.~Yin, X.~Cai \emph{et~al.}, ``{Measurement-based massive MIMO channel modeling for outdoor LoS and NLoS environments},'' \emph{IEEE Access}, vol.~5, pp. 2126--2140, Jan. 2017.

\bibitem{C1_GTY}
T.~Gao, P.~Tang, L.~Tian \emph{et~al.}, ``{A 3GPP-like channel simulation framework considering near-field spatial non-stationary characteristics of massive MIMO},'' in \emph{Proc. IEEE Global Commun. Workshops (GC Wkshps)}, Dec. 2023, pp. 1493--1498.

\bibitem{C1_VR_APP_1}
A.~Tang, J.-B. Wang, Y.~Pan \emph{et~al.}, ``Joint visibility region and channel estimation for extremely large-scale mimo systems,'' \emph{IEEE Trans. Commun.}, vol.~72, no.~10, pp. 6087--6101, Oct. 2024.

\bibitem{C1_VR_APP_2}
Y.~Han, S.~Jin, C.-K. Wen \emph{et~al.}, ``Channel estimation for extremely large-scale massive mimo systems,'' \emph{IEEE Wireless Commun. Lett.}, vol.~9, no.~5, pp. 633--637, May 2020.

\bibitem{C1_VR_APP_3}
B.~Xu, J.~Zhang, J.~Li \emph{et~al.}, ``Jac-pcg based low-complexity precoding for extremely large-scale mimo systems,'' \emph{IEEE Trans. Veh. Technol.}, vol.~72, no.~12, pp. 16\,811--16\,816, Dec. 2023.

\bibitem{C1_VR_APP_4}
Y.~Chen and L.~Dai, ``Non-stationary channel estimation for extremely large-scale mimo,'' \emph{IEEE Trans. Wireless Commun.}, vol.~23, no.~7, pp. 7683--7697, Jul. 2024.

\bibitem{C1_SaLi}
J.~Liu \emph{et~al.}, ``{A spatially non-stationary fading channel model for simulation and (semi-) analytical study of ELAA-MIMO},'' \emph{IEEE Trans. Wireless Commun.}, vol.~23, no.~5, pp. 5203--5218, May 2024.

\bibitem{C1_YZQ_RT}
Z.~Yuan, J.~Zhang, V.~Degli-Esposti \emph{et~al.}, ``Efficient ray-tracing simulation for near-field spatial non-stationary mmwave massive mimo channel and its experimental validation,'' \emph{IEEE Trans. Wireless Commun.}, vol.~23, no.~8, pp. 8910--8923, Aug. 2024.

\bibitem{C1_THz_Dir}
D.~M. Bodet and J.~M. Jornet, ``{Directional antennas for sub-THz and THz MIMO systems: bridging the gap between theory and implementation},'' \emph{IEEE Open J. Commun. Soc.}, vol.~4, pp. 2261--2273, 2023.

\bibitem{C1_THz_SNS}
Y.~Lyu, Z.~Yuan, F.~Zhang \emph{et~al.}, ``Virtual antenna array for w-band channel sounding: design, implementation, and experimental validation,'' \emph{IEEE J. Sel. Top. Signal Proces.}, vol.~17, no.~4, pp. 729--744, Jul. 2023.

\bibitem{C2_UPD}
H.~Lu and Y.~Zeng, ``{Communicating with extremely large-scale array/surface: unified modeling and performance analysis},'' \emph{IEEE Trans. Wireless Commun.}, vol.~21, no.~6, pp. 4039--4053, Jun. 2021.

\bibitem{C3_Kurner}
R.~Piesiewicz, C.~Jansen, D.~Mittleman \emph{et~al.}, ``{Scattering analysis for the modeling of THz communication systems},'' \emph{IEEE Trans. Antennas Propag.}, vol.~55, no.~11, pp. 3002--3009, Nov. 2007.

\bibitem{C3_Corr_Sunshu}
S.~Sun, H.~Yan, G.~R. MacCartney \emph{et~al.}, ``{Millimeter wave small-scale spatial statistics in an urban microcell scenario},'' in \emph{Proc. IEEE Int. Conf. Commun. (ICC)}, 2017, pp. 1--7.

\bibitem{C3_Coupla}
R.~B. Nelsen, \emph{An introduction to copulas}.\hskip 1em plus 0.5em minus 0.4em\relax Springer, 2006.

\bibitem{C4_Veri_Rapp}
S.~Ju, Y.~Xing, O.~Kanhere \emph{et~al.}, ``{Millimeter wave and sub-terahertz spatial statistical channel model for an indoor office building},'' \emph{IEEE J. Sel. Areas Commun.}, vol.~39, no.~6, pp. 1561--1575, Jun. 2021.

\bibitem{C3_PSS}
X.~Gao, O.~Edfors, F.~Rusek \emph{et~al.}, ``{Massive MIMO performance evaluation based on measured propagation data},'' \emph{IEEE Trans. Wireless Commun.}, vol.~14, no.~7, pp. 3899--3911, Jul. 2015.

\end{thebibliography}

\end{document}